\documentclass[aps,prx,twocolumn,floatfix,nofootinbib,amsmath,amssymb,preprintnumbers]{revtex4-2}

\usepackage{dcolumn}
\usepackage{bm}
\usepackage{booktabs}
\usepackage{color,verbatim}
\usepackage{braket,dsfont}
\usepackage{physics, wasysym }
\usepackage[colorlinks=true,citecolor=blue,filecolor=blue,linkcolor=blue,urlcolor=blue,pdftex]{hyperref}
\usepackage{graphicx}
\usepackage{booktabs}
\usepackage{gensymb}
\usepackage[normalem]{ulem}
\usepackage{soul}

\newcommand{\DarkSphere}{{\sc DarkSphere }}
\newcommand{\tjW}[6]{ \begin{pmatrix}
   #1 & #2 & #3 \\
   #4 & #5 & #6 
  \end{pmatrix}}

\begin{document}

\preprint{KCL-PH-TH-2021-31}

\title{Fuelling the search for light dark matter--electron scattering\\ with spherical proportional counters}

\author{Louis Hamaide}
 \email{louis.hamaide@kcl.ac.uk}
\author{Christopher McCabe}
\affiliation{Theoretical Particle Physics and Cosmology Group, Department of Physics, King's College London, Strand, London, WC2R 2LS, United Kingdom}%

\begin{abstract}
Dark matter (DM) detectors employing a Spherical Proportional Counter (SPC) have demonstrated a single-electron detection threshold and are projected to have small background rates. 
We explore the sensitivity to DM--electron scattering with SPC detectors in the context of {\sc DarkSphere}, a proposal for a 300~\!cm diameter fully-electroformed SPC.
SPCs can run with different gases, so we investigate the sensitivity for five targets: helium, neon, xenon, methane, and isobutane. 
We use tools from quantum chemistry to model the atomic and molecular systems, and calculate the expected DM induced event rates.
We find that \DarkSphere has the potential to improve current exclusion limits on DM masses above 4~MeV by up to five orders of magnitude. 
Neon is the best all-round gas target and provides good sensitivity to scenarios with both light and heavy mediators. Gas mixtures, where methane or isobutane is added to a noble gas, can extend the sensitivity at lower masses.
Our study highlights the currently untapped potential of SPCs to search for DM-electron scattering in the MeV-to-GeV DM mass range.
\end{abstract}

\maketitle

\section{Introduction}\label{introduction}

Inferring more about the particle physics properties of dark matter (DM) remains one of the most important open challenges in astro-particle physics~\cite{APPEC,Bertone:2016nfn}.
It is in this context that the experimental program of DM direct detection operates, since if a signal from DM is observed
in a direct detection experiment, we will learn about the value of the DM mass and DM interaction with normal matter~\cite{Bertone:2004pz,Edwards:2018lsl,Billard:2021uyg}.
Traditionally, direct detection experiments have focused on the search for nuclear recoils induced by scattering from
DM particles with the canonical WIMP mass range of around 5-1000 GeV (see, e.g.,~\cite{Baudis:2015mpa, Undagoitia:2015gya,Schumann:2019eaa}).
However, the lack of any conclusive DM measurement has meant that during the past decade the mass range of interest has broadened
to encompass DM particles with a mass below and above this narrow mass window~\cite{Battaglieri:2017aum}.

The shift to searching for DM in the MeV to few-GeV mass range is challenging experimentally, and a number
of different avenues are being pursued.
The search for nuclear recoils induced by sub-5-GeV DM has continued by using detectors with significantly smaller energy-detection thresholds, 
often in combination with the utilisation of lighter target nuclei 
(see, e.g.,~\cite{Arnaud:2017usi, Agnese:2018gze, Abdelhameed:2019szb, Abdelhameed:2019hmk,  Alkhatib:2020slm, Aguilar-Arevalo:2020oii}).
Alternatively, observable signals from sub-5-GeV DM can be generated from rare processes that involve the emission of photons or `Migdal' electrons from the recoiling atom 
(see, e.g.,~\cite{Kouvaris:2016afs,McCabe:2017rln,Ibe:2017yqa,Dolan:2017xbu,Akerib:2018hck,Armengaud:2019kfj,Bell:2019egg, Liu:2019kzq,Baxter:2019pnz,Essig:2019xkx,Liang:2019nnx,Knapen:2020aky, Araujo:2022wjh, Cox:2022ekg}), or from the small flux of boosted DM
that is non-galactic in origin, or arises from interactions between DM and cosmic rays, the Sun, or mesons (see, e.g.,~\cite{Kouvaris:2015nsa, An:2017ojc, Emken:2017hnp, Bringmann:2018cvk,Ema:2018bih,Alvey:2019zaa,Cappiello:2019qsw,Herrera:2021puj}).

A complementary approach is to search for DM-electron scattering processes that excite or ionise an electron from
an atom or molecule, excite roto-vibrational states or break chemical bonds in molecules, or excite an electron across the band-gap in semi-conductors or other 
quantum materials (see, e.g.,~\cite{Essig:2011nj,Graham:2012su,Hochberg:2015pha,Essig:2015cda,Hochberg:2015fth,
Derenzo:2016fse,Essig:2016crl,Cavoto:2017otc,Hochberg:2017wce,Kurinsky:2019pgb,Essig:2019kfe, Geilhufe:2019ndy, Blanco:2021hlm, Kahn:2021ttr,Essig:2016crl,Essig:2019kfe}).
These searches have the advantage that only a small amount of energy, of order the ionisation energy or band-gap energy,
needs to be transferred to the electron to produce an observable signal.
There already exists an impressive range of existing constraints on the DM-electron cross section in the MeV-to-GeV mass range,
ranging from detectors designed specifically for this search channel to detectors that have been designed first and foremost
for nuclear recoil searches but have been repurposed to search for DM-electron 
scattering~\cite{Essig:2012yx, Essig:2017kqs, Agnes:2018oej, Agnese:2018col, Aprile:2019xxb, Aguilar-Arevalo:2019wdi, Blanco:2019lrf, Arnaud:2020svb, Barak:2020fql}.

It is in the latter context that we come to the main topic of this paper.
The {\it New Experiments With Spheres-Gas} (NEWS-G) detectors
employ a spherical proportional counter (SPC)~\cite{Giomataris:2003bp, Giomataris_2008} 
and can be filled with different species of atomic or molecular gases.
Limits on nuclear-recoil signals from sub-GeV DM have already been placed
with a 60~cm diameter SPC~\cite{Arnaud:2017bjh}, while a 140~cm diameter detector, `S140' (or sometimes {\sc SnoGlobe}), which utilises ultra-radiopure copper
to reduce the background rate, is being commissioned at SNOLAB~\cite{Balogh:2020nmo,NEWS-G:2022kon}.
Funding has also been secured for {\sc ECuME}, a planned follow-up 140~cm diameter detector that will benefit from a fully electroformed vessel to lower the background rate, and a feasibility study of hosting a 300~cm diameter fully electroformed SPC dubbed \DarkSphere at the Boulby Underground Laboratory is underway~\cite{Balogh:2023hba}.

\begin{figure}[t!]\label{fig:ideas}
\includegraphics[width=0.88\columnwidth]{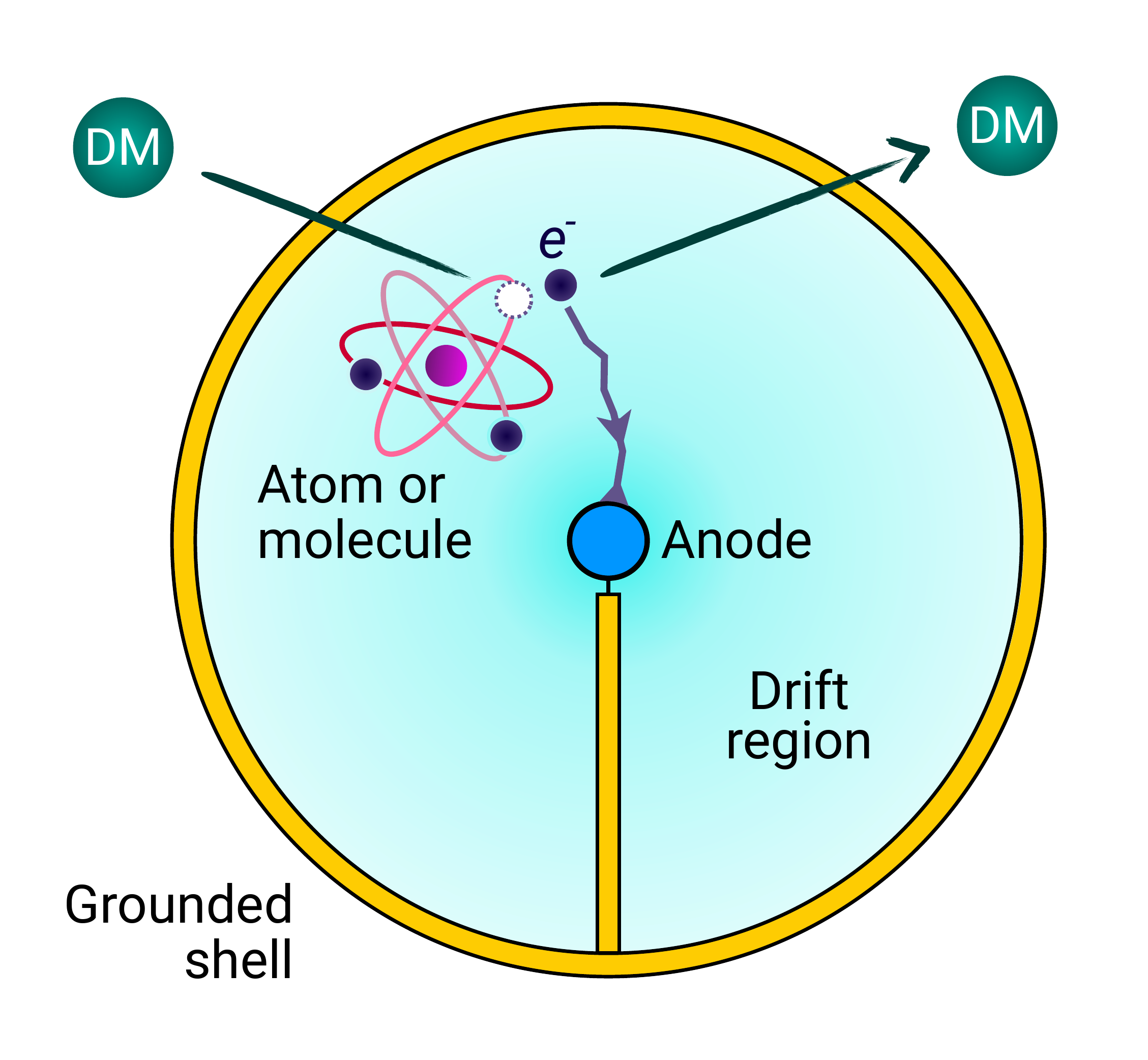} 
\caption{A schematic of the detection principle of {\sc DarkSphere},
which can be filled with a variety of atomic and molecular gases:
we consider helium, neon, xenon, methane, and isobutane.
The outer sphere is a grounded, metallic shell, and an anode is placed at the centre.
DM scattering can ionise an electron from an atom or molecule. After ionisation, the electron drifts to the anode where an avalanche occurs,
and the signal is read out.
This setup has a single-electron threshold, which is key to measuring the DM-induced signal.
}
\end{figure}

In its simplest form, the operation of the SPC is illustrated in fig.~\ref{fig:ideas}.
The detector consists of a grounded, spherical, metallic
vessel filled with a gas mixture and a spherical anode of radius
$\sim 1$~mm is placed at the centre. 
The anode produces an electric field that varies as~$r^{-2}$ in an ideal SPC
and allows the ionisation electrons produced through particle interactions in the gas volume to drift to the anode. 
When the electron reaches the anode, an avalanche occurs, providing signal amplification, which is read out by a grounded metallic~rod.
Larger SPCs will employ a multi-anode sensor (an `ACHINOS'~\cite{Giganon:2017isb,Giomataris:2020rna}),
which improves detector stability by decoupling the drift field from the avalanche field,
but the basic principle of operation is similar to the simple SPC shown in fig.~\ref{fig:ideas}.\footnote{The electric field induced by a 60-anode ACHINOS, as envisaged for \DarkSphere, varies from $\sim 0.1\,\mathrm{V/cm}$ at the outer sphere to $\sim 10\,\mathrm{V/cm}$ near the anodes~\cite{Katsioulas:2022cqe}.}

Owing to the low capacitance of the central sensor and the high amplification gain near the anode, 
it has been demonstrated that this detector setup has a single ionisation electron threshold~\cite{Bougamont:2010mj, Arnaud:2019nyp}.
This threshold is key to the $\mathcal{O}(15)$~eV threshold envisaged for nuclear recoil searches~\cite{Arnaud:2019nyp,Balogh:2023hba}, 
where the recoiling nucleus produces electrons as it moves through the gas volume.
However, crucially, a single ionisation electron threshold also allows for the possibility of searching for DM-electron interactions,
where the signal induced by the DM is a small number of electrons that are produced when an atom or molecule in the gas mixture is ionised.

In this paper, we quantify the range of DM masses and DM-electron scattering cross sections that could be probed by {\sc DarkSphere}, 
which we assume gives the ultimate sensitivity from a large-SPC experiment.
The \DarkSphere SPC can easily be filled with different gases;
therefore, we study the response to five different gases: 
helium, neon, and xenon, which are noble atoms commonly used for DM experiments,
and methane and isobutane (also known as 2-methylpropane), 
which may be more commonly known for their use as fuel.
The chemical formulae are He, Ne, Xe, CH$_{4}$ and C$_{4}$H$_{10}$, respectively.
Helium, neon, and molecules with hydrogen and carbon atoms have been proposed because the light nucleus of these atoms enhances the sensitivity to light DM nuclear-recoil interactions. Meanwhile, xenon has been proposed as a target for supernova-neutrino detection and neutrinoless double beta decay SPC searches~\cite{Vergados:2005ny, Meregaglia:2017nhx}.
For the noble gases, $\mathcal{O}(10\%)$ of methane or isobutane will in general be added to the mixture as a `quench gas'. Quench gases are used in proportional counters to improve detector stability by absorbing UV photons that can be produced following the excitation of neutral atoms by collisions between recoiling electrons or ions with the atoms in the gas. Without the quench gas, the UV photons may lead to a loss of proportionality and/or induce spurious signals (see e.g.~\cite{Knollbook1981} for a more extended discussion). In contrast to the noble gases, the detectors could operate stably with pure methane or pure isobutane as these gases can act as both the proportional and  quench gas.
In addition to studying pure gas mixtures, 
which is advantageous to clearly assess the advantages
and disadvantages of each gas, we will also study gas mixtures with $\mathcal{O}(10\%)$ of methane and isobutane as this is the more likely gas-mixture to be employed in an operating detector. 

Although the primary motivation for adding methane and isobutane comes from the requirement of adding a quench gas,
there are several reasons to believe that these molecular targets
may improve the sensitivity to DM-electron interactions compared to atomic targets.
Firstly, the ionisation energies in molecules are often lower than the ionisation energies in the atomic targets used in DM searches. For example, considering neon and methane which both have 10 bound-electrons, the ionisation energy of the outer six electrons in methane is 13.6~eV compared to 23.1~eV for the outer six electrons in neon.
Isobutane has an even lower ionisation energy at approximately 11.1~eV, compared with xenon's 12.7~eV.
Secondly, for a gas at the same pressure, volume, and temperature, so that the number of moles is the same, a molecular gas can benefit from a larger number of target electrons. 
For example, counting the number of electrons with an ionisation energy less than 100~eV, we find that neon has eight electrons,
xenon has 18 electrons, while isobutane has 26 electrons (although xenon has 54 atomic electrons in total, as we will discuss, 
electrons with large ionisation energies cannot be ionised through DM scattering).
In this case, this means that the ratio of target electrons is 8:18:26 per mole of neon, xenon, and isobutane, respectively.

To the best of our knowledge, there has been no attempt to compare atomic and molecular targets to see whether these apparent advantages are borne out. 
Therefore, we will explore whether any improvements in the sensitivity to DM masses and DM-electron scattering cross sections can be improved through the use of methane and isobutane. 
While DM-electron scattering from atomic xenon~\cite{Essig:2015cda,Roberts:2016xfw,Pandey:2018esq} has previously been considered, 
no results exist for helium, neon, methane and isobutane.

The rest of this paper is structured as follows. In section~\ref{sec:DMeScattering}, we review the formalism of electron-DM scattering 
and present our calculations of the bound- and continuum-electron wave functions, as well as the ionisation form factors for the atomic and molecular gases under consideration. 
In section~\ref{sec:DM_ER}, we present the DM-electron scattering rates and compare with the anticipated \DarkSphere background rate.
Section~\ref{section:sensitivities} presents our sensitivity
projections for {\sc DarkSphere} and we discuss the possible advantages of gas mixtures over single-species gases. 
We summarise our results and give our conclusions in section~\ref{sec:conclusions}. 
Appendices provide more details on the numerical implementation of our calculations, provide alternative calculations of the molecular
 ionisation form factors, and compare our xenon calculations with others in the literature.

Finally, many of our numerical results are available online. This includes expressions for the bound-electron wave functions in position and momentum space, interpolation tables for the ionisation form factors, and code to calculate the scattering rates~\cite{Git_results}.

\section{Dark matter scattering with atomic and molecular electrons \label{sec:DMeScattering}}

We begin by presenting our calculation of the DM -- electron scattering event rate. In units of counts/keV/kg/year,
the differential scattering event rate for a DM particle of mass $m_\mathrm{DM}$ to ionise an electron from an atom or molecule~is
\begin{equation}\label{ER}
    \dv{R}{E_e}=\frac{1}{m_A} \frac{\rho_\mathrm{DM}}{m_\mathrm{DM}}\sum_{i}w_{i}\,\dv{\langle\sigma^{i\to f}_{\rm{ion}} v_{\mathrm{DM}}\rangle}{E_e}\;,
\end{equation}
where $m_{A}$ is the mass of a target atom or molecule, $\rho_\mathrm{DM}$ is the local DM density, 
$w_{i}$ is the occupation number for electrons in the atomic or molecular orbital labelled by the quantum number(s) $i$,
$\langle\sigma^{i \to f}_{\rm{ion}} v_{\rm{DM}}\rangle$ is the velocity-averaged cross section for DM that scatters with an 
electron in the state labelled by $i$ to the state $f$, and finally, $E_\mathrm{e}$ is the kinetic energy of the continuum-electron.
For an atom, $i$ would be the familiar $n,l,\cdots$ quantum numbers while for molecules, the quantum numbers depend on the discrete symmetries of the system.

We follow the convention established in ref.~\cite{Essig:2011nj} and parameterise the differential cross section according to
\begin{equation}   
\label{eq:dsigmadE} 
    \dv{\langle\sigma^{i\to f}_\mathrm{ion}v\rangle}{\ln E_e} = \frac{\bar{\sigma}_e}{8\mu^2_{e}} \int_{q^i_-}^{q^i_+} q\, dq\, \abs{f^{i\to f}_\mathrm{ion}}^2 |F_{\rm{DM}}|^2 g(v^{i}_\mathrm{min})\;.
\end{equation}
In this equation, $\bar{\sigma}_e$ is a model-independent reference cross section,
 $\mu_{e}$~is the DM-electron reduced mass, $q$~is the amplitude of the momentum transferred by the DM to 
the electron with upper and lower limits~$q^i_+$ and~$q^i_-$, $F_{\rm{DM}}(q)$~is the dimensionless DM `form factor', $g(v^{i}_\mathrm{min}(q))$ is the DM velocity integral,
and $f^{i \to f}_{\mathrm{ion}}(E_\mathrm{e},q)$~is the dimensionless ionisation form factor.
For gas atoms or molecules in the \DarkSphere SPC, the orientation of the atoms and molecules will continuously change
so we will always use the rotationally averaged ionisation form factor in our calculations.
Explicitly, the general form of $f^{i \to f}_{\mathrm{ion}}(E_\mathrm{e},q)$ that we will use is
\begin{equation}\label{eq:apfion_el}
\begin{split}
    &\left| f^{i\to f}_{\rm{ion}}(E_e, q) \right|^2 \\
    &\quad =\left\langle \int d \Omega_{k_e} \frac{2 k_e^3}{8 \pi^3}  \left| \int d^3x \psi^*_f(\mathbf{x},\mathbf{k_e}) e^{i \mathbf{q}\cdot \mathbf{x}} \psi_i (\mathbf{x}) \right|^2 \right\rangle,
    \end{split}
\end{equation}
where $k_e=\sqrt{2m_e E_e}$ is the momentum of the unbound electron described by $\psi_f(\mathbf{x})$, the initial bound wave function is $\psi_i(\mathbf{x})$,
and the angled-brackets indicates that we are averaging uniformly over all orientations of the atom or molecule. 
As described further in appendix~\ref{sec:app:formfactors}, the rotational averaging procedure means that $|f^{i \to f}_{\mathrm{ion}}|^2$
depends on the scalar $q=|\mathbf{q}|$ rather than the vector~$\mathbf{q}$.
We use the standard convention for the bound-state wave functions that $\int d^3x\,\psi^*_i (\mathbf{x})\psi_i (\mathbf{x})=1$
 while the continuum wave functions are normalised such that $\int d^3x \,\psi^*_f(\mathbf{x},\mathbf{k}) \psi_f(\mathbf{x},\mathbf{k'}) = (2 \pi)^3 \delta^3(\mathbf{k}-\mathbf{k'})$.
 Many authors work in the spherical-wave basis where the integral over the solid angle of $k_e$
 does not appear. 
 However, we prefer to work in a basis where the solid angle is explicit
 as it makes some of the manipulations in appendix~\ref{sec:app:formfactors} more transparent.
The final results are of course independent of the choice of basis.
It is also worth stating clearly that with our convention, $f^{i \to f}_{\mathrm{ion}}(E_\mathrm{e},q)$ is the form factor for a single electron (e.g.\ one electron in the $5p$-shell of xenon). This may be different from other definitions in the literature
that absorb the occupation number $w_i$ into the definition of the form factor.

To provide the most straightforward comparison of the \DarkSphere sensitivity with other constraints, we use the Standard Halo Model (SHM) for the DM velocity distribution with $v_0=235$~km/s~\cite{Reid:2004rd,Gravity2019,Gravity2021,Baxter:2021pqo}, $v_{\rm{esc}}=544$~km/s (consistent with the latest estimates~\cite{Deason2019}), the Solar peculiar velocity from ref.~\cite{Schoenrich:2009bx}, and the average Earth velocity from ref.~\cite{McCabe:2013kea}. The minimum speed parameter~is
\begin{equation}\label{eq:vmin}
    v^{i}_{\rm{min}}(q) = \frac{E_e+I^{i}}{q} + \frac{q}{2m_{\rm{DM}}}\;,
\end{equation}
where $I^{i}$ is the ionisation energy of an electron in the initial-state with quantum number(s) $i$, and~$q^i_{\pm}$ are determined by solving eq.~\eqref{eq:vmin} for~$q$ with $v^{i}_{\rm{min}} = v^{\rm{max}}_{\rm{DM}}$, where $v^{\rm{max}}_{\rm{DM}}$ is the maximum DM speed in the detector frame of reference. This leads to
\begin{equation}\label{eq:qmaxmin}
    q^i_{\pm} = m_{\rm{DM}} v^{\rm{max}}_{\rm{DM}}\left(1 \pm \sqrt{1-\frac{E_e+I^i}{\tfrac{1}{2}m_{\rm{DM}} (v^{\rm{max}}_{\rm{DM}})^2}} \right).
\end{equation}

For our choice of parameters $v^{\rm{max}}_{\rm{DM}} = 794~\mathrm{km/s}$.
We note that the SHM is of course an idealised model while the real Milky Way DM halo contains much sub-structure not included within the SHM (e.g.\ see discussion in refs.~\cite{OHare:2018trr,Evans:2018bqy,OHare:2019qxc,Buch:2020xyt,Radick:2020qip}). 
Finally, we set $\rho_\mathrm{DM}=0.3~\mathrm{GeV}\mathrm{cm}^{-3}$ to facilitate comparison with other experimental constraints.

\subsection{Bound-state wave functions}

As we have just discussed, determining the scattering rate requires the ionisation form factor $f^{i \to f}_{\mathrm{ion}}(E_\mathrm{e},q)$, 
which in turn depends on the bound and continuum wave functions. We will begin by discussing our calculation of the bound-state wave functions.
There are many code packages capable of computing bound-state wave functions in atomic and molecular systems using self-consistent field (SCF) methods.
In this work, we make use of PySCF~\cite{PYSCF2017, PYSCF2020}, a versatile open-source python-based quantum chemistry 
package to calculate the bound-state wave functions for helium, neon and xenon atoms and for methane and isobutane molecules. 
Although PySCF allows for various SCF techniques to obtain the wave functions, we will stick to the approach most often used in the DM literature: the non-relativistic Hartree-Fock method.

\begin{figure*}[t!]
\includegraphics[width=1.8\columnwidth]{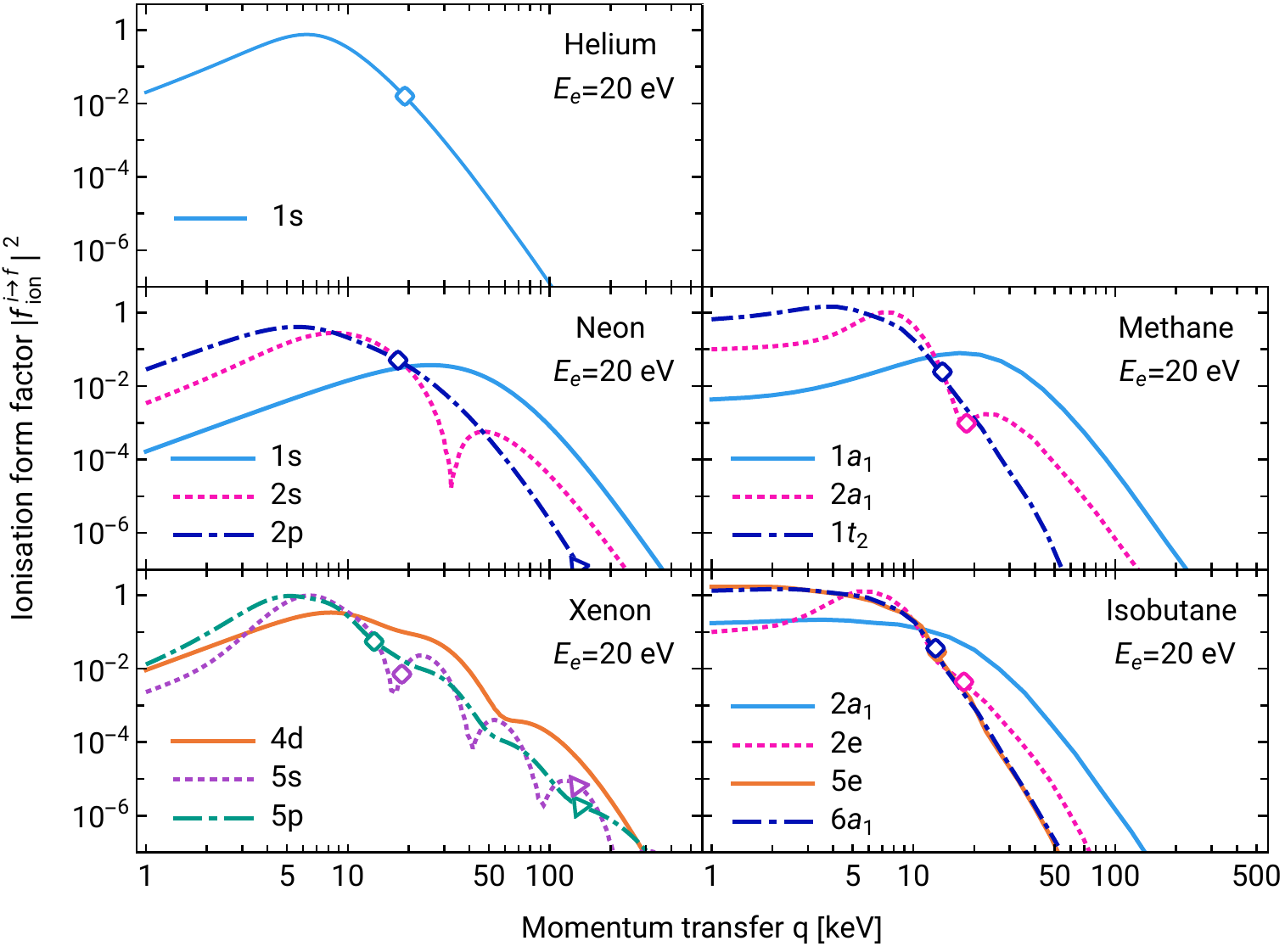}
\caption{\label{fig:fion}  
The dimensionless ionisation form factor for helium (top-left), neon (middle-left), xenon (bottom-left),
methane (middle-right) and isobutane (bottom-middle).
In all cases, we show results when the continuum-electron kinetic energy $E_e$ is 20~eV. 
For clarity in the xenon panel, we only show the results for the three outermost shells, $4d$, $5s$ and $5p$,
as these usually give the dominant contribution.
In the isobutane panel, for clarity, we show two results from the outer valence orbitals ($6a_1$ and $5e$)
to demonstrate that they are very similar, one inner valence ($2e$) and one core orbital ($2a_1$).
The diamonds and triangles on some lines indicate the $q_{-}$ and $q_{+}$ values, respectively,
for $m_{\mathrm{DM}}=30$~MeV.
There are similarities in terms of the peak position and overall shape between the form factors for the outermost orbitals
in helium ($1s$), neon ($2p$) and xenon ($5p$), and between all three form factors in neon and methane.
The shape of the methane and isobutane orbitals are also somewhat similar.
Note however that the form factors for methane and isobutane drop-off more rapidly than for neon as~$q$ increases.
}
\end{figure*}

It should be noted that the non-relativistic Hartree-Fock approach may have limitations.
Firstly, relativistic effects have been shown to become important when the electron recoil energy is $\mathcal{O}(\mathrm{keV})$,
which requires that $q \langle r \rangle \gg1$, where $\langle r \rangle \sim a_0/Z_{\rm{eff}}$
is the average position of the bound electron under consideration, $a_0$ is the Bohr radius 
and $Z_{\rm{eff}}$ is the effective charge felt by the electron. In this regime,
 non-relativistic calculations can underestimate the rate by an order of magnitude or more~\cite{Roberts:2015lga, Roberts:2016xfw}.
 Using the comparison between relativistic and non-relativistic calculations in ref.~\cite{Roberts:2019chv},
 we can estimate that relativistic effects begin to become significant (meaning there is a factor two or more difference) when
$q \gtrsim  35 Z_{\rm{eff}}$~keV.
In this work, we will consider electron recoil energies $\mathcal{O}(10)~\mathrm{eV}$ and our 
calculations will be dominated by $q$ values $\sim 20~\mathrm{keV}$ (as we will see in fig.~\ref{fig:fion})
 so we do not expect any significant error from not incorporating relativistic effects in our calculations.
Secondly, the Hartree-Fock method is limited in its accuracy because it does not fully account for electron correlation.
For instance, in photoionisation cross section calculations with full shell atoms, the absence of correlation effects limits the
accuracy of Hartree-Fock calculations to around~30\%~\cite{Kennedy1972}. 
We therefore expect there to be a similar error in our calculations.

PySCF expresses the wave functions in terms of a finite-dimensional Gaussian basis set. 
In general, it is important to choose a large enough Gaussian basis set to obtain accurate 
forms for the wave functions, particularly at small and large values of $r$.
 For helium and neon, we therefore use the rather large aug-cc-pV5Z 
 basis set~\cite{aug-cc-pv5Z-Woon1994,aug-cc-pv5Z-Ne-doi:10.1063/1.456153},
 while for xenon, we use the Jorge-QZP basis set~\cite{Xe_qzp_Ceolin2013} as aug-cc-pV5Z is not available for xenon. For methane and isobutane, 
 we use the 6-31G(d,p) set~\cite{ditchfield1971a, hehre1972a, hariharan1973a}, which is sufficiently large to 
 capture the effects of the chemical bonds between the atoms. We have used numerical coefficients from the Basis Set Exchange database~\cite{feller1996a,schuchardt2007a,pritchard2019a}.
 For the methane and isobutane calculations, we also need to specify the structure of the molecules. 
 In our calculations, we use the geometric data from the National Institute of Standards and Technology Computational Chemistry Comparison and Benchmark Database (NIST CCCBDB)~\cite{NISTCCCBDB}. 
 
\begin{table*}[t!]
\centering
\begin{tabular*}{1.45\columnwidth}{ c c c  c c c  c c c }
\toprule
 \multicolumn{3}{c}{Helium (He)} & 
 \multicolumn{3}{c}{Neon (Ne)} & \multicolumn{3}{c}{Methane (CH$_4$)}\\ 
\cmidrule(r){1-3} \cmidrule(r){4-6} \cmidrule(r){7-9}
\multicolumn{3}{c}{Basis: aug-cc-pV5Z} &
\multicolumn{3}{c}{Basis: aug-cc-pV5Z} &
\multicolumn{3}{c}{Basis: 6-31G(d,p)}  \\
\multicolumn{3}{c}{Tot.\ energy [au]$^\dagger$: $-2.8616$ \;} &  
\multicolumn{3}{c}{Tot.\ energy [au]$^\dagger$: $-128.5467$ \;} & 
\multicolumn{3}{c}{Tot.\ energy [au]$^\dagger$: $-40.2016$ \;}\\
\cmidrule(r){1-3} \cmidrule(r){4-6}\cmidrule(r){7-9}
Orbital & $I_{\mathrm{HF}}$ [eV] & $I_{\mathrm{exp}}$ [eV] &
Orbital & $I_{\mathrm{HF}}$ [eV] & $I_{\mathrm{exp}}$ [eV] &
Orbital & $I_{\mathrm{HF}}$ [eV] & $I_{\mathrm{exp}}$ [eV] \\
$1s^2$ & 24.98 & 24.6  & $2p^6$ & 23.14 & 21.7  & $1t_2^6$ & 14.80 & 13.6 \\
&   &   & $2s^2$ & 52.53 & 48.5 & $2a_1^2$ & 25.66 & 22.9  \\
&   &   & $1s^2$ & 891.79 & 870.2 & $1a_1^2$ & 304.96 & 290.8 \\ 
\bottomrule
\end{tabular*}
\begin{tabular*}{1.45\columnwidth}{ cc c c c  c c c c }
\toprule
\hspace{2.5em} & \hspace{3em} & \multicolumn{3}{c}{Isobutane (C$_4$H$_{10}$)} & \multicolumn{3}{c}{Xenon (Xe)} & \\ 
\cmidrule(r){3-5} \cmidrule(r){6-8} 
& & \multicolumn{3}{c}{Basis: 6-31G(d,p)} & \multicolumn{3}{c}{Basis: Jorge-QZP} & \\
& & \multicolumn{3}{c}{Tot.\ energy [au]$^\dagger$: $-157.3123$ \;} & \multicolumn{3}{c}{Tot.\ energy [au]$^\dagger$: $-7229.7195$\; } & \\
\cmidrule(r){3-5} \cmidrule(r){6-8}
& & Orbital & $I_{\mathrm{HF}}$ [eV] & $I_{\mathrm{exp}}$ [eV] & Orbital & $I_{\mathrm{HF}}$ [eV] & $I_{\mathrm{exp}}$ [eV] &  \\
& & $6a^2_1$ & 12.34 & 11.13 & $5p^6$ & 12.45 & 12.7 &\\
& &$5e^4$ & 12.44 & 11.75 & $5s^2$ & 25.54 & 23.3  &\\
& &$1a_2^2$ & 13.86 & 12.85 & $4d^{10}$ & 75.72 & 68.5 &\\ 
& &$4e^4$ & 14.54 & 13.71 & $4p^6$ & 163.56 & 146.1 &\\
& &$3e^4$ & 16.04 & 15.03 & $4s^2$ & 212.69 & 213.2 &\\ 
& &$5a^2_1$ & 17.15 & 15.91 & $3d^{10}$  &  711.26 & 682.7  &\\
& &$4a^2_1$ & 20.62 & 18.58 &  $3p^6$ &  958.02 & 971.4 & \\
& &$2e^4$ & 25.17 & 21.83 & $3s^2$  & 1087.7  &  1149 &\\
& &$3a^2_1$ & 29.44 & 24.83 & $2p^6$   & 4839.8  & 4947   &\\
& &$2a^2_1$ & 305.01 & --- & $2s^2$  &  5132.0 & 5453  &\\
& &$1e^4 $ & 305.01 & --- & $1s^2$  &  33321 & 34561 & \\
& &$1a^2_1$ & 305.30 & --- &   &   &   &\\
\bottomrule
\end{tabular*}
\caption{A summary of the most pertinent details from our PySCF calculations. In the second and third rows, 
we give the basis-set used and the total converged energy in atomic units (au) for each atomic or molecule species.
The values $I_{\rm{HF}}$ and $I_{\rm{exp}}$ give the ionisation energy in eV from the PySCF Hartree-Fock calculation 
and the experimental value from refs.~\cite{Darrah1970,BIERI1980,Deng2001,lbl_xray_booklet}, respectively. The superscript in the orbital name gives the number of 
electrons in that orbital. For isobutane, we do not have a reference for the ionisation energies of the 
inner core orbitals (denoted by ---). Following ref.~\cite{Deng2001}, we group the isobutane orbitals into three categories: 
the outer valence orbitals ($6a_1$, $5e$, $1a_2$, $4e$, $3e$,~$5a_1$), the inner valence orbitals ($4a_1$, $2e$,~$3a_1$),
and the core orbitals ($2a_1$, $1e$,~$1a_1$).
The large number of orbitals (and therefore number of electrons) in isobutane
with ionisation energies less than $30~\mathrm{eV}$ is much higher than in the other species.
\newline $^\dagger$ To provide for a more straightforward comparison of the total energy with other computational approaches, we quote the total energy in atomic units, where $1\,\mathrm{au}=27.211\,\mathrm{eV}$.
}
\label{table_energies}
\end{table*}
 
The main results of our PySCF calculations for helium, neon, methane, isobutane and xenon are summarised 
in tab.~\ref{table_energies}.\footnote{An even more detailed summary of the PySCF output is given on the 
GitHub repository~\cite{Git_results}, which also includes the full analytic forms of the wave functions in position and momentum space.}
The atomic orbitals are labelled by the principal and angular momentum $n$ and $l$ quantum numbers.
The labels for the methane and isobutane orbitals follow from the discrete symmetry of the molecule.
Methane orbitals are eigenfunctions of irreducible representations of the discrete tetrahedral group T$_d$
while the isobutane orbitals are eigenfunctions of irreducible representations of the $C_{3v}$ point group.

For the atomic results, we can compare directly with the tabulated computational results in ref.~\cite{BUNGE1993113}, which use 
Slater-type orbitals, and against output from the ATSP atomic code~\cite{ASTP_FISCHER2000635}. 
We find that the total energy and the individual orbital ionisation energies from our PySCF calculation, $I_{\rm{HF}}$,
agree very well, at the level of $0.001\%$ or better for helium and neon, 
and at the level $0.03\%$ or better for xenon. We find excellent agreement between the atomic wave functions when comparing them directly.
We also find very good agreement between the calculated methane total energy and the computed ionisation energies in ref.~\cite{Brundle1970}, agreeing at the level of $0.5\%$ or better.
A similar level of agreement is found more generally between our methane and isobutane total energy calculations and the 
suite of calculations summarised in the NIST CCCBDB database~\cite{NISTCCCBDB}. 
Finally, we also find good agreement between our spherically-averaged methane and isobutane momentum-space wave functions and the various calculations presented in refs.~\cite{HAMEL2002} and~\cite{Deng2001}.

From tab.~\ref{table_energies}, we can also compare our calculated $I_{\rm{HF}}$ values with the experimental values $I _{\exp}$.
For atoms, we take the values from ref.~\cite{lbl_xray_booklet} and quote experimental values averaged over the fine-structure levels.
The experimental values for methane and isobutane are from refs.~\cite{BIERI1980, Darrah1970, Deng2001}, respectively, where we have quoted the vertical ionisation energies.
We do not have a reference for the experimental values of the inner-most core orbitals of isobutane; 
however, these orbitals either do not contribute at all or provide a sub-dominant contribution.
The theoretical values are, however, consistent with the inner-most methane values, which is what we would expect given these are all carbon $1s$-like states. 

We can take the analogy between methane and isobutane further
by following ref.~\cite{Deng2001}, which groups the isobutane orbitals into three groups:
the outer valence orbitals ($6a_1$, $5e$, $1a_2$, $4e$, $3e$,~$5a_1$), the inner valence orbitals ($4a_1$, $2e$,~$3a_1$),
and the core orbitals ($2a_1$, $1e$,~$1a_1$).
After this grouping, the average of the outer valence ionisation energies is similar to~$1t_2$ in methane,
the inner valence average is somewhat similar to~$2a_1$ in methane,
and as already noted the core orbital average is similar to~$1a_1$.

In general, we see that the theoretical values and experimental values do not match exactly.
This is the normal outcome of a non-relativistic Hartree-Fock calculation since this approach misses many of the effects due to electron correlation, 
which are required to obtain more accurate results. 
Electron correlation can be accounted for with methods such as the multi-configurational self-consistent field, 
which can also incorporate relativistic effects (for a detailed discussion, see, e.g., ref.~\cite{GrantBook}).
There are also mature code packages such as the atomic code GRASP~\cite{Jonsson:2007,Fischer:2019} or the molecular code BERTHA~\cite{Quiney:1998}
that can calculate these effects.
However, we leave the exploration of multi-configurational self-consistent field techniques to future work and in our calculations, 
we use the PySCF wave functions but, where available, use the experimental values of the ionisation energy from tab.~\ref{table_energies}.

\begin{figure}[t!]
\includegraphics[width=0.95\columnwidth]{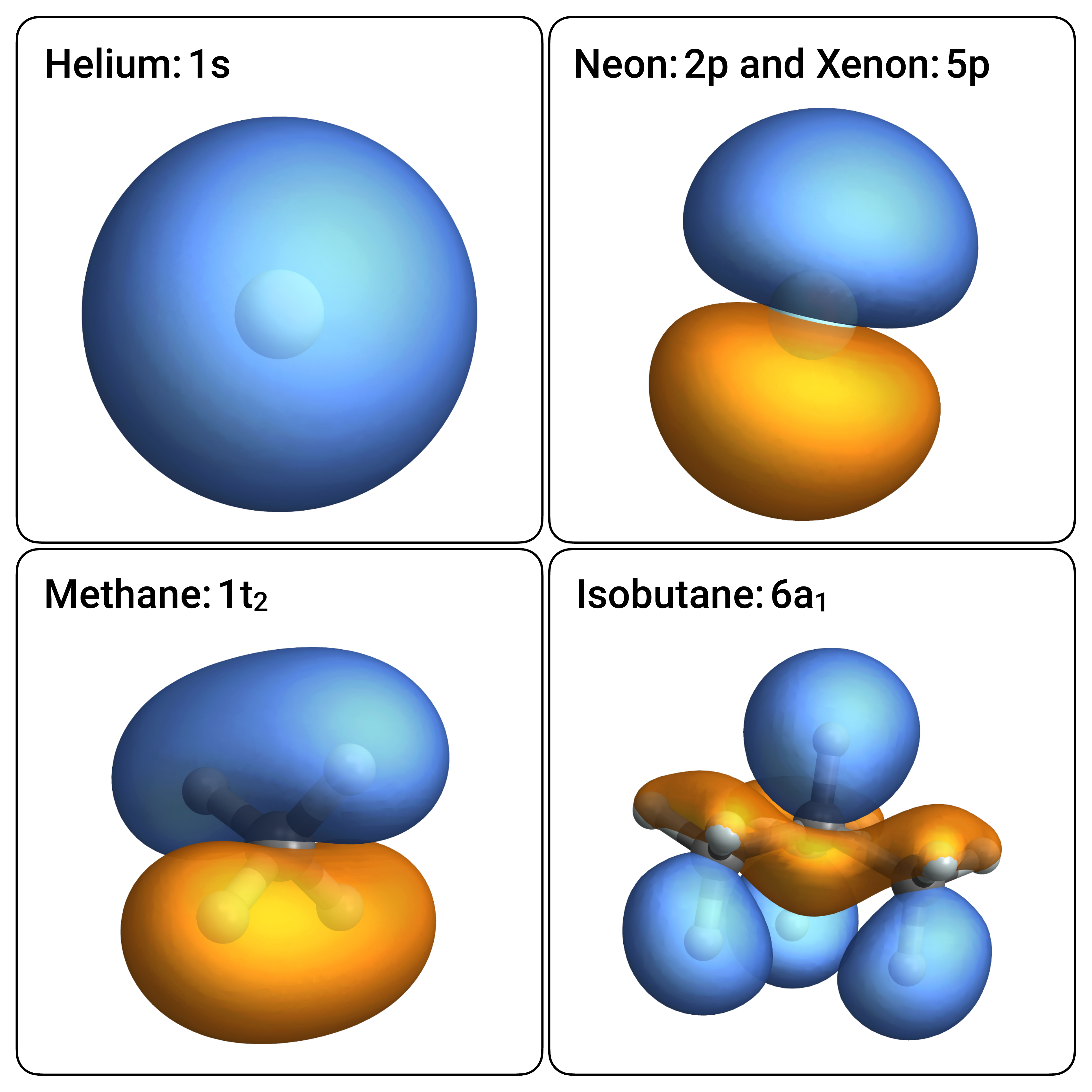}
\caption{\label{fig:orbitals}
A visualisation of the outermost orbitals for helium, neon and xenon, methane, and isobutane.
The blue and orange regions indicate the spatial extent of the electron orbital.
The white and grey nodes represent the positions of the atomic nuclei,
and the connections between the nuclei in the lower panels indicate the bonds.
This visualisation was generated by Mathematica~\cite{Mathematica} using the cubegen output from PySCF.
}
\end{figure}

Finally, as the molecular orbitals are not usually encountered in the DM literature, in fig.~\ref{fig:orbitals}, 
we show a visualisation of the outermost orbitals for helium, neon (and xenon), methane, and isobutane. 
The blue and orange shaded regions indicate the typical spatial extent of the wave function, while the smaller spheres
indicate the position of the atomic nuclei. In the lower two panels, the bonds between the hydrogen and carbon atoms are also shown.
Methane is a tetrahedron with four equivalent bonds between carbon at the centre and the four outer hydrogen atoms.
The tetrahedral geometry of the highest occupied
molecular orbital of methane is the result of $sp^3$-hybridisation between electrons originally associated to the carbon ($2p$) and hydrogen ($1s$) states.
The carbon atoms in isobutane form a trigonal pyramid, where the carbon atom at the centre is bonded with 
three carbon atoms at the base of the pyramid and one hydrogen atom at the top of the pyramid.
Each base carbon atom then has bonds with three hydrogen atoms, forming tetrahedrons.
The outermost orbital in isobutane is again formed as a result of the sharing of electrons between the carbon $2p$ and hydrogen $1s$ electrons. 

\subsection{Continuum-electron wave functions}

We now have the bound-state wave functions that enter eq.~\eqref{eq:apfion_el}.
 Next, we describe our calculations of the continuum-electron wave functions.
For both atomic and molecular systems, we look for outgoing states that have been expanded into partial waves,
\begin{equation}
\label{eq:ionisedwavefunction}
\psi_f(\mathbf{x},\mathbf{k_e}) = \sum_{l,m} i^l \frac{P_{k_e l}(r)}{r}Y^*_{l m}(\hat{\mathbf{x}}) Y_{lm}(\hat{\mathbf{k}}_e)\;,
\end{equation}
where $P_{k_e l}(r)$ is the radial part of the wave function and $Y_{lm}$ are the spherical harmonics.
The normalisation condition $\int d^3x \,\psi^*_f(\mathbf{x},\mathbf{k}) \psi_f(\mathbf{x},\mathbf{k'}) = (2 \pi)^3 \delta^3(\mathbf{k}-\mathbf{k'})$
holds if the radial wave functions satisfy $\int dr P_{k l}(r)P_{k' l'}(r)  = (2\pi)^3\delta_{l l'} k^{-2}\delta(k-k')$.
We impose this by following the standard procedure of ensuring that at large values of $r$ our solution $P_{ k_e l}(r)$ 
asymptotes to a sine function with amplitude $4\pi/k_e$~\cite{Bates1949}.

We first discuss our numerical approach to finding atomic states, where, owing to the spherical symmetry of atoms, the expansion
in eq.~\eqref{eq:ionisedwavefunction} is exact. The continuum-electron wave functions $P_{nl\to k_e l_e}(r)$ for an electron initially in the state $n$, $l$ 
that has been promoted to an continuum state with quantum numbers $k_e$, $l_e$ are obtained by numerically solving an 
approximate form of the one-electron Hartree-Fock equation in the mean field of the remaining electrons. 
Explicitly, the equation that we solve, written in atomic units, is
\begin{equation}\label{eqHF1}
\begin{split}
&\left[  -\frac{1}{2}\frac{d^2}{dr^2} +  \frac{l_e(l_e+1)}{2r^2} 
+ V_{nl\to k_e l_e}(r)\right] P_{nl\to k_e l_e}(r) \\
&\qquad \qquad = E_e P_{nl\to k_e l_e}(r)+\sum_{n' l'} \delta_{l_e l'}\lambda_{k_e n'}P_{n' l'}(r)\;,
\end{split}
\end{equation}
where~$\lambda_{k_e n'}$ are Lagrange multipliers that are introduced to ensure that radial wave functions with the same~$l$ quantum numbers obey the correct orthogonality relations. 

The potential energy in the full Hartree-Fock equations contains both a local and non-local term~\cite{Amusiabook1997}. 
For the continuum electron, we employ Cowan's Hartree-plus-statistical-exchange (HX) method to 
approximate the non-local exchange potential with a local potential, which dramatically reduces the 
complexity of the problem while producing results in good agreement with the full Hartree-Fock method~\cite{Cowan1967,Cowanbook1981}.
Cowan's HX method is particularly suited to ionisation problems since the potential energy is guaranteed to have the expected $-1/r$ relation for an electron far from the atom.

In the HX method, the potential energy is approximated as
\begin{equation}
\label{eq:HXpotential}
V_{nl\to k_e l_e}(r) = -\frac{Z}{r} + V^{\rm{H}}_{nl\to k_e l_e}(r) + V^{\rm{HX}}_{nl\to k_e l_e}(r)\;,
\end{equation}
where the first term is the familiar Coulomb interaction of the electron with a nucleus of 
charge $Z$, $V^{\rm{H}}(r)$ is the classical potential energy of an electron in the mean 
field of the other atomic electrons and $V^{\rm{HX}}(r)$ is the local approximation to the exchange potential.

The classical potential energy in the frozen core approximation~is
\begin{equation}
    V^{\rm{H}}_{nl\to k_e l_e}(r)=\sum_{n'l'}(w_{n'l'} - \delta_{nl,n'l'})\int_0^{\infty} \frac{dr'}{r_{>}}P^2_{n'l'}(r')\,,
\end{equation}
where $r_>=\text{min}(r,r')$ and $w_{nl}$ is the electron occupation number. The sum 
is over all bound electrons, $P_{nl}(r)$ are the bound wave functions from 
PySCF (i.e.\ in the ionisation calculation, we `freeze' the wave functions of the bound core electrons), 
and the Kronecker delta excludes the bound electron that has been ionised from contributing to the potential.

The HX exchange potential energy in the frozen core approximation for continuum states~is
\begin{equation}\label{eq:VHX}
    V^{\rm{HX}}_{nl\to k_e l_e}(r)= -\frac{k_x}{2} \left(\frac{\rho'(r)}{\rho(r)} \right) \left(\frac{24 \rho(r)}{\pi} \right)^{1/3},
\end{equation}
where $k_x$ is an $\mathcal{O}(1)$ numerical factor, and
\begin{align}\label{eqHFlast}
\rho(r) &= \sum_{n'l'} w_{n'l'}\frac{P_{n'l'}^2(r)}{4 \pi r^2}\;,\\
\rho'(r)&= \sum_{n'l'}\left[ w_{n'l'}-\text{min}(2,w_{nl})\delta_{nl,n'l'}\right] \frac{P_{n'l'}^2(r)}{4 \pi r^2} \;.
\end{align}
 Again, the sum is over all bound electrons, while the $\text{min}(2,w_{nl})$ term ensures that the 
 exchange potential vanishes for any two-electron configuration, e.g., as found in helium, as 
 exchange terms only arise between electrons with parallel spin.
 As with the classical potential energy calculation, $P_{nl}(r)$ are PySCF bound wave functions,
 and the Kronecker delta excludes the self-interaction of the electron that has been ionised. We fix $k_x=0.65$ as this leads to the most accurate results~\cite{Cowan1967,Cowanbook1981, Condonbook1980}.

Next, we discuss our approach to calculating continuum states from molecules.
A rigorous calculation of the continuum-electron states in a molecule is a much more challenging calculation.
 Firstly, the geometry of methane and isobutane means that the potential energy is no longer exactly spherically symmetric,
 and, secondly, the potential contains multiple poles at the positions of the atomic nuclei.
In this work, we take a simplified approach and model the outgoing electron with a single-centred Coulomb wave function,
which is an analytic solution to the Schroedinger equation for a $-Z/r$ potential. 
This simplified approach has previously been used in the DM literature for atomic states (see, e.g.,~\cite{Agnes:2018oej,Catena:2019gfa}).
In atomic units, the Coulomb function takes the form
\begin{equation}\label{eq:PCoulomb}
\begin{split}
    P_{kl}(r)&= \frac{4 \pi}{2 k} \frac{\left|\Gamma \left(\ell+1-\frac{iZ}{ k} \right) \right| e^{\frac{\pi Z}{2 k}}  }{(2 \ell+1)!}  (2 k r)^{\ell+1}  \\
   & \times  e^{-i kr} \, M \left(\ell+1+\frac{i Z}{ k}, 2\ell+2;2ikr\right)\;,
    \end{split}
\end{equation}
where $M(a,b;x)$ is Kummer's confluent hypergeometric function and $\Gamma(z)$ is the Gamma function.

The charge $Z$ remains a free parameter in the Coulomb function.
We again opt for a simple and transparent method using the ionisation energies to choose $Z$ for the methane and isobutane orbitals.
Using the hydrogen-like scaling for the ionisation energy, $I= Z^2/n^2 \times 13.6~\mathrm{eV}$ with 
$n=1$ for $1a_1$ and $1t_2$ and $n=2$ for $2a_1$,
 leads to $Z$ values 
of $4.7$, $2.6$ and $1.0$ for the $1a_1$, $2a_1$ and $1t_2$ states in methane, respectively.
For isobutane, we follow ref.~\cite{Deng2001} and use the grouping of orbitals into inner core states ($1a_1$, $1e$, $2a_1$), 
inner valence states ($3a_1$, $2e$, $4a_1$), and outer valence states (all remaining orbitals).
From the ionisation energies and using $n=1$ for the inner core and outer valence states and $n=2$ for the inner valence states, 
we find that $Z=4.7$ for the inner core orbitals, $Z=2.7$ for the inner valence orbitals, and $Z=1$ for the outer valence orbitals.
We also have to specify the centre of the Coulomb wave functions. In our calculations, the centre is
defined by the coordinate system from the NIST CCCBDB database~\cite{NISTCCCBDB},
which takes the centre-of-nuclear-charge as the origin.
In methane, this corresponds to functions centred on the carbon nucleus,
while in isobutane, the centre is in the orange region directly below
the central carbon nucleus at the top of trigonal pyramid in the lower right panel of fig.~\ref{fig:orbitals}.

The Coulomb wave functions have some obvious limitations.
Firstly, the $-Z/r$ Coulomb potential misses the 
$r$-dependent contributions from $V^{\mathrm{H}}$ and $V^{\mathrm{HX}}$ in eq.~\eqref{eq:HXpotential}.
Secondly,  the assignment of the $Z$ value for molecules is less robust than for atomic states.
Thirdly, the Coulomb functions are not guaranteed to have the correct orthogonality conditions
for the bound-state functions, as imposed by the Lagrange multipliers in eq.~\eqref{eqHF1}.
Fourthly, the Coulomb potential is spherically symmetric, while the methane and isobutane potentials are non-spherical.
And finally, the Coulomb waves may not be centred at the centre-of-nuclear-charge but instead at one of the nuclei.

Employing a spherically averaged potential is a commonly used approximation
even in more advanced treatments of scattering problems
involving molecules (see, e.g.,~\cite{Jain1986, Baluja_1992, Das2014, Er_Jun_2007, Mahato2019}) 
and typically leads to results with an accuracy
of approximately $50\%$ or better compared to more rigorous methods.
We explore the uncertainty induced by changing the $Z$ value and by moving the origin of the Coulomb waves
in appendix~\ref{app:comparions_molecules}.

\subsection{Ionisation form factors}

Having obtained the bound- and continuum-electron wave functions, we are now able to calculate the
rotationally averaged ionisation form factor $|f^{i\to f}_{\rm{ion}}(E_e, q)|^2$.
The general expression was given in eq.~\eqref{eq:apfion_el}; however, to generate the results in this section, we use
the equivalent forms derived in appendix~\ref{sec:app:formfactors} as they are more amenable to numerical calculations.
We have made extensive use of the GNU Scientific Library-2.7~\cite{GSL}, Cubature~\cite{Cubature}, and GNU Parallel~\cite{tange_2022_6891516} 
in our numerical calculations.

In fig.~\ref{fig:fion}, we plot the results of our ionisation form factor calculation for atoms (left column) and molecules (right column).
For clarity, we show only a subset of the orbitals for xenon and isobutane (bottom panels).
In the xenon panel, we show the outer three orbitals as these are the most important for DM scattering.
In the isobutane panel, we show two of the outer valence orbitals ($6a_1$ and $5e$) to show that they are almost
identical, differing only slightly at $q\lesssim5$~keV. The other outer valence orbitals (not plotted) are also very similar.
Similarly, we plot one inner valence orbital ($2e$) and one core orbital ($2a_1$) as the other inner valence and core orbitals
are similar.\footnote{Tabulated values of the ionisation form factors for all of the helium, neon, xenon, methane and isobutane orbitals are available on the 
GitHub repository~\cite{Git_results}.}

In fig.~\ref{fig:fion}, we show results when the continuum electron's kinetic energy is $E_e=20$~eV.
 At smaller values of $E_e$, the ionisation form factors have a similar shape but with a 
smaller amplitude and a small shift by $\sim\mathcal{O}(\mathrm{keV})$ to smaller values of $q$. 
This change in the amplitude occurs because of the dependence of the electron kinetic energy in the definition 
of the dimensionless ionisation form factor, eq.~\eqref{eq:apfion_el} (recall that $E_e$ enters through $k_e$), 
while the shift to smaller values of $q$ occurs because a smaller continuum-electron kinetic energy requires a smaller kick from the momentum transfer. 
Conversely, for larger values of $E_e$, the amplitude of the form factors is larger and the peak shifts toward 
larger values, e.g., when $E_e\sim 1$~keV, the peak of the outer orbitals shifts to the range $q\sim 30$~to~$50$~keV.

Comparing the atomic form factors, we see that the results of the outermost orbitals 
($1s$, $2p$ and $5p$ for helium, neon and xenon, respectively) share similarities in 
the height and position of the peak values.
This should not be too surprising, given that outer-shell atomic electrons of different atoms
 find themselves in a similar screened-Coulomb potential.
Comparing neon to methane and isobutane, we also see similarities between all three form factors. 
Finding some similarities between neon and methane may not be too surprising as both are systems with 10 electrons. 
Furthermore, after grouping the isobutane orbitals into the outer valence, inner valence, and core orbitals,
we already observed similarities between the methane and isobutane ionisation energies, and here
we see similarities between the form factors.
However, there are two important differences particularly at small and large values of $q$ that deserve further comment.

The first important difference, at $q\lesssim10$~keV, is that the methane and isobutane form factors 
do not show the steady increase with~$q$ observed with the atomic results.
Instead, in the molecular form factors there is a gradual flattening to a finite value as $q\to 0$.
This is a result of using Coulomb wave functions in the methane and isobutane calculations, which only approximately obey
the orthogonality conditions expected between the bound and continuum wave functions; it should be the case that when $q=0$ the overlap integral
in eq.~\eqref{eq:apfion_el} is zero. Instead, when using Coulomb functions for the continuum-electron, 
the overlap integral gives a constant non-zero value when $q=0$.

However, for the DM scattering rates, this anomalous behaviour at small $q$ is not important.
We can understand why by considering the diamonds and triangles marked on some of the lines in fig.~\ref{fig:fion}.
The diamonds show the $q_{-}$ value, while the triangles show the $q_{+}$ values, 
which are the limits of integration that appeared in eq.~\eqref{eq:dsigmadE}, for $m_{\mathrm{DM}}=30$~MeV.
Orbitals that do not have a diamond or triangle do not contribute to the differential cross section at this value of $m_{\mathrm{DM}}$.
For some orbitals (e.g., helium 1s), the triangle does not appear simply as a result of our plotting scale for the form factor.
For all targets, we see that the typical $q_{-}$ value is approximately 10~to 20~keV and therefore above the region
where the Coulomb functions could introduce an error in the molecular calculations from the anomalous orthogonality conditions.
Figure~\ref{fig:fion} shows this explicitly for $m_{\mathrm{DM}}=30$~MeV and 
a single value of $E_e$, but we can use eq.~\eqref{eq:qmaxmin} to show 
that the smallest value of $q_{-}$ is approximately $11~\mathrm{keV}\cdot (E_e+I^i)/(30~\mathrm{eV})$, from which we conclude that
the anomalous region $q\lesssim5$~keV does not enter the 
integration over $q$ in eq.~\eqref{eq:dsigmadE}.

The second and more important difference between neon, methane and isobutane is at larger values of~$q$, 
where fig.~\ref{fig:fion} shows that the form factors drop-off more rapidly as~$q$ increases.
This is particularly evident when comparing the outer shell orbitals ($2p$, $1t_2$, and $6a_1$).
This effect can be understood as follows.
Rewriting the form factor (eq.~\eqref{eq:apfion_el}) in momentum space, we would find that $|f_{\rm{ion}}^{i \to f}|^2 \sim$ $\abs{\tilde{\psi}_i(\mathbf{k_e}-\mathbf{q})}^2$,
where $\tilde{\psi}$ is the momentum-space wave function (see appendix~\ref{sec:app:formfactors:PW} for an explicit 
realisation of this when the outgoing electron is modelled by a plane-wave).
This implies that large values of~$q$ correspond to a probe of the large-momentum tail of the momentum-space wave function.
The outermost momentum-space wave functions fall away more quickly for methane and isobutane at large values of the momentum
compared to neon, helium, and xenon, so the molecular form factors fall away more rapidly.

\section{Comparison of scattering rates}
\label{sec:DM_ER}

\begin{figure}[t!]
\includegraphics[width=0.95\columnwidth]{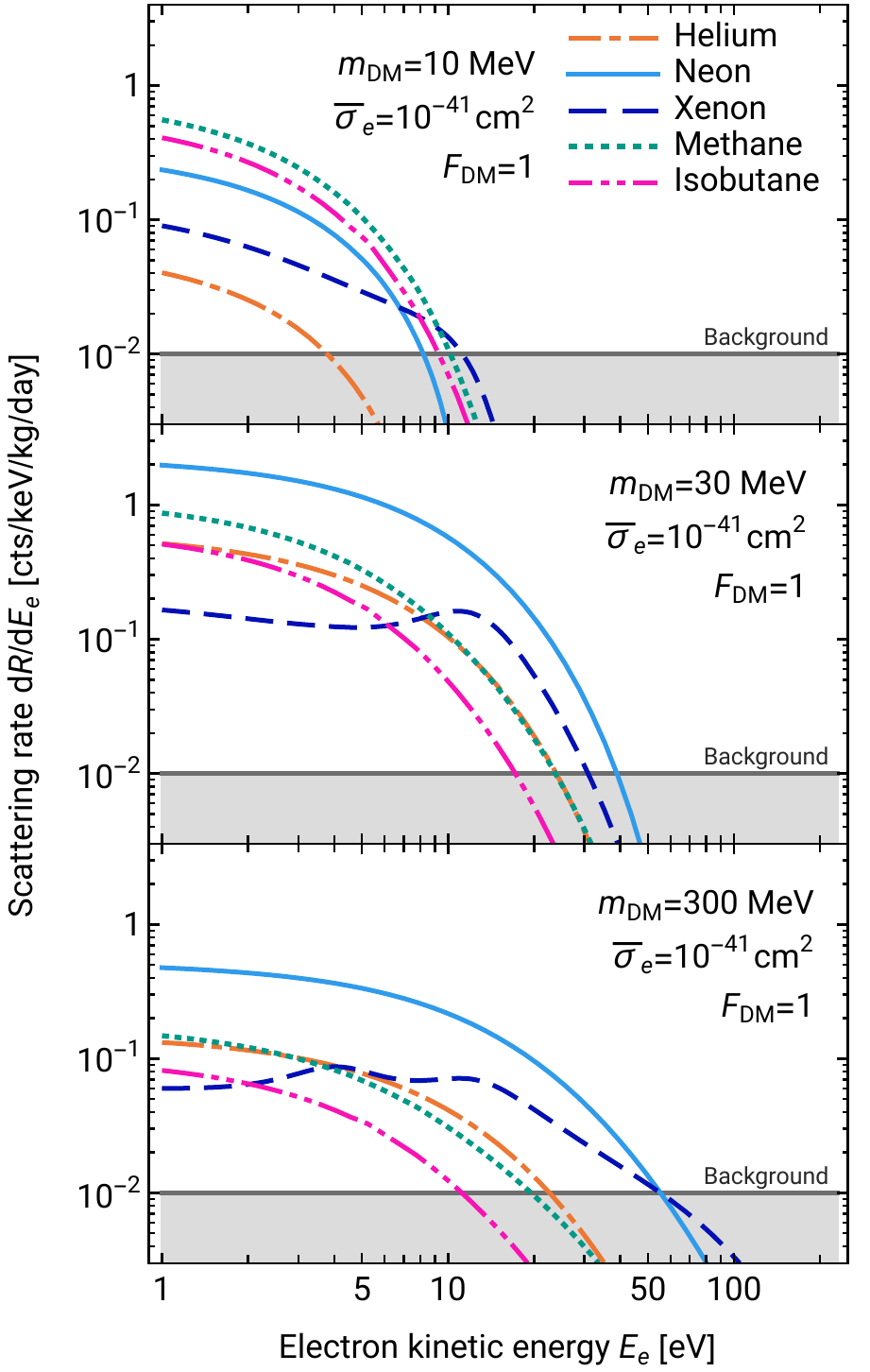}
\caption{\label{fig:dRdEFDM1}  
The differential event rates for DM-induced ionisation for DM with mass 10~MeV (top), 30 MeV (middle), and 300~MeV (bottom).
Rates are shown for helium (dot-dashed orange), neon (light blue), xenon (dashed dark-blue), methane (dotted green), and isobutane (dot-dot-dashed pink).
In all three cases, we show results for $\bar{\sigma}_e=10^{-41}$~cm$^2$ and $F_{\mathrm{DM}}=1$. 
Neon tends to have the largest rate, except at the lowest mass (top panel), where both neon and helium suffer a
 suppression from the relatively high ionisation energy of the outer shell.
 The features in the xenon curve arise from the interplay of the $5p$, $5s$ and $4d$ orbitals.
 The grey shaded region is an estimate of the background rate for \DarkSphere assuming that
 it is situated in the Large Experimental Cavern at the Boulby Underground Laboratory.
}
\end{figure}

With the ionisation form factors in hand, it is now straightforward to calculate the
differential cross section using eq.~\eqref{eq:dsigmadE} and the differential scattering rate using eq.~\eqref{ER}.
In fig.~\ref{fig:dRdEFDM1}, we plot the scattering event rates as a function of the ionised electron's kinetic energy $E_e$ for the
 five different gases assuming $\bar{\sigma}_e=10^{-41}~\mathrm{cm}^2$, $F_{\rm{DM}}=1$ and for three values of the DM mass: 
$m_{\rm{DM}}=10$~MeV (top), $m_{\rm{DM}}=30$~MeV (middle), and $m_{\rm{DM}}=300$~MeV (bottom).
For all targets except helium, more than one orbital contributes to the scattering rate.
This is most apparent in the xenon rate (dashed dark-blue line) in the bottom panel,
where the features at different energies arise from the interplay of the $5p$, $5s$, and $4d$ orbitals, which have been summed 
together.\footnote{Appendix~\ref{app:comparions_rates} provides a comparison of our xenon scattering rate calculation with other xenon calculations in the literature.}

For $m_{\rm{DM}}=30$~and 300~MeV, we see that the largest rate occurs for neon (solid blue line),
except for the high-energy tail above $E_e\sim60$~eV for $m_{\mathrm{DM}}=300$~MeV where xenon takes over.
Neon has a number of advantages compared to the other gases. 
Compared to helium, there are three times as many electrons available in the outer-shell.
Compared to methane and isobutane, its form factor falls off more slowly at large~$q$,
as discussed in the previous section.
And compared to xenon, the rate suffers less of a suppression by the~$1/m_A$ factor in the scattering rate (cf.\ eq.~\eqref{ER}).

\begin{figure}[t!]
\includegraphics[width=0.95\columnwidth]{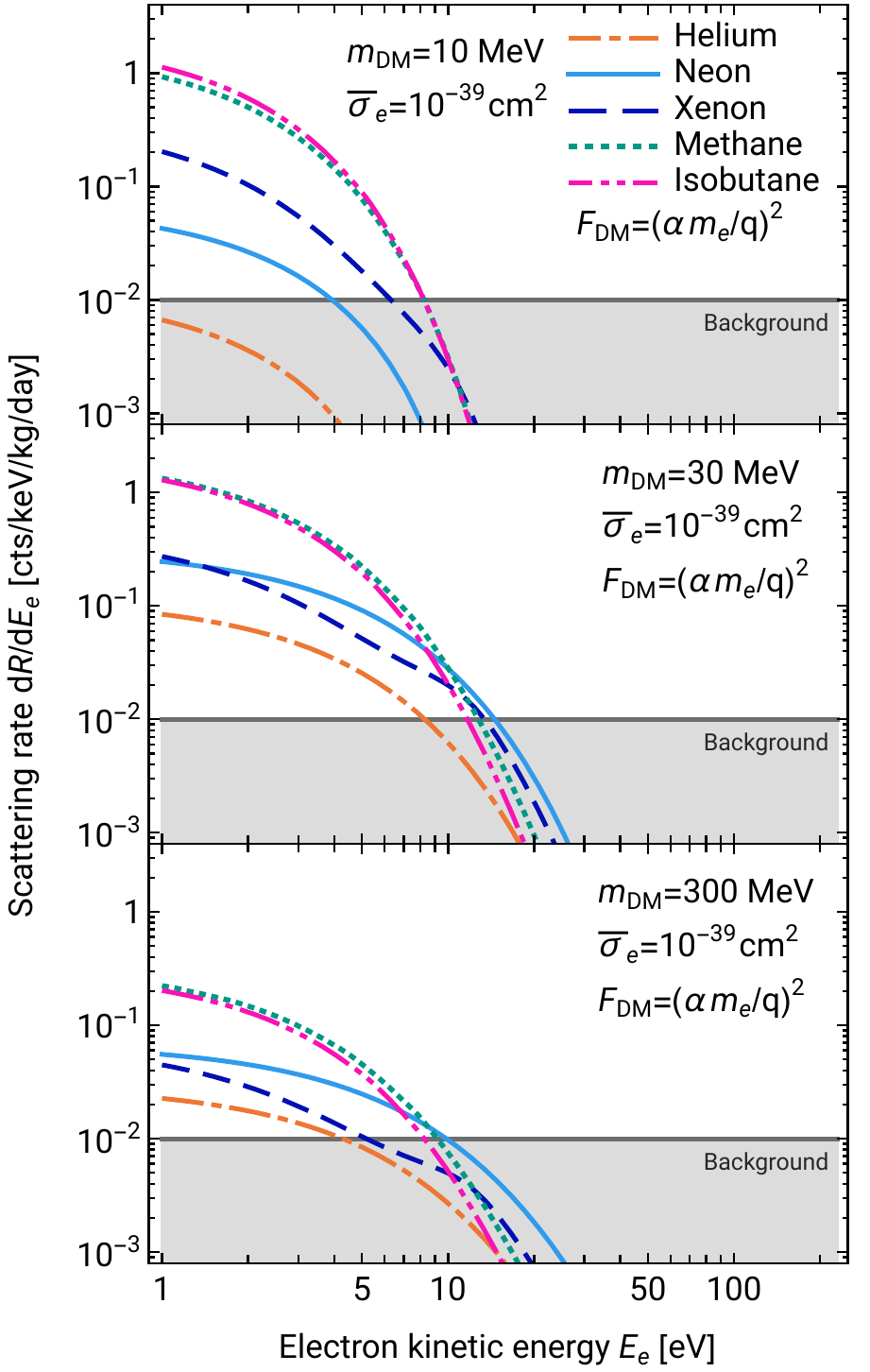}
\caption{\label{fig:dRdEFDMq}  
Similar to fig.~\ref{fig:dRdEFDM1} but with $F_{\rm{DM}}=(\alpha m_e/q)^2$ and $\bar{\sigma}_e=10^{-39}~\mathrm{cm}^2$.
For this form factor, the rates fall off much more rapidly as $E_e$ increases compared to $F_{\rm{DM}}=1$.
Methane and isobutane tend to dominate at low-energies (below $\sim10$~eV) while above this energy, the rate for neon or xenon is larger.
}
\end{figure}

For DM masses between 300~MeV and 1~GeV, the scattering rates have the same shape but scale with the $1/m_{\mathrm{DM}}$ factor in eq.~\eqref{ER}.
At $m_{\rm{DM}}=10$~MeV (top panel), the situation is different.
Both the neon and helium rates (solid blue and dashed-dot orange lines, respectively) are more suppressed relative to the other gases. 
This follows as a result of the larger outer-shell ionisation energies in neon and helium compared to the other targets (cf.\ tab.~\ref{table_energies}).
Comparing methane and isobutane (dashed green and dot-dot-dashed pink lines, respectively), 
the relative scattering rates again depend on the competition between more electron targets in isobutane versus the larger suppression from~$1/m_A$.

In fig.~\ref{fig:dRdEFDMq}, we plot the scattering event rates for the five gases
but this time for $F_{\rm{DM}}=(\alpha m_e/q)^2$, where~$\alpha$ is the fine-structure constant.
This form for $F_{\rm{DM}}$ is appropriate for models that have a light-particle mediating the interaction between DM and electrons.
In fig.~\ref{fig:dRdEFDMq}, we also choose a larger value of the reference cross section, $\bar{\sigma}_e=10^{-39}~\mathrm{cm}^2$,
but this is still below current bounds for this form of $F_{\rm{DM}}$ (cf.\ fig.~\ref{fig:sensitivityall_q-2}).
The top, middle, and bottom panels again show the scattering rates for $m_{\rm{DM}}=10$, 30, and 300~MeV, respectively.

Comparing figs.~\ref{fig:dRdEFDM1} and~\ref{fig:dRdEFDMq}, there are several obvious differences.
Firstly, the rates in fig.~\ref{fig:dRdEFDMq} all drop off more quickly as $E_e$ increases. This is especially noticeable at $m_{\rm{DM}}=300$~MeV.
The rapid fall-off occurs because an increase in $E_e$ leads to an increase in $q_{-}$,
and when the ionisation form factors are multiplied by the additional $1/q^4$ factor, the cross section integrands drop off even more quickly with $q$,
so the scattering rates are even more sensitive to $q_{-}$. 
Secondly, in fig.~\ref{fig:dRdEFDMq} the largest rates at small values of $E_e$ occur for methane and isobutane at all three values of the mass shown,
while for $F_{\rm{DM}}=1$, methane and isobutane were the largest only for $m_{\rm{DM}}=10$~GeV.
This is again directly related to the sensitive dependence on $q_{-}$. 
As neon and helium have larger values of the ionisation energy, this in turn means that the $q_{-}$ values are larger (cf.\ eq.~\eqref{eq:qmaxmin}).
Although xenon also has a low ionisation energy, the larger~$1/m_A$ suppression means that its rate is below methane and isobutane.

The sensitivity to DM-electron scattering will ultimately be limited by the background rate.
The S140 and {\sc ECuME} SPC experiments are projected to achieve background rates of $\mathcal{O}(1)~\!\!\text{cts/keV/kg/day}$ and 
$\mathcal{O}\left(10^{-1}\right)~\!\!\text{cts/keV/kg/day}$, respectively, in the energy range below 1~\!keV~\cite{NEWS-G:2022kon}.
\DarkSphere aims to achieve a background rate of $\mathcal{O}\left(10^{-2}\right)~\!\text{cts/keV/kg/day}$ below 1~\!keV.
The grey line and shaded region in figs.~\ref{fig:dRdEFDM1} and~\ref{fig:dRdEFDMq} show an energy-independent 
background rate of $10^{-2}~\!\text{cts/keV/kg/day}$ 
below 1~\!keV, which we will use in our sensitivity calculations for all gas mixtures.

A preliminary study~\cite{Balogh:2023hba} shows that a background rate of $10^{-2}~\!\text{cts/keV/kg/day}$ could be achieved
with a fully electroformed-underground copper shell and by shielding the $3$\,m diameter of \DarkSphere
with a $2.5$\,m thick water tank in the Large Experimental Cavern (LEC) at the Boulby Underground Laboratory.
After fully electroforming the copper shell underground, it is expected that contaminants from $^{210}$Pb and progeny decays 
would contribute a background rate $\sim 2\times10^{-5}~\!\text{cts/keV/kg/day}$ below 1~\!keV~\cite{Balogh:2023hba}, which is significantly smaller than the targeted rate.
Instead, the dominant background is expected to arise from environmental backgrounds, specifically from neutrons, photons and muons
that scatter in the SPC. Muons are present in cosmic rays while the neutrons and photons are generated in the cavern rock, 
either from radioactivity or from cosmic-ray interactions
(see refs.~\cite{Kudryavtsev:2003aua, Smith:2005se, Malczewski:2013lqy, brossard:tel-02923528} for related studies).
A Geant4~\cite{GEANT4} simulation assuming a $2.5$\,m thick water tank in the LEC at Boulby surrounding the \DarkSphere SPC, and assuming a 90\%--10\% He--C$_4$H$_{10}$ 
gas mixture, predicts a photon, neutron and muon-induced environmental background rate below 1~keV of $\sim 5 \times 10^{-3}~\!\text{cts/keV/kg/day}$~\cite{Balogh:2023hba}.
The dominant contribution arises from 2-to-3~\!MeV photons, which produce an almost-flat spectrum below 1~keV.

Although both the contaminants in the copper shell and the environmental backgrounds are below the targeted rate,
there are two further sources of background that could increase the background rate above $10^{-2}~\!\text{cts/keV/kg/day}$ below 1~\!keV.
The first is from potential radioactive contaminants in the gas mixture, with $^{14}\mathrm{C}$ and $^{3}\mathrm{H}$ being of 
particular concern (see, e.g., ref.~\cite{Amare:2017roa}). 
Tritium can be produced in neon when exposed to cosmic rays and
estimates in~ref.\cite{brossard:tel-02923528} show that the gas should remain on the surface for less than one~month
to ensure that the background from $^{3}\mathrm{H}$ remains sub-dominant.
Methane or isobutane could provide more of a challenge since they are composed solely of~C and~H 
isotopes and either gas will generally be present in their role as a quench gas.
Both methane and isobutane are produced from underground natural gas deposits, which will have been underground for geological timescales,
so the intrinsic contamination from $^{14}\mathrm{C}$ and $^{3}\mathrm{H}$ should only be introduced when extracted, manipulated 
on the surface, or transported. 
The NEWS-G Collaboration will need to work with gas-manufacturing companies to ensure that sufficiently pure gas is procured and that time on the surface is minimised.

The second background source that could increase the rate beyond $10^{-2}~\!\text{cts/keV/kg/day}$
 is the excess low-energy background below $\mathcal{O}(100)~\!\mathrm{eV}$ that has been observed in 
 many DM experiments~\cite{Fuss:2022fxe}. Indeed, a test run of the S140 detector with methane gas 
 found a large excess of events in the single-electron channel 
(corresponding to an energy below  $\sim28$\!~eV), 
with an event rate orders of magnitude larger than in the two-electron channel. 
The single-electron events cannot currently be removed by cuts, so physics searches may be limited to the 
two-electron channel (corresponding to an energy $\gtrsim28$\!~eV)~\cite{Fuss:2022fxe}.

With these caveats in mind, we will proceed assuming the energy-independent background rate of $10^{-2}~\!\text{cts/keV/kg/day}$ 
below 1~\!keV as shown in figs.~\ref{fig:dRdEFDM1} and~\ref{fig:dRdEFDMq}.
The smallest value of the cross section that can be constrained will be approximately determined by the point at which 
the signal rate is the same magnitude as the background rate
(we will use a more rigorous likelihood-based method in the next section to more carefully quantify the sensitivity).
The reference cross section $\bar{\sigma}_e=10^{-41}~\mathrm{cm}^2$ used in fig.~\ref{fig:dRdEFDM1}
 is below current constraints for the three values of the mass
shown (cf.~fig.~\ref{fig:sensitivityall_q1}), so since all of the rates in fig.~\ref{fig:dRdEFDM1} are above the background rate for some values of~$E_e$,
this indicates that \DarkSphere has the potential to probe new regions of 
parameter space. 
Similarly, we see that the signal rates in fig.~\ref{fig:dRdEFDMq} are 
above the background rate for a cross section below current bounds, so we anticipate that \DarkSphere has the potential to probe new regions of 
parameter space for the $F_{\rm{DM}}=(\alpha m_e/q)^2$ form~factor.

\begin{figure*}[t!]
\includegraphics[width=0.95\columnwidth]{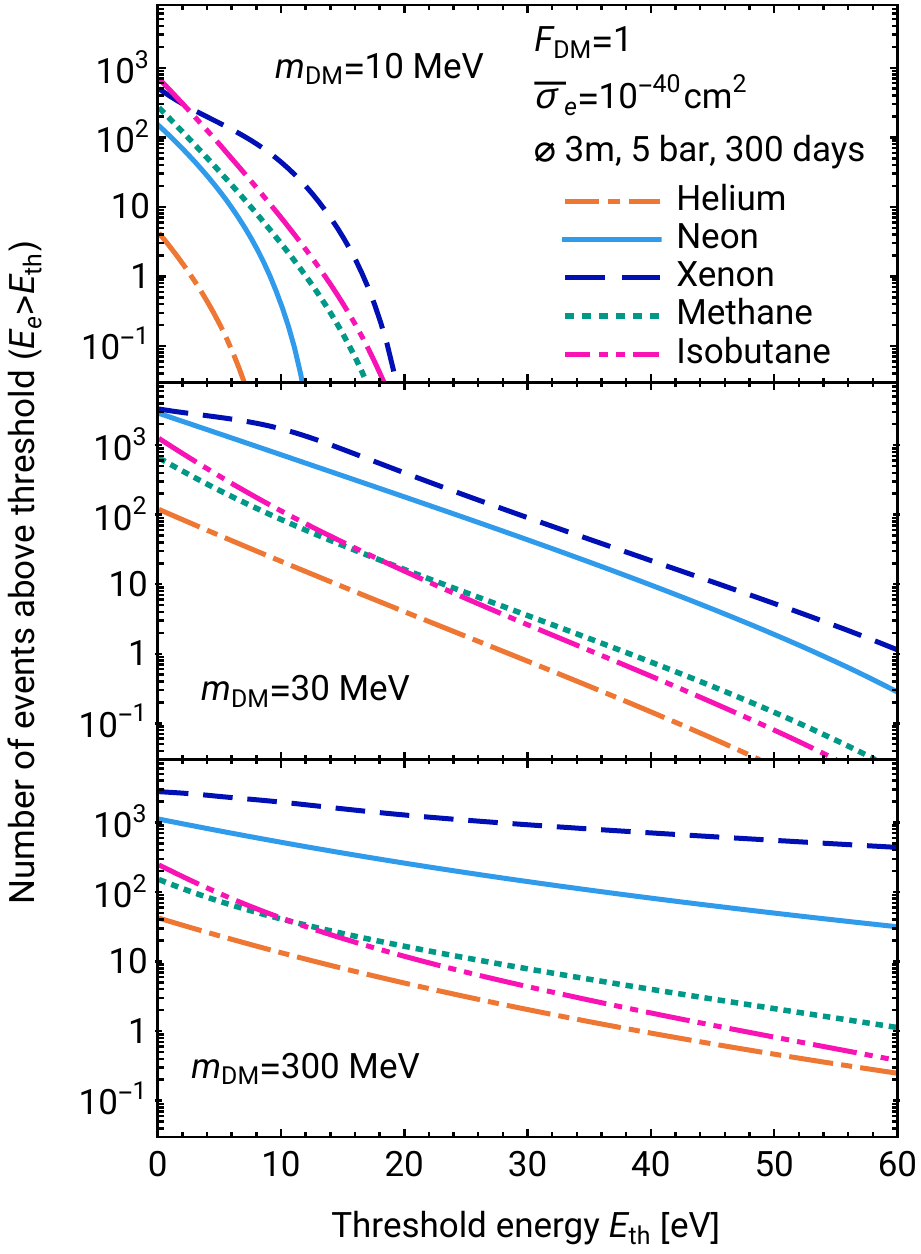}\hspace{5mm}
\includegraphics[width=0.95\columnwidth]{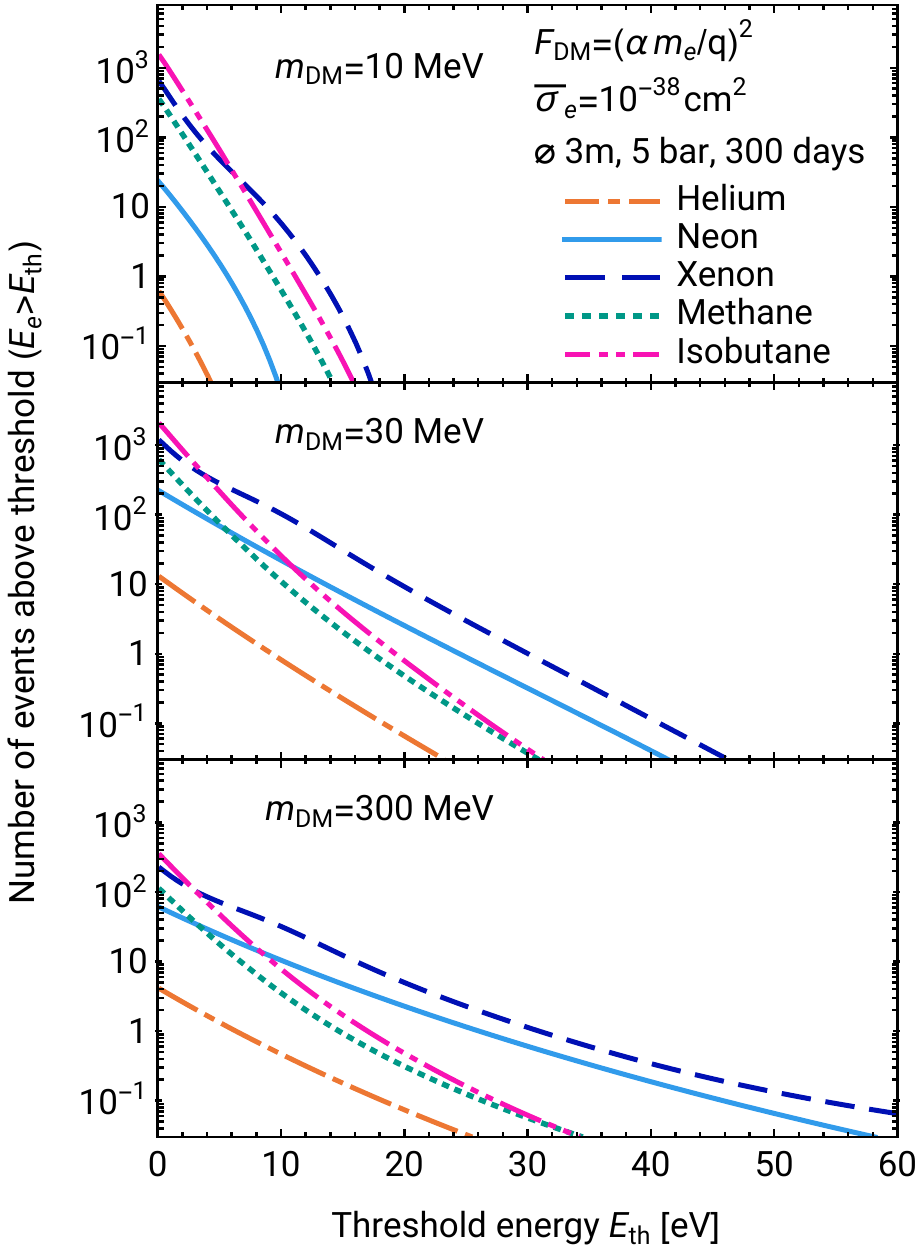}
\caption{\label{fig:events}  
The total (integrated) number of events above a given threshold energy as a function of the threshold energy
for DM with mass 10 (top), 30 (middle), and 300~MeV (bottom).
The number of events is shown for $F_{\mathrm{DM}}=1$ in the left panel and $\bar{\sigma}_e=10^{-40}$~cm$^2$, and for
$F_{\rm{DM}}=(\alpha m_e/q)^2$ and $\bar{\sigma}_e=10^{-38}~\mathrm{cm}^2$ in the right panel.
We have assumed the baseline \DarkSphere parameters: an SPC with a 3~\!m diameter filled with the target gas at a pressure of 
5~bar, temperature of $20^\circ$C, and taking data for 300~days. Over most of the parameter space xenon gives the largest number of events as
it has the largest target mass (380.8~kg), while helium gives the smallest number of events as
it has the smallest target mass (11.6~kg). The lines in the right panel drop off more rapidly as $E_{{\rm{th}}}$ increases
compared to the left panel, indicating that a smaller threshold energy is required to observe events when $F_{\rm{DM}}=(\alpha m_e/q)^2$.
}
\end{figure*}

\section{Sensitivity projections}\label{section:sensitivities}

In this section, we will present projections for \DarkSphere in the DM mass -- cross section parameter space.
Before showing the results, we should clarify the expected behaviour
inside the \DarkSphere detector after ionisation from the atom or molecule has occurred
(see ref.~\cite{Knollbook1981} for a more in-depth discussion).
After ionisation, the electric field will cause the ionised electron to drift to the anode,
where a Townsend avalanche is triggered when the electron is within a few millimetres of the anode.
This avalanche produces a large number of secondary electron-ion pairs that are detected as a charge pulse.
However, there are two additional processes that may induce additional primary electrons,
which if produced, will also drift toward the anode and increase the size of the pulse.
Firstly, if the kinetic energy of the ionised electron is larger than the energy required to
create a new electron-ion pair in the gas, additional primary electrons can be produced.
Secondly, if an electron is ionised from an inner-orbital of the atom or molecule, 
a higher-energy bound electron will transition to the hole, and additional energy will be released, 
e.g.,\ in the form of a photon or Auger electron.
An Auger electron is obviously an additional electron, while
more primary electrons can be produced if the photon or Auger-electron energy is larger than the electron-ion pair threshold energy.

Ideally, we would show scattering rates in terms of the experimentally measurable parameters 
such as the number of primary electrons created or the number of photoelectrons detected.
However, this requires a precise characterisation of the detector
response at very low energies, which is not yet available for all of the gases and gas mixtures that we will consider (although important steps 
in this direction have been taken in ref.~\cite{Arnaud:2019nyp,brossard:tel-02923528}).
Therefore, we will instead adopt a simpler approach and discuss the experimental sensitivity only in terms of the kinetic energy of the ionised electron ($E_e$).
This means that our treatment is somewhat conservative as it ignores the contribution from any 
additional primary electrons that could be produced, e.g.,~in the case that an inner-orbital is ionised.
This is likely to lead to an under-estimate of the sensitivity, particularly for xenon where
the difference between the orbital ionisation energies can be relatively large ($\gtrsim 100$~eV).

Characterising the sensitivity estimates in terms of $E_e$ also means that we should 
define the analysis threshold in terms of $E_e$.
To investigate the dependence of the number of signal events on
the threshold energy, in fig.~\ref{fig:events} we have plotted 
the number of events
above a given threshold energy as a function of the threshold energy ($E_{\rm{th}}$).
The left panels are for $F_{\rm{DM}}=1$ and $\bar{\sigma}_e = 10^{-40}~\mathrm{cm}^2$, while the right panels are for 
$F_{\rm{DM}}=(\alpha m_e /q )^2$ and $\bar{\sigma}_e = 10^{-38}~\mathrm{cm}^2$. The top, middle, and lower panels are for
$m_{\mathrm{DM}}=10,~30$, and~$300~\mathrm{MeV}$, respectively.
To calculate the total number of events, we have assumed the baseline
detector and operating parameters for \DarkSphere from~\cite{Balogh:2023hba}; this assumes
an SPC with a 3~\!m diameter (\diameter) filled with the target gas at a pressure of 
5~bar, temperature of $20^\circ$C and  taking data for 300~days. 
The same pressure for the different gases implies that the total mass
 inside the detector is different for each gas. We find that
there should be approximately 2900~mol for each gas, corresponding to
11.6, 58.5, 380.8, 46.5 and 168.6~kg for helium, neon, xenon,
methane, and isobutane, respectively. 

From fig.~\ref{fig:events} we observe the general pattern that xenon
gives the largest number of events and helium gives the smallest number.
This follows primarily from the large difference in the target mass of the two targets.
We also see that for the chosen cross sections, which lie below current
exclusion bounds, the total number of events can be rather large ($\gtrsim 1000$~events).
When $m_{\rm{DM}}=10$~MeV (top panels), we see for both choices of $F_{\rm{DM}}$
that the number of events falls off rapidly as the threshold energy increases, and falls to zero completely for $E_{\rm{th}}\gtrsim20$~eV.
For higher masses and when $F_{\rm{DM}}=1$, the left-middle and left-lower panels show that the drop off in the number of events as the threshold increases is much shallower,
such that the number if events is still above one (and still significantly above one for neon and xenon gases) at $E_{\mathrm{th}}\simeq30$~eV.
Conversely, the right panels show that when $F_{\rm{DM}}=(\alpha m_e /q )^2$, the drop off in the number of events remains significant, even at larger masses. 
This is a reflection of the behaviour observed previously in the discussion of fig.~\ref{fig:dRdEFDMq}.

\subsection{Single-species projections}

\begin{figure}[t!]
\includegraphics[width=0.95\columnwidth]{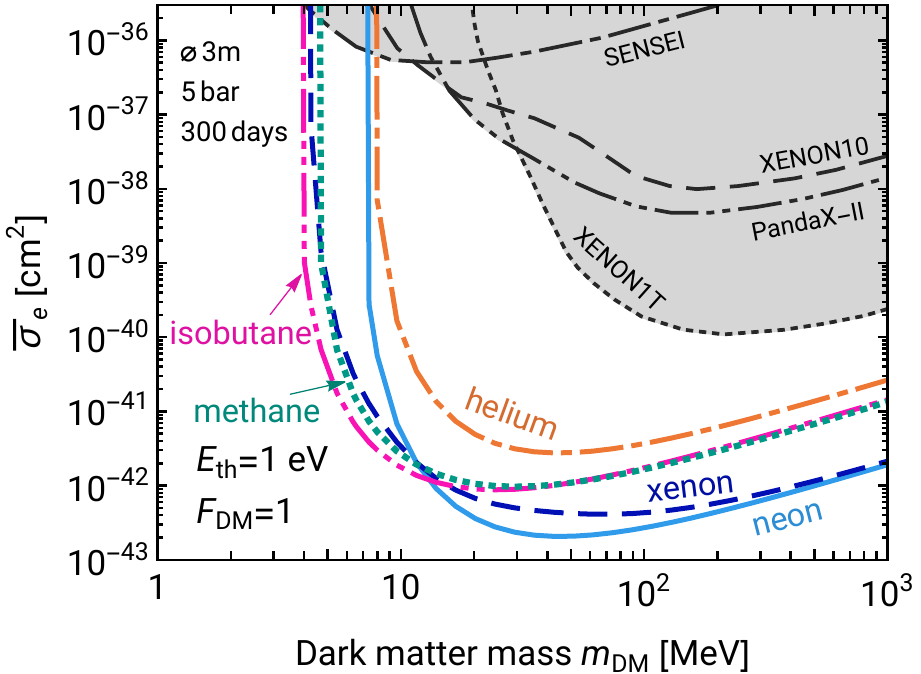}\\
\hspace{10mm}
\includegraphics[width=0.95\columnwidth]{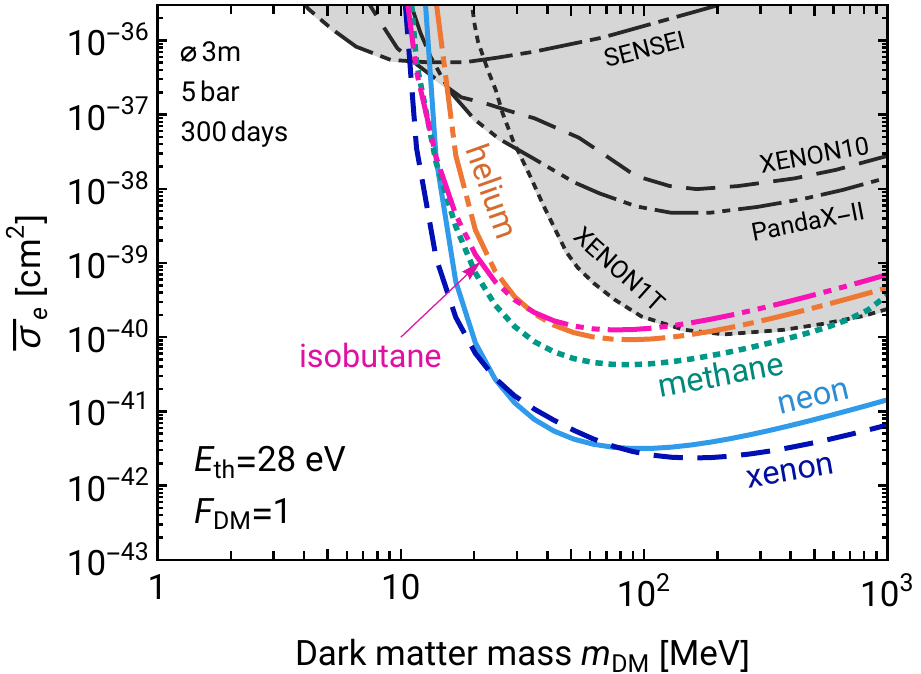}
\caption{\label{fig:sensitivityall_q1}  
Projected 90\% C.L.\ exclusion limits for $F_{\rm{DM}}=1$ when pure mixtures of only helium, neon, xenon, methane, or isobutane are used.
We have assumed the background rate shown in fig.~\ref{fig:dRdEFDM1} together with the baseline \DarkSphere parameters: 
an SPC with a 3~\!m diameter filled with the target gas at a pressure of 5~bar, temperature of $20^\circ$C, and taking data for 300~days.
The upper panel assumes a threshold energy $E_{\rm{th}}=1$~eV, chosen to mimic a single-electron threshold, while the
bottom panel assumes $E_{\rm{th}}=28$~eV, to mimic a two-electron threshold.
Neon performs well at higher DM masses but is worse at lower DM masses where xenon, methane and isobutane benefit from lower ionisation energies.
Grey bands show current exclusion limits, so we see that \DarkSphere has the potential to exclude unexplored parameter space with any of
the five gas targets shown.
}
\end{figure}

We show in fig.~\ref{fig:sensitivityall_q1} our projections for the 90\% confidence limit (C.L.) exclusion limits for pure 
gas mixtures assuming $F_{\rm{DM}}=1$ for the baseline
detector and operating parameters for \DarkSphere from~\cite{Balogh:2023hba}.
As a reminder, this assumes an SPC with a 3~\!m diameter filled with the target gas at a pressure of 
5~bar, temperature of $20^\circ$C and  taking data for 300~days with the background rate shown in figs.~\ref{fig:dRdEFDM1} and~\ref{fig:dRdEFDMq}.

The two panels in fig.~\ref{fig:sensitivityall_q1} show projections under two assumptions for the detection threshold.
The upper panel in fig.~\ref{fig:sensitivityall_q1} assumes a threshold energy $E_{\rm{th}}=1$~eV.
This means that essentially all electrons that have been ionised
will be pulled by the electric field in the \DarkSphere drift region towards the anode, where the signal will be read out.
This threshold is chosen to mimic a single-electron search threshold since
electrons with a kinetic energy around $10$~eV or smaller will not have sufficient energy to generate additional electrons in the gas.
This is an idealised scenario as in a real detector, there will be an imperfect efficiency to detect single electrons. 
For instance, the test run of the NEWS-G S140 detector achieved a 50\% detection efficiency for single electrons~\cite{Fuss:2022fxe}.
This calculation also assumes that the Geant4 background model is sufficiently accurate
all the way to 1~eV and includes all sources that contribute to the single-electron background.
As discussed in section~\ref{sec:DM_ER}, this is also an idealised scenario and therefore the results in the upper panel of fig.~\ref{fig:sensitivityall_q1} 
should be understood as an indication of the absolute best-case sensitivity given the \DarkSphere detector and operating parameters.

Arguably a more realistic scenario is shown in the lower panel in fig.~\ref{fig:sensitivityall_q1}, which assumes a threshold energy $E_{\rm{th}}=28$~eV.
This energy is chosen to mimic a detection threshold of at least two electrons.
As discussed in section~\ref{sec:DM_ER}, the test run of the S140 experiment found a large excess of events in the single-electron channel, with an event rate orders of magnitude larger than in the two-electron channel, so the two-electron channel may be preferable for physics searches~\cite{Fuss:2022fxe}.
The justification for $28$~eV as a proxy for a two-electron threshold follows from a consideration of the $W$-values.
Recall that the $W$-value is the mean energy to create a new electron-ion pair and its value
depends on the gas under consideration.\footnote{The $W$-value is the mean energy and not the minimum energy to create a new electron-ion pair. 
The minimum energy is lower than~$W$~\cite{Combecher1980},
and is around the value of the ionisation energies in tab.~\ref{table_energies}.}
The asymptotic value for xenon is $W\sim22$~eV~\cite{Combecher1980}, 
for methane and isobutane is $W\sim 28$~eV~\cite{Combecher1980}, and for neon is $W\sim37$~eV~\cite{Parks1972}, 
while for helium, it is slightly higher at $W\sim48$~eV~\cite{Dalgarno1999}.\footnote{The difference between the $W$-value at approximately $30$~eV and the asymptotic value is $\lesssim 10\%$ for a wide range of gases (including methane)~\cite{Katsioulas:2021sgl}.             }
Although in fig.~\ref{fig:sensitivityall_q1} we show results for a single-species gas, in reality, the noble gases
would always be mixed with a small component of methane or isobutane to act as quench gases.
For gas mixtures, the $W$-value for the mixture is much closer 
to the lowest $W$-value of the gases in the mixture~\cite{Parks1972,Vinagre2000}. For example, $W\approx28$~eV for a $\mathrm{Ne}+\mathrm{CH}_4 \,(2\%)$ mixture~\cite{Arnaud:2019nyp}.
Therefore, we expect that at or above approximately $28$~eV for all gases the ionised electron has enough energy to be above the mean energy to create a new electron-ion pair. In the lower panel, we again assume the idealised scenario where the efficiency factor is 100\% above the threshold energy.

In the calculation of the 90\% C.L.\ limits, we use a binned-likelihood approach and employ the exclusion-limit test-statistic
and `Asimov' data approach from ref.~\cite{Cowan:2010js}. We bin the data into equally sized bins of width 14~eV 
from the threshold energy up to 99 (98)~eV for $E_{\rm{th}}=1\, (28)$~eV. 
We use an energy resolution of the form 
$\sigma(E^{\rm{eV}}_e)/E^{\rm{eV}}_e = \alpha+\beta/\sqrt{E^{\rm{eV}}_e}$, where $\sigma(E^{\rm{eV}}_e)$ and $E^{\rm{eV}}_e$ have units 
of eV, and $\alpha=4.4\times 10^{-3}$ and $\beta=5.28$ are estimated by fitting to the 270 and 2822~eV lines in~\cite{Arnaud:2019nyp}.
In the case of $E_{\rm{th}}=28$\,eV, 
since $\sigma(E^{\rm{eV}}_e=28\,\mathrm{eV})\simeq 28\,\mathrm{eV}$, to prevent 
upward fluctuations biasing the limits, we multiply the rate by the rapidly falling error function,
$\tfrac{1}{2}(1+\erf{[(E^{\rm{eV}}_e - 28\,\mathrm{eV})/(\sqrt{2}\cdot 3\,\mathrm{eV})}])$, 
which only allows upward fluctuations from a small energy range below the threshold.
Finally, we also include a 10\% uncertainty in the overall normalisation of the
background rate.

At DM masses above 100~MeV in both panels of fig.~\ref{fig:sensitivityall_q1}, we see that neon and xenon 
constrain smaller values of the cross section compared to helium, methane, and isobutane.
At smaller masses, gases that have the lowest ionisation energies perform best, so around 10~MeV,
 isobutane and methane perform better than neon.
Comparing the upper and lower panels, we see that the effect of the higher threshold is an overall shift to higher cross sections
by a factor $\sim3$ for xenon, $\sim7$ for neon, $\sim18$ for helium, $\sim27$ for methane, and $\sim49$ for isobutane,
accompanied by a more severe weakening in the limits for DM masses below around 20~MeV.
 The grey regions in fig.~\ref{fig:sensitivityall_q1} show the current constraints from PandaX-II~\cite{PandaX-II:2021nsg}, SENSEI~\cite{Barak:2020fql}, XENON10~\cite{Essig:2017kqs}, and XENON1T~\cite{Aprile:2019xxb}, 
so we see that, even for the higher threshold scenario, the projected limits from all gases exclude regions of unexplored parameter space.

\begin{figure}[t!]
\includegraphics[width=0.95\columnwidth]{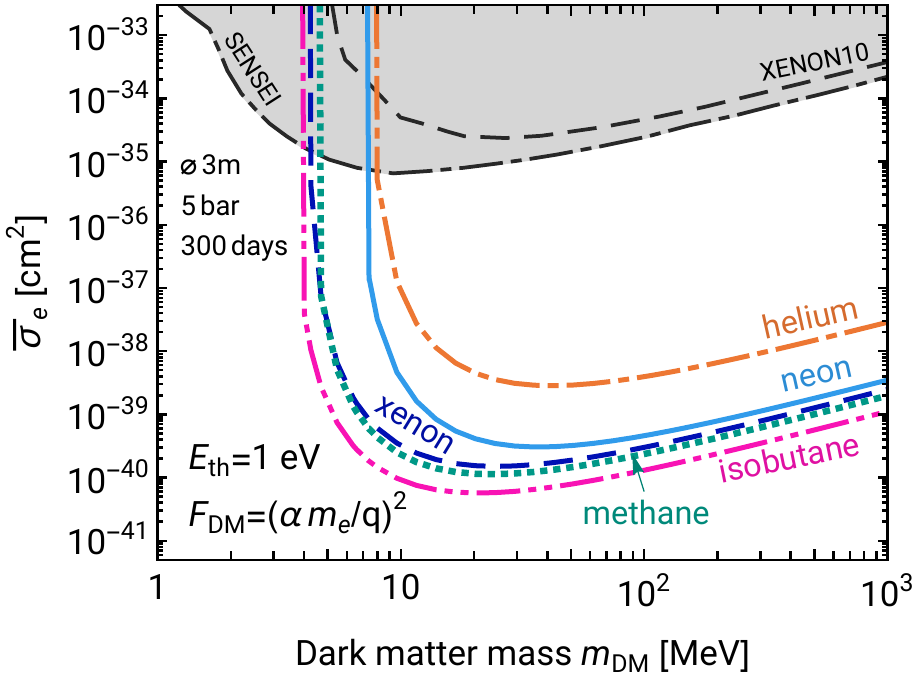}\\
\hspace{10mm}
\includegraphics[width=0.95\columnwidth]{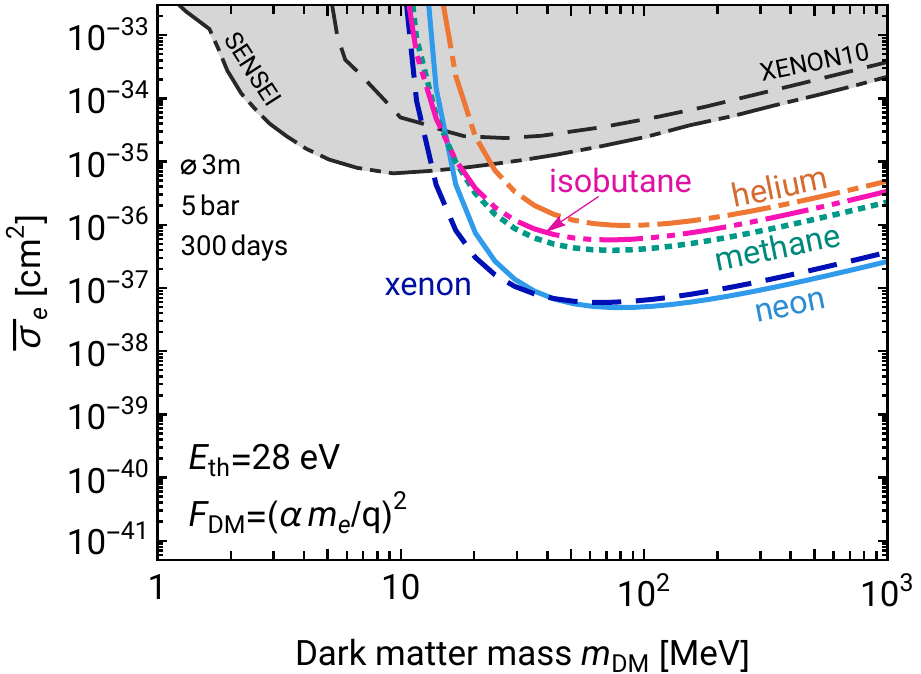}
\caption{\label{fig:sensitivityall_q-2}  
Similar to fig.~\ref{fig:sensitivityall_q1} but with $F_{\rm{DM}}=(\alpha m_e /q )^2$.
When $E_{\rm{th}}=1$\,eV, we see that methane and isobutane can outperform the noble atom targets but suffer a larger suppression factor
when moving to $E_{\rm{th}}=28$\,eV.
Grey bands show current exclusion limits, so we again see that \DarkSphere has the potential to exclude regions of unexplored parameter space with any of
the five gas-targets shown.
}
\end{figure}

In fig.~\ref{fig:sensitivityall_q-2}, we show the analogous 90\% C.L.\ limits for $F_{\rm{DM}}=(\alpha m_e /q )^2$.
Again, we show projections for $E_{\rm{th}}=1$~eV (top panel) and $E_{\rm{th}}=28$~eV (bottom panel)
to mimic a single-electron and multi-electron search.
The same assumptions about the background and resolution and the same binned-likelihood method as used in the $F_{\rm{DM}}=1$
calculation are used to generate these limits.
In the upper panel of fig.~\ref{fig:sensitivityall_q-2}, we see that neon, xenon, methane, and isobutane all have a similar performance for DM masses above 100~MeV,
while helium is higher by a factor $\sim 10$.
Comparing the upper and lower panels, we find a more dramatic reduction in the sensitivity relative to the  $F_{\rm{DM}}=1$ scenario.
Here, the shift to higher cross sections is a factor $\sim100$ for xenon, neon, and helium and a factor $\sim1000$ for methane and isobutane.
The larger reduction in sensitivity follows from the more rapid drop-off in the scattering rate with energy when $F_{\rm{DM}}=(\alpha m_e /q )^2$,
as was highlighted previously in the discussion surrounding figs.~\ref{fig:dRdEFDMq} and~\ref{fig:events}.
The grey-shaded regions show the current constraints from SENSEI~\cite{Barak:2020fql} and XENON10~\cite{Essig:2017kqs} on this parameter space, 
and we again observe that the projected limits from all gases exclude regions of unexplored parameter space.

\subsection{Mixed-species projections}

\begin{figure}[t!]
\includegraphics[width=0.95\columnwidth]{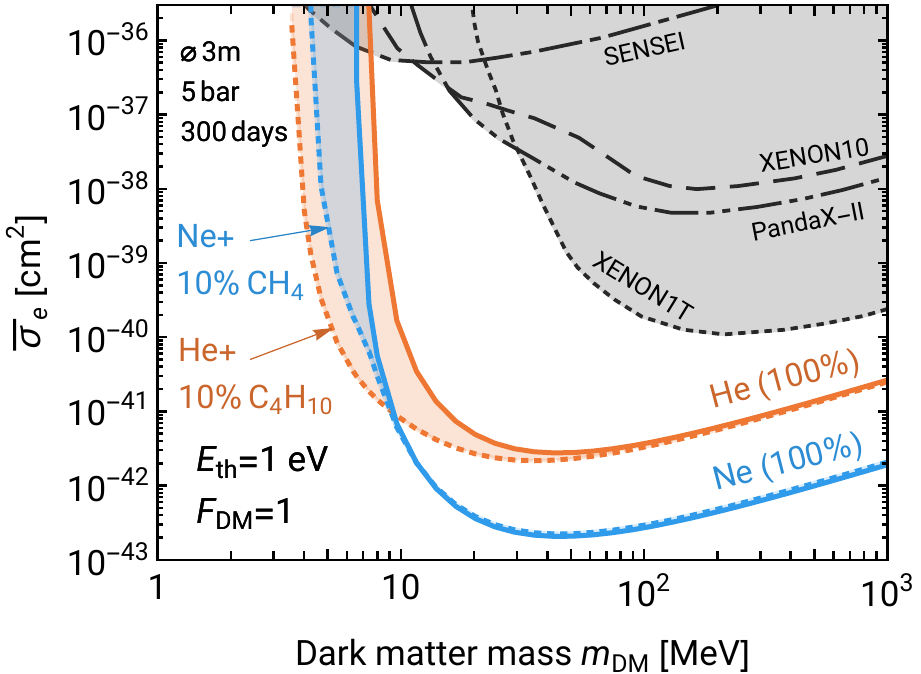}\\
\hspace{10mm}
\includegraphics[width=0.95\columnwidth]{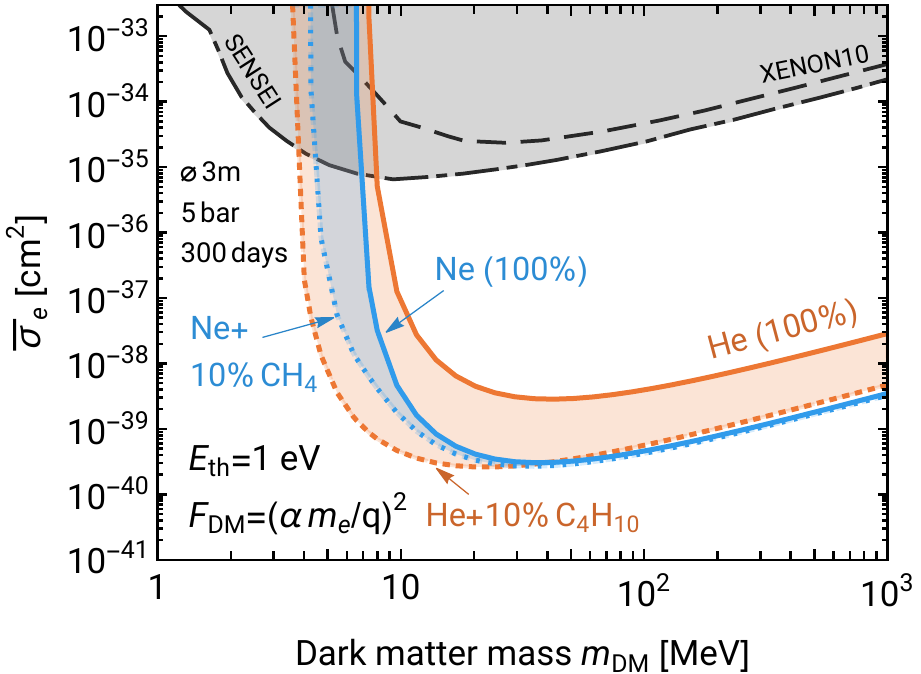}
\caption{\label{fig:sensitivityall_mixes}  
Projected 90\% C.L.\ exclusion limits for gases consisting of pure-helium (solid orange) and pure-neon (solid blue), 
as well as the more realistic scenario of gas mixtures: 90\% helium and 10\% isobutane in dashed orange 
and 90\% neon and 10\% methane in dashed blue. The coloured shading indicates the exclusions limits for a fraction of the
molecular gases between 0\% and 10\%.
In both panels, we assume the baseline \DarkSphere parameters and $E_{\rm{th}}=1$~eV.
The upper panel has $F_{\rm{DM}}=1$, while the lower panel has $F_{\rm{DM}}=(\alpha m_e /q )^2$.
Adding a 10\% component of isobutane to helium significantly improves the sensitivity over helium alone,
especially in the lower panel.
Adding a 10\% component of methane to neon has a smaller impact,
although still leads to a small improvement at low DM masses.
}
\end{figure}

In this final section, we show projections involving mixtures of gases. As we have stated previously, this is the most likely mode of operation
in a real experiment as the noble gases need to run with some amount of methane or isobutane, since they act as quenchers within the proportional counter~\cite{Knollbook1981}.
Although it is in principle possible to run with a pure mixture of methane or isobutane as they can act as both the `fill' and quench gas, 
as both methane and isobutane are flammable,
in reality a pure mixture of either is unlikely to pass the safety constraints in an underground laboratory.

In fig.~\ref{fig:sensitivityall_mixes}, we show the projected 90\% C.L.\ exclusion limits for helium and neon (solid orange and blue lines) 
as well as for two gas mixtures: 90\% helium and 10\% isobutane (`$\mathrm{He}+10\%\mathrm{C}_4\mathrm{H}_{10}$') as dashed orange; 
and 90\% neon and 10\% methane (`$\mathrm{Ne}+10\%\mathrm{C}\mathrm{H}_{4}$') as dashed blue. A smaller percentage of methane or
isobutane in the mixture will lie in the shaded regions between the dashed and solid lines. 
We have chosen these mixtures as it is anticipated that this neon--methane mixture 
will be used in S140 and {\sc ECuME}, while the initial \DarkSphere studies have assumed this helium--isobutane mixture.
These mixtures imply a total mass of 57.3~kg for neon--methane and 27.3~kg for helium--isobutane in the \DarkSphere spherical chamber.

The upper panel in fig.~\ref{fig:sensitivityall_mixes} shows the projected limits for $F_{\rm{DM}}=1$, 
while the lower panel shows $F_{\rm{DM}}=(\alpha m_e /q )^2$. 
For $F_{\rm{DM}}=1$, the benefit of adding isobutane to helium starts to become apparent for DM masses below about 30~MeV.
Above this mass, the pure-helium and helium--isobutane mixture lines are almost indistinguishable.
The benefit of adding methane to neon is much smaller, with only a noticeable change in a relatively narrow window 
of DM masses between approximately~4 and 8~MeV. At higher DM masses, the difference between neon
and a neon--methane mixture is almost indistinguishable.
The situation changes for $F_{\rm{DM}}=(\alpha m_e /q )^2$. In this case, the sizeable increase in sensitivity of the
helium--isobutane mixture compared to pure-helium
is apparent over the full range of DM masses.
 Even for neon, we see in the lower panel that there is more of an advantage of running with a neon--methane mixture
for DM masses below approximately 20~MeV.

The limits in fig.~\ref{fig:sensitivityall_mixes} both assume $E_{\rm{th}}=1$~eV, however, a similar pattern is
also observed for gas mixtures with $E_{\rm{th}}=28$~eV (not shown): we again find there is an
advantage to the helium--isobutane mixture over a pure-helium gas, while for neon, the benefit of adding methane
is smaller.

For clarity, we have only shown two gas mixtures. However, a helium--methane mixture will be similar to a helium--isobutane mixture, while
a neon--methane mixture will be similar to a neon--isobutane mixture.
This is for two reasons. Firstly, the benefits of methane and isobutane arise largely because of their lower ionisation energies relative to helium and neon,
and importantly, the smallest ionisation energies for both methane and isobutane are relatively similar (cf.\ tab.~\ref{table_energies}).
Secondly, we have demonstrated explicitly that methane and isobutane have a similar level of sensitivity in the analysis of pure gas targets (cf.\ figs.~\ref{fig:sensitivityall_q1} and~\ref{fig:sensitivityall_q-2}).

With these considerations in mind, we can also determine that the benefit of adding methane or isobutane to xenon will be even smaller than for neon.
At higher DM masses, neon and xenon behave similarly (cf.\ figs.~\ref{fig:sensitivityall_q1} and~\ref{fig:sensitivityall_q-2}), so,
 as was the case with neon, we find little increase in sensitivity to a xenon mixture compared to pure xenon at higher DM masses.
 Meanwhile, at lower DM masses, xenon already benefits from a relatively low ionisation energy that is comparable to 
 the methane and isobutane ionisation energies.
 These considerations are borne out by explicit calculations (not shown) in which the 90\% xenon and 10\% methane or isobutane 
 projected limits are almost indistinguishable from the pure-xenon projected limit.
 Therefore, for constraining DM-electron scattering, there is no meaningful increase in sensitivity when adding methane or isobutane to xenon
 (although they are still useful experimentally to act as quenchers in the proportional counter).

\section{Summary and outlook \label{sec:conclusions}}

The NEWS-G Collaboration employs a large spherical proportional counter (SPC)
filled with atomic or molecular gases to search for dark matter (DM).
It has been demonstrated that this detector technology has a single-ionisation electron threshold,
which allows it to set competitive constraints on DM-nucleon interactions in the few-GeV mass range.
However, a single ionisation electron threshold should allow for the possibility of constraining DM-electron interactions (cf.\ fig.~\ref{fig:ideas}).
In this paper, we investigated this possibility in 
the context of {\sc DarkSphere}, a proposal for a 300~cm diameter SPC with a fully electroformed-underground copper shell
located in the Boulby Underground Laboratory.
Proportional counters are most commonly operated with a mixture of gases. Therefore, we considered several
 noble gases -- helium, neon and xenon -- plus methane and isobutane, all of which have been proposed as the fill and quench gases for operation in the SPC.

While several independent DM-electron event rate calculations involving xenon atoms have
been presented in the literature, new calculations for helium, neon, methane, and isobutane were required.
The crucial quantity for calculating the event rate is the dimensionless ionisation form factor,
which is calculated from the bound- and continuum-electron wave functions.
We made use of the PySCF quantum chemistry package to calculate the bound-state wave functions, 
and the output was summarised in tab.~\ref{table_energies} and fig.~\ref{fig:orbitals}.
To find the continuum-electron wave functions, we used the HX-method for the noble atoms,
while for methane and isobutane, we employed the simpler and less accurate (analytic) approximation of a Coulomb wave
with effective charges determined using a hydrogenic scaling of the ionisation energies.
The resulting dimensionless ionisation form factors were shown in fig.~\ref{fig:fion},
and the DM-electron scattering rates were given in figs.~\ref{fig:dRdEFDM1} and~\ref{fig:dRdEFDMq}.

Projected exclusion limits for the \DarkSphere proposal, presented in figs.~\ref{fig:sensitivityall_q1},~\ref{fig:sensitivityall_q-2}, and~\ref{fig:sensitivityall_mixes},
show that this detector technology has the potential to constrain large swathes of new parameter space.
Existing constraints on the DM-electron cross section for DM masses above 4~MeV can be improved by up to five orders of magnitude
in the case of a low-background, single electron search.

We found that neon is the best all-round gas target although it suffers somewhat at low DM mass relative to xenon because of the comparatively large outer-shell ionisation energy.
In this context, adding a 10\% component of methane or isobutane to neon helps somewhat, as the lower ionisation energies of methane and isobutane allow
the gas-mixture to match xenon's sensitivity at low DM mass.
In contrast, we find that adding methane or isobutane to helium can result in an increase in sensitivity. Helium has the largest ionisation energy of any atom,
and with only two electrons, it has the lowest sensitivity of the five gases that we considered.
Adding a 10\% component of methane or isobutane to helium can potentially lead to a significant improvements in the sensitivity,
especially for the $F_{\rm{DM}}=(\alpha m_e/q)^2$ scenario (cf.\ fig.~\ref{fig:sensitivityall_mixes}).

Given their ability to aid in the search for DM-electron interactions, further refinements of the methane and isobutane calculations are warranted.
Our approach of treating the continuum-electron as a single-centred Coulomb wave could be refined, for instance,
by solving the Schroedinger equation in the spherically averaged potential of the molecule, along
the lines of the method discussed in ref.~\cite{Mahato2019} or by rewriting the single particle states and electron density with an auxiliary single-centre basis set that could also incorporate non-spherical contributions.

Our discussion has been centred around {\sc DarkSphere}, the largest of the proposed SPC experiments. 
We end by speculating on the sensitivity to DM-electron scattering that may be achievable with S140,
 an experiment that uses a smaller diameter SPC and where the background rate is higher.
The S140 detector is currently being commissioned at SNOLAB, and the background rate is expected to be $\mathcal{O}(1)$~cts/kg/keV/day~\cite{brossard:tel-02923528}.
Using figs.~\ref{fig:dRdEFDM1} and~\ref{fig:dRdEFDMq}, we can estimate the sensitivity by finding the cross section when the background and signal rates are comparable.
For $F_{\rm{DM}}=1$, we therefore anticipate that S140 could achieve sensitivity to $\bar{\sigma}_e\sim10^{-40}$ or $10^{-41}$~cm$^2$,
while for $F_{\rm{DM}}=(\alpha m_e/q)^2$,  $\bar{\sigma}_e\sim10^{-36}$ or $10^{-37}$~cm$^2$ may be possible, depending
on the precise value of the detection efficiency that can be achieved.
For both forms of $F_{\rm{DM}}$, this would be an improvement on current exclusion limits for DM heavier than approximately 10~MeV.
This underlines the currently untapped potential of SPCs to search for DM-electron scattering by sub-GeV DM,
and we advocate that this potential should be explored in greater depth.


\noindent\\
\textbf{Data Access Statement:}
The data supporting the findings reported in this paper are openly available from the GitHub repository in~\cite{Git_results}.

\acknowledgments 
We are particularly indebted to Kostas Nikolopoulos, Ioannis Katsioulas, Patrick Knights, 
and Jack Matthews for discussions about the \DarkSphere detector and for providing the simulated background rates in the Boulby Underground Laboratory; George Booth and Robert Anderson for help with running PySCF and for lending us books on quantum chemistry;
Peter Cox, Matthew Dolan, and Harry Quiney for discussions on atomic and molecular physics; and to Thomas Edwards for comments on the manuscript. L.H.\ is supported by the Cromwell Scholarship at King's College London. C.M.\ is supported by the Science and Technology Facilities Council (STFC) Grants No.\ ST/N004663/1 and No.\ ST/T000759/1. 
For the purpose of open access, the authors have applied a Creative Commons Attribution (CC BY) licence to any Author Accepted Manuscript version arising from this submission.

\appendix

\section{Expressions for the ionisation form factor}\label{sec:app:formfactors}

In section~\ref{sec:DMeScattering}, we gave the general form for the rotationally averaged ionisation form factor for a single electron.
To recap, the general expression for the form factor is
\begin{equation}
\label{eq:app:fiongen}
\begin{split}
    \left|f^{i\to f}_{\rm{ion}}(E_e,\mathbf{q})\right|^2 &=\int d \Omega_{k_e} \frac{2 k_e^3}{8 \pi^3} \\
    & \times \left| \int d^3x\, \psi^*_f(\mathbf{x},\mathbf{k_e}) \,e^{i \mathbf{q}\cdot \mathbf{x}} \,\psi_i (\mathbf{x}) \right|^2,
    \end{split}
\end{equation}
where $k_e=|\mathbf{k_e}|=\sqrt{2 m_e E_e}$ is the momentum of the continuum-electron 
and $d \Omega_{k_e}$ is the integration over the direction of the outgoing electron. 
Note that this expression depends on $\mathbf{q}$. However, it does not take into 
account that the atoms or molecules in a gas will be found at an arbitrary and constantly changing orientation.
In our results, we therefore average over all possible orientations of the atom of molecule.
By averaging uniformly over all orientations encoded by the Euler angles $\alpha$, $\beta$ and $\gamma$,
\begin{align}
\label{eq:app:fionrot}
\left|f^{i\to f}_{\rm{ion}}(E_e,q)\right|^2 &= \left\langle \left|f^{i\to f}_{\rm{ion}}(E_e,\mathbf{q})\right|^2 \right\rangle\\ 
\begin{split}
&= \frac{1}{8 \pi^2 } \int_{-1}^{1}  \!\!d\cos \alpha \int_0^{2\pi}\!\! d\beta  \int_0^{2\pi} \!\!d\gamma\\
&\times\left|f^{i\to f}_{\rm{ion}}(E_e,\mathbf{q},\alpha,\beta,\gamma)\right|^2 \;.
\end{split}
\end{align}
As a result of this averaging procedure, the ionisation form factor now only depends on $q$, the magnitude of~$\mathbf{q}$.

Although eqs.~\eqref{eq:app:fiongen} and~\eqref{eq:app:fionrot} are the most general expressions, in practice
there are more convenient expressions that can be used in calculations.
In this appendix, we provide derivations of the more convenient expressions. 

\subsection{Ionisation form factor with a plane wave \label{sec:app:formfactors:PW}}

Although we only consider scattering rates where the outgoing electron is a plane wave in appendix~\ref{app:comparions_molecules},
it is a useful starting point as the expressions are the most straightforward to derive.
We begin with the function
\begin{equation}
    \psi_f(\mathbf{x},\mathbf{k_e})=e^{i \mathbf{k}_e \cdot \mathbf{x}}\;,
\end{equation}
which trivially satisfies our normalisation convention for continuum wave functions: $\int d^3x \,\psi^*_f(\mathbf{x},\mathbf{k}) \psi_f(\mathbf{x},\mathbf{k'}) = (2 \pi)^3 \delta^3(\mathbf{k}-\mathbf{k'})$.
Substituting this into eq.~\eqref{eq:app:fiongen}, we recognise that we can express the form factor in terms of the momentum-space wave function of the initial state,
\begin{equation}\label{eq:apfion_p}
\left|f^{i\to f}_{\rm{ion}}(E_e,\mathbf{q})\right|^2 =\int d \Omega_{k_e} \frac{2 k_e^3}{8 \pi^3}
    \ \left| \tilde{\psi}_i (\mathbf{k}_e-\mathbf{q}) \right|^2,
\end{equation}
where, to be explicit about our convention, 
\begin{equation}
\tilde{\psi}(\mathbf{p})=\int d^3x\, e^{-i \mathbf{p}\cdot \mathbf{x}} \psi_i (\mathbf{x})\;.
\end{equation}
In practice, it more convenient to rewrite this by first noting that
\begin{equation}
   \left| \tilde{\psi}_i (\mathbf{k}_e-\mathbf{q})\right|^2 = \int d^3 p\; \delta^3(\mathbf{p}-(\mathbf{k}_e-\mathbf{q}))\left|\tilde{\psi}_i(\mathbf{p})\right|^2\;.
\end{equation}
and then integrating the delta-function over  $d\Omega_{k_e}$ and $d\cos \theta_p$ to obtain
\begin{equation}\label{eq:apfionmom}
\begin{split}
    \left|f^{i\to f}_{\rm{ion}}(E_e,\mathbf{q})\right|^2 &= \frac{1}{4 \pi^3} \frac{k_e^2}{q} \int_{0}^{2\pi} \!\! d \phi_p \int_{p^-}^{p^+} \!\! p \,dp  \left|\tilde{\psi}_i (p, \phi_p) \right|^2  
    \end{split}.
\end{equation}
In this expression, we have aligned $\mathbf{q}$ in the $z$-direction and $\cos\theta_p$ is the angle between $\mathbf{p}$ and $\mathbf{q}$. 
The integration over the delta-function enforces the relation $\cos \theta_p = (k_e^2-p^2-q^2)/(2 p q)$ and sets the limits of integration, $p^{\pm}=|k_e\pm q|$.
For the rotationally averaged molecular plane wave form factors shown in appendix~\ref{app:comparions_molecules},
we calculate
\begin{equation}
\begin{split}
\left|f^{i\to f}_{\rm{ion}}(E_e,q)\right|^2 &= \frac{1}{8 \pi^2 } \int_{-1}^{1}  \!\!d\cos \alpha \int_0^{2\pi}\!\! d\beta  \int_0^{2\pi} \!\!d\gamma\\
&\times \frac{1}{4 \pi^3} \frac{k_e^2}{q} \int_{0}^{2\pi} \!\! d \phi_p \int_{p^-}^{p^+} \!\!p \,dp \\ &  \times   \left|\tilde{\psi}_i (p, \phi_p,\alpha,\beta,\gamma) \right|^2 \;,
\end{split}
\end{equation}
where the integration over the Euler angles is carried out using Monte Carlo techniques.
As a result of this averaging procedure, the ionisation form factor now only depends on $q$, the magnitude of~$\mathbf{q}$.

For full shell atoms, averaging over the $m$ quantum number is equivalent to rotational averaging.
Therefore before proceeding, we pause to check that we recover the result in ref.~\cite{Essig:2011nj} for a spherically symmetric atom with full shells.
Writing $\tilde{\psi}_{nlm}(\mathbf{p})=\chi_{nl}(p) Y_{lm}(\theta_p,\phi_p)$, where $\chi_{nl}$ is the radial part of the momentum wave function and $Y_{lm}(\theta_p,\phi_p)$ is the spherical harmonic, we find, after summing over the spin and magnetic quantum numbers, 
\begin{align}
\begin{split}
\sum_{m s}\left|f^{i \to f}_{\rm{ion}}(E_e,\mathbf{q})\right|^2 &= \frac{2}{4 \pi^3} \frac{k_e^2}{q} \int_{p^-}^{p^+} \!\! p\, dp   \left|\chi_{nl} (p) \right|^2  \\   
&\times  \int_{0}^{2\pi} \!\!d \phi_p \sum_m \left|Y_{lm} (\theta_p,\phi_p) \right|^2 
\end{split}\\
    &= \frac{2l+1}{4 \pi^3} \frac{k_e^2}{q} \int_{p^-}^{p^+} \!\!p \,dp   \left|\chi_{nl}(p) \right|^2, \label{eq:apatom}
\end{align}
which is the formula quoted in ref.~\cite{Essig:2011nj}.

\subsection{Ionisation form factor with a Coulomb wave  \label{sec:app:formfactors:CW}}

For the Coulomb wave result, we start by writing the partial wave expansion in the form
\begin{equation}
\label{eq:app:partialwave}
\psi_f(\mathbf{x},\mathbf{k_e}) = \sum_{l}\sum_{m=-l}^{l} \psi_{k_e l m}(\mathbf{x}) Y_{lm}(\hat{\mathbf{k}}_e)\;.
\end{equation}
Substituting this form into eq.~\eqref{eq:app:fiongen}
and using the key result 
\begin{equation}
 \int d \Omega_{k_e}  Y^*_{lm}(\hat{\mathbf{k}}_e) Y_{l'm'}(\hat{\mathbf{k}}_e) = \delta_{l l'} \delta_{m m'}
\end{equation}
allows us to move the sum over $l$ and $m$ outside the modulus squared, and we obtain
\begin{equation}\label{eq:app:coulomb}
    \left|f^{i\to f}_{\rm{ion}}(E_e,\mathbf{q})\right|^2 =\frac{2 k_e^3}{8 \pi^3} \sum_{lm} \left| \int d^3x \,  \psi^*_{k_elm}(\mathbf{x})  e^{i \mathbf{q}\cdot \mathbf{x}} \psi_i (\mathbf{x}) \right|^2.
\end{equation}
Concretely then, for the Coulomb wave, we use the expression 
\begin{equation}
 \psi_{k_e l m}(\mathbf{x}) = \frac{P_{kl}(r)}{r} Y^*_{lm}(\hat{\mathbf{x}})
 \end{equation}
where $P_{kl}(r)$ is given in eq.~\eqref{eq:PCoulomb}. Any phase factors (including the $i^l$ term) 
are removed by the absolute magnitude squared in eq.~\eqref{eq:app:coulomb}.
In our calculations, we sum over $l$ and $m$ until the result converges to a few percent accuracy.
Typically, this occurs for $l\lesssim5$; however, in some cases, we need to sum up to $l\sim15$.
The rotationally averaged result is again obtained by averaging over the Euler angles, which we do using Monte Carlo integration.

In passing, we note that eq.~\eqref{eq:app:coulomb} can also be used to calculate the form factor for plane waves
and this approach bypasses the need to calculate the momentum-space wave function.
From the plane wave expansion, we have that
\begin{equation}
 \psi_{k_e l m}(\mathbf{x}) = 4 \pi j_l (k_e r) Y^*_{lm}(\hat{\mathbf{x}})\;,
 \end{equation}
where $j_l(kr)$ is the spherical Bessel function.
In our calculations however, we found it easier to use the expressions in appendix~\ref{sec:app:formfactors:PW},
especially at large values of $q$.

\subsection{Ionisation form factor for atoms \label{sec:app:formfactors:atoms}}

The expressions in eqs.~\eqref{eq:apfionmom} and~\eqref{eq:app:coulomb} make no assumption about the symmetry of the bound-state wave function.
For atomic systems, we can exploit the spherical symmetry of the atom to arrive at another compact expression that is in use in the literature.
We start by substituting the expressions
\begin{align}
\psi_{i} (\mathbf{x}) &= \frac{P_{nl}(r)}{r} Y_{lm}(\hat{\mathbf{x}})\\
\psi^*_{k_e l' m'}(\mathbf{x}) &= \frac{P_{k_e l'} (r)}{r} Y_{l' m'}(\hat{\mathbf{x}})\\
e^{i \mathbf{q}\cdot \mathbf{x}} & = \sum_{LM}4 \pi i^L j_L (qr) Y_{LM} (\hat{\mathbf{x}}) Y^*_{LM}  (\hat{\mathbf{q}})
\end{align}
into eq.~\eqref{eq:app:coulomb}. Note that here we have assumed that $P_{k_e l'} (r)$ is real.
We can simplify the resulting expression by using various identities of the spherical harmonics,
\begin{align}
\begin{split}&\int d \Omega_x \, Y_{l m}(\hat{\mathbf{x}})  Y_{l' m'}(\hat{\mathbf{x}})  Y_{LM}(\hat{\mathbf{x}})  \\
&= \sqrt{\frac{(2l+1)(2 l'+1)(2L+1)}{4 \pi}} \\
&\qquad \times \tjW{l}{l'}{L}{0}{0}{0} \tjW{l}{l'}{L}{m}{m'}{M}\;,
\end{split}\\
& \sum_M Y_{LM}(\hat{\mathbf{q}}) \,Y^*_{LM}(\hat{\mathbf{q}}) = \frac{2L+1}{4 \pi}\;,
\end{align}
where the term in brackets is the Wigner 3-j symbol, which satisfies
\begin{equation}
\sum_{m m'}\tjW{l}{l'}{L}{m}{m'}{M} \tjW{l}{l'}{L'}{m}{m'}{M'} = \frac{\delta_{L L'} \delta_{M M'}}{2L+1}
\;.
\end{equation}

\begin{figure*}[t!]
\includegraphics[width=1.8\columnwidth]{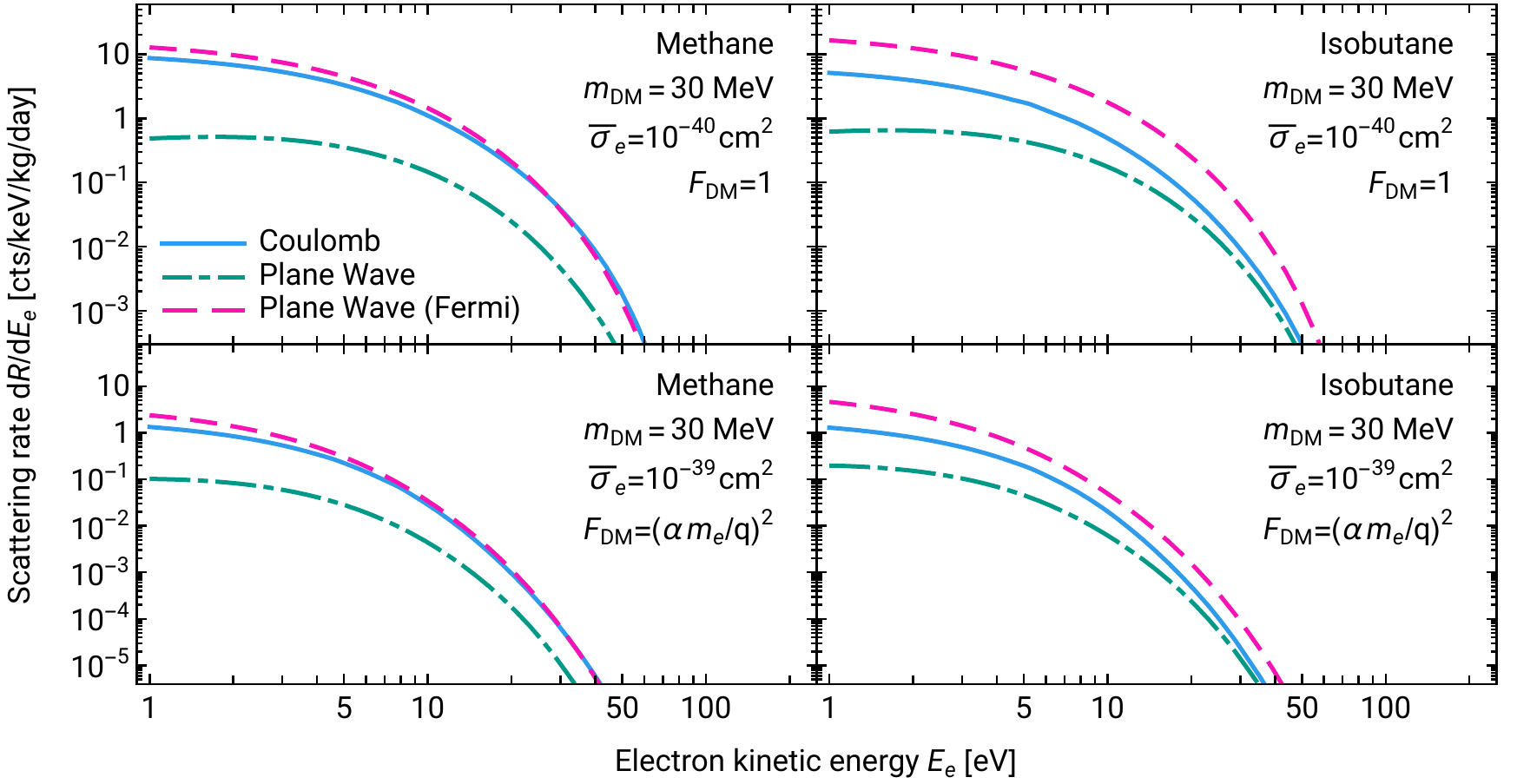}
\caption{\label{fig:ER_comparison_mol}
Differential event rates for DM-induced ionisation from methane (left panels) and isobutane (right panels) for various continuum-state approximations: 
Coulomb wave (solid blue), plane wave (dot-dashed teal), and plane wave with the Fermi function using the same $Z_{\rm{eff}}$ values that enter the Coulomb calculation (pink dashed). 
The DM mass is fixed to 30~MeV. The top panels show $F_{\rm{DM}}=1$, while the bottom panels show $F_{\rm{DM}}=(\alpha m_e/q)^2$.
Note the change in the scattering rate scale between the top and bottom panels.}
\end{figure*}

Pulling all of these results together, we find that the form factor averaged over the initial-state $m$ quantum number is
\begin{align}
\left|f^{nl \!\to \!f}_\mathrm{ion} (E_e,q)\right|^2 \!&= \frac{1}{2l +1}\sum_{m=-l}^{l}  \left|f^{nl \to f}_{\rm{ion}}(E_e,\mathbf{q})\right|^2\\
\begin{split}\label{eq:app:fionatom}
& = \frac{k_e^{3}}{4\pi^3} \sum_{l' L} \left[\!\tjW{l}{l'}{L}{0}{0}{0}\!\right]^2 (2 l'+1) (2L+1) \\
& \times  \left|\int dr P_{k_e l'}(r) P_{nl}(r)j_L(qr)\right|^2.
\end{split}
\end{align}
This matches the result in ref.~\cite{Essig:2012yx} after multiplying by the occupation number $w_{nl}= 2(2l+1)$
since ref.~\cite{Essig:2012yx} includes $w_{nl}$ in their definition of the form factor.
In our atomic physics calculations, we use this result and sum both $l'$ and $L$ in eq.~\eqref{eq:app:fionatom} up to 20, 
which gives a result that converges to better than 2\% accuracy for helium and neon, and better than 3\% for xenon.
As the $P_{k_e l'}(r)$ and $j_L(qr)$ terms can oscillate very rapidly for large values of $k_e$ and $q$, 
we are careful to determine all of the real roots and supply these
to our integration algorithm to ensure accurate numerical results.

\section{Methane and isobutane rates: alternative approaches \label{app:comparions_molecules}}

In section~\ref{sec:DMeScattering} we calculated the methane and isobutane rates with the assumption that the
continuum-electron wave functions were Coulomb waves centred on the centre-of-nuclear-charge of the molecule.
We also made a simplifying assumption for the charge~$Z$, which is a free parameter in the Coulomb function.
In this appendix, we investigate three alternative approximations used in the continuum-electron wave functions and consider
their impact on the scattering rate. 
Firstly, we consider a more basic approximation where we treat the continuum-electron as a plane wave. 
Secondly, as isobutane is a large molecule, we consider the impact of centring the Coulomb waves at one of the carbon atoms at the base of the trigonal pyramid rather than on the centre-of-nuclear-charge of the molecule.
Finally, we consider the impact of changing the values of the charge~$Z$ in the Coulomb function.

\subsection{Plane wave approximation}

In atomic calculations at low electron energies, the plane wave approximation gives results that are less accurate 
than calculations with a Coulomb wave or that solve the Schroedinger equation in the potential generated by the 
bound-state electrons (see e.g., ref.~\cite{Roberts:2019chv}). Nonetheless, a comparison of the plane wave results with the Coulomb 
wave results can give a crude estimate of the accuracy of the calculations.

In atomic calculations, the plane wave calculations are improved somewhat by including the Fermi function as a prefactor in eq.~\eqref{eq:dsigmadE}. 
The Fermi function is defined as
\begin{align}
    F(k_e)&=\left(\frac{\psi^{\rm{Coulomb}}_f(\mathbf{x}\to 0,\mathbf{k_e})}{\psi^{\rm{Plane}}_f(\mathbf{x}\to 0,\mathbf{k_e})}\right)^2\\
    &= \frac{2\pi \eta(k_e)}{1-\exp[-2\pi\eta(k_e)]}\;,
\end{align}
which is the square of the ratio of the Coulomb wave to the plane wave at the origin~\cite{Abramowitzbook},
and where
 \begin{equation}
 \label{eq:Sommerfeldeta}
\eta(k_e)=Z_{\rm{eff}}\frac{\alpha\, m_e}{k_e}
\end{equation}
is the Sommerfeld parameter (here given in natural units, while we gave the Coulomb function in eq.~\eqref{eq:PCoulomb} in atomic units).

As with the Coulomb wave calculation presented in section~\ref{sec:DMeScattering}, we centre the plane wave on the centre-of-nuclear-charge of the molecule, and
we must supply a value of $Z_{\rm{eff}}$ that enters eq.~\eqref{eq:Sommerfeldeta}.
We follow the same procedure used in section~\ref{sec:DMeScattering}, which leads to $Z_{\rm{eff}}$ values 
of $4.7$, $2.6$, and $1.0$ for the $1a_1$, $2a_1$, and $1t_2$ states in methane, respectively,
and $Z_{\rm{eff}}=4.7$ for $1a_1$, $1e$, $2a_1$, $Z_{\rm{eff}}=2.7$ for $3a_1$, $2e$, $4a_1$, and 
$Z_{\rm{eff}}=1$ for the remaining outer valence orbitals in isobutane.

In fig.~\ref{fig:ER_comparison_mol}, we plot the methane (left panels) and isobutane (right panels) scattering event rates in the Coulomb wave and plane wave approximations (with and without the Fermi function). 
The results are shown for $F_{\rm{DM}}=1$ (top panels) and $F_{\rm{DM}}=(\alpha m_e/q)^2$ (bottom panels) with $m_{\rm{DM}}=30$~MeV.
Similar relative differences are found between the rates for DM masses in the range 5 to 1000~MeV.

In all cases, the Fermi function improves the agreement between the plane wave and the Coulomb wave.
For methane, the agreement over most of the energy range between the Coulomb rate and plane wave rate with the Fermi function is
at about the level of 30\%.
 Without the Fermi function, the Coulomb and plane wave rates differ by more than an order of magnitude.
The differences are larger for isobutane, where there is a factor of approximately 2-to-3 difference between the Coulomb and plane wave rate with the Fermi function.
As the plane wave with Fermi function rates are all larger than the Coulomb rates, the
corresponding projected exclusion limits that would be obtained using the plane wave with Fermi function 
would be a factor of approximately two-to-three lower than the limits shown in section~\ref{section:sensitivities}.

\begin{figure}[t!]
\includegraphics[width=0.9\columnwidth]{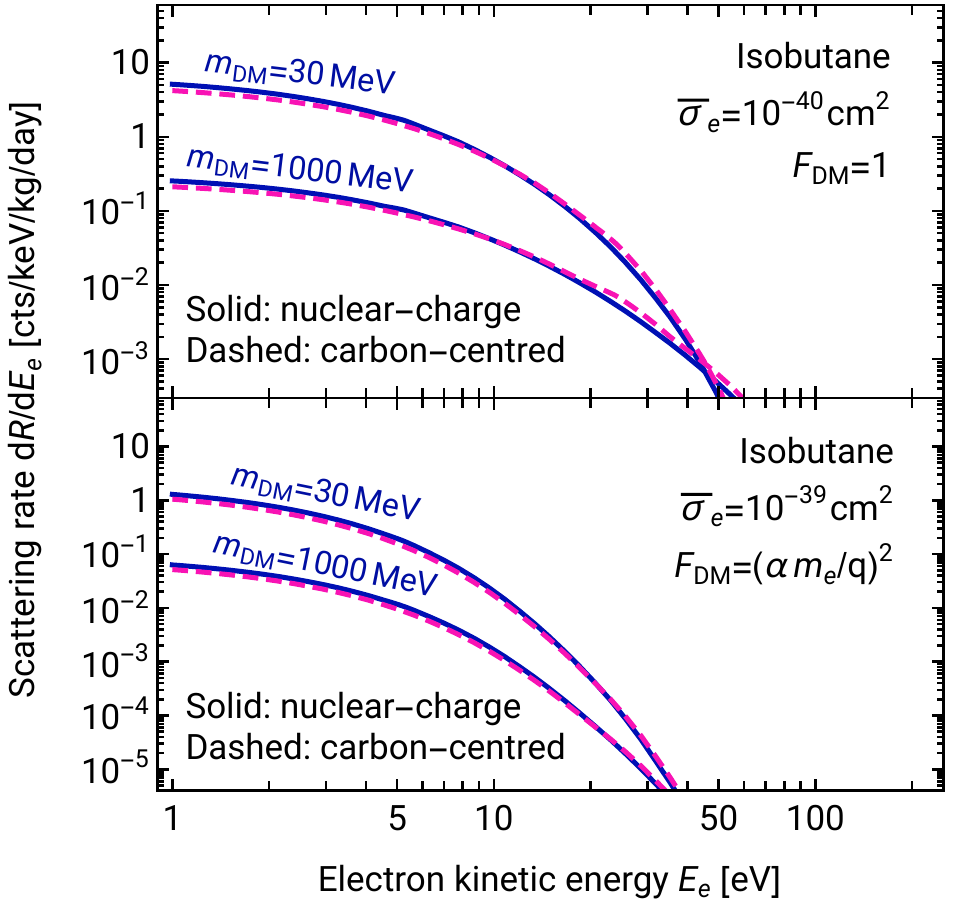}
\caption{\label{fig:CWvsCarbon}
Comparison of the differential event rates for DM-induced ionisation from isobutane when the continuum-electron Coulomb waves are centred
at the centre-of-nuclear-charge (solid dark-blue lines), and at a carbon atom at the base of the trigonal pyramid (dashed pink lines). 
The comparison is made for a DM mass of 30 and 1000~MeV.  The top panel shows $F_{\rm{DM}}=1$, while the bottom panel shows $F_{\rm{DM}}=(\alpha m_e/q)^2$.
Note the change in the scattering rate scale between the top and bottom panels.}
\end{figure}

\subsection{Alternative centre for the Coulomb waves}

In section~\ref{sec:DMeScattering} the Coulomb waves were centred at the centre-of-nuclear-charge. 
While in methane the centre-of-nuclear-charge corresponds to the position of the carbon atom, in isobutane the centre-of-nuclear-charge is a point in space between the carbon atoms.
In the case of isobutane, it is therefore possible that after ionisation the positive charge could be localised on a specific carbon atom rather than the centre-of-nuclear-charge.
To investigate the impact of this possibility on the scattering rate prediction, we have recalculated the ionisation form factors when the continuum-electron Coulomb waves are centred on a carbon atom at the base of the trigonal pyramid. 
Specifically, we centre the Coulomb waves on the carbon nucleus at $(2.38, -1.37, -0.19)$~Bohr in our Cartesian coordinate system, as defined by the NIST CCCBDB database~\cite{NISTCCCBDB}.

In fig.~\ref{fig:CWvsCarbon}, we show the differential scattering rate for two DM masses, 30 and 1000~MeV, and for $F_{\rm{DM}}=1$ (top) and $F_{\rm{DM}}=(\alpha m_e/q)^2$ (bottom). 
The solid lines show the result of the calculation in section~\ref{sec:DMeScattering} while the dashed lines show the rates when the centre is the carbon atom at the base of the trigonal pyramid. 
The difference between the two calculations is 40\% or smaller for both masses and both choices of $F_{\rm{DM}}$. Similar results are found for all other DM masses considered in this work.

\begin{figure}[t!]
\includegraphics[width=0.9\columnwidth]{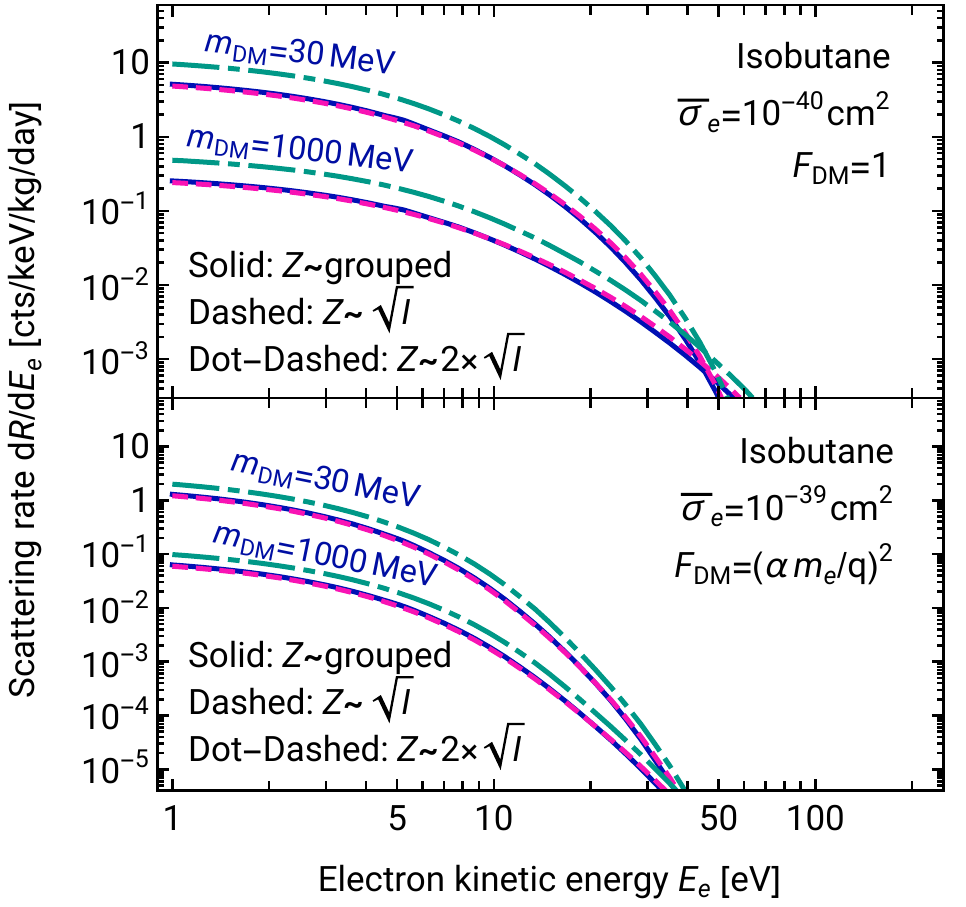}
\caption{\label{fig:CWvsZ}
Comparison of the differential event rates for DM-induced ionisation from isobutane for different values of $Z$,
which is a free parameter in the Coulomb function. Solid blue shows the rates when common $Z$ values are assigned to the groups of inner core, inner valence and outer valence orbitals;
dashed pink shows the rates when $Z = n \sqrt{I/13.6~\mathrm{eV}}$, where $I$ is the ionisation energy; and dot-dashed teal is when $Z = 2\times n \sqrt{I/13.6~\mathrm{eV}}$.
The comparison is made for a DM mass of 30~MeV and 1000~MeV.  The top panel shows $F_{\rm{DM}}=1$ while the bottom panel shows $F_{\rm{DM}}=(\alpha m_e/q)^2$.
Note the change in the scattering rate scale between the top and bottom panels.}
\end{figure}

\begin{figure*}[t!]
\includegraphics[width=1.8\columnwidth]{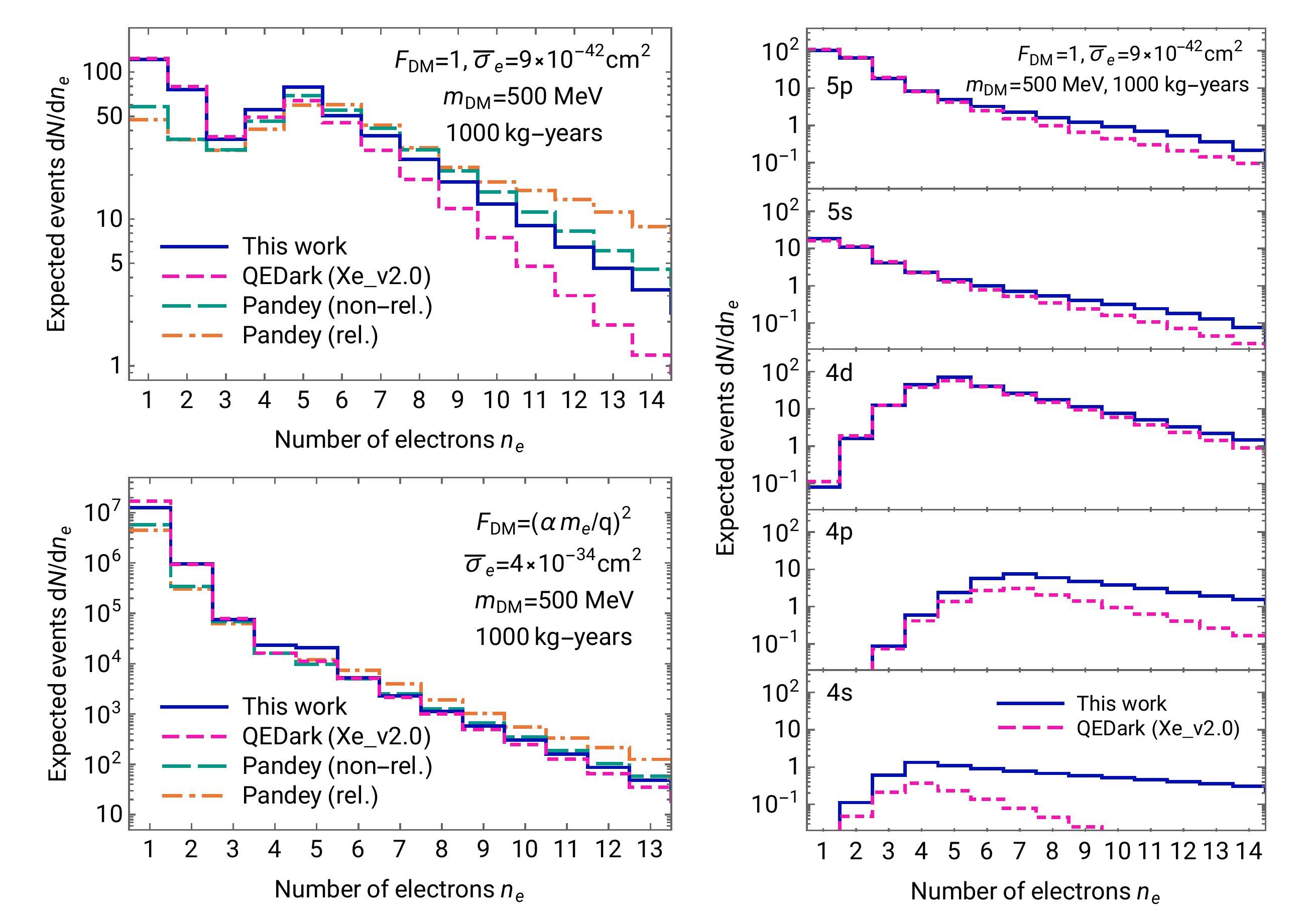}\\
\caption{\label{fig:dNdnFDMcomparison}
Left panels: Comparison of the expected number of events from this work, QEDark~\cite{QEDark_Git} and Pandey et al~\cite{Pandey:2018esq} for DM-electron scattering in xenon, for $m_{\rm{DM}}=500$\,MeV, a 1000 kg-years exposure, and for $F_{\rm{DM}} = 1$ (top) and $F_{\rm{DM}}=(\alpha m_e/q)^2$ (bottom). Right panel: An orbital-by-orbital comparison of the expected number of events from this work and QEDark for $m_{\rm{DM}}=500$\,MeV and $F_{\rm{DM}} = 1$.
}
\end{figure*}

\subsection{Alternative values of the effective charge}

The charge $Z$ remains a free parameter in the Coulomb function. We used the values~$Z=4.7$, $Z=2.6$, and~$Z=1.0$ for the~$1a_1$, $2a_1$, and~$1t_2$ states in methane, which were obtained
using the formula for the hydrogen-like scaling for the ionisation energy.
For isobutane we made a further simplifying approximation, which was to assign $Z=4.7$ 
for the inner core orbitals ($1a_1$, $1e$, $2a_1$), $Z=2.7$ for the inner valence orbitals ($3a_1$, $2e$, $4a_1$), and $Z=1$ for the outer valence states (all remaining orbitals).

Here, we investigate two alternative choices for the $Z$ values that enter the isobutane calculation. Firstly, we assign different $Z$ values to each orbital, calculated using $I= Z^2/n^2 \times 13.6~\mathrm{eV}$, where $I$ is the experimental ionisation energy (when available or theoretical value otherwise) and $n=1$ for the inner core and outer valence states, and $n=2$ for the inner valence states. 
The resulting scattering rates for two DM masses, 30 and 1000~MeV, are shown as the dashed pink lines in fig.~\ref{fig:CWvsZ}. 
These are compared with the rates from section~\ref{sec:DMeScattering} shown in solid blue, where common $Z$ values were assigned to the groups of inner core, inner valence, and outer valence orbitals.
The rates differ by 25\% or less for both DM masses shown and for both choices of $F_{\rm{DM}}$.
Secondly, we consider the (possibly unphysical) scenario where the value of $Z$ is arbitrarily doubled, such that $Z = 2 \times n \sqrt{I/13.6~\mathrm{eV}}$, in order to obtain intuition for how the rates scale with $Z$. 
The resulting rates are shown as the dot-dashed teal lines in fig.~\ref{fig:CWvsZ}. Comparing with the dashed pink lines, we find that the scattering rates where $Z$ has been doubled are between a factor of $1.5$ and $2$ larger

\section{Comparison with other calculations \label{app:comparions_rates}}

The literature contains several calculations of the DM-electron scattering rate for xenon.
In this appendix, we provide a comparison with the results from QEDark (specifically, Xe\_v2.0)~\cite{QEDark_Git}, a commonly used package in limit-setting calculations (see, e.g.,~\cite{Aprile:2019xxb}), and with Pandey et al~\cite{Pandey:2018esq}, which provides both a relativistic and non-relativistic calculation and reports some deviations from the QEDark results.

In making the comparisons, we use the same astrophysical parameters employed in those papers: $\rho_{\rm{DM}} = 0.4 \,\mathrm{GeV}/\mathrm{cm}^3 $, $v_0 = 220\,\mathrm{km/s}$, and $v_{\mathrm{esc}} = 544\,\mathrm{km/s}$. We also follow the same prescription used in those works to calculate the expected number of electrons $n_e$ from a given $dR/dE_e$ recoil spectrum, namely, we use the xenon binding energies and the input parameters $W=13.8\,\mathrm{eV}$, $f_e=0.83$, and $f_R=0$ from ref.~\cite{Essig:2017kqs}. 

The solid blue lines in fig.~\ref{fig:dNdnFDMcomparison} show the spectrum of the expected number of events from our non-relativistic calculations for $F_{\rm{DM}}=1 $ (top-left panel) and $F_{\rm{DM}}=(\alpha m_e/q)^2 $ (bottom-left panel). We have assumed a DM mass of 500\,MeV and a total exposure of 1000\,kg-years. In the same panels, we show the non-relativistic result from QEDark (Xe\_v2.0), together with the  non-relativistic and relativistic calculations from Pandey et~al.

Comparing firstly with QEDark (pink dashed lines), we find good agreement for a small number of electrons ($n\lesssim4$) but a deviation at larger values. This is most obvious when $F_{\rm{DM}}=1$ where the difference reaches a factor of approximately three for $n=14$. The difference at larger $n_e$ is less pronounced when $F_{\rm{DM}}=(\alpha m_e/q)^2 $. To better understand the difference in the $F_{\rm{DM}}=1$ case, in the right panel of fig.~\ref{fig:dNdnFDMcomparison}, we have plotted the contributions from the individual orbitals for our results and for QEDark. This clearly shows that the difference arises from the $4p$ shell. While our calculation shows that $4d$ and $4p$ both contribute at a similar level at  $n_e\sim14$, the QEDark result for $4p$ is approximately an order of magnitude smaller. There is also a substantial difference with $4s$, but this orbital gives a sub-dominant contribution to the total rate. 

Turning next to a comparison with Pandey et al (green dashed and orange dot-dashed lines), we find agreement with the non-relativistic calculation at the level of 30\% or better for $n_e\gtrsim3$. Both our result and the Pandey et al non-relativistic calculation deviate from QEDark at larger values of $n_e$. The relativistic calculation shows an enhancement over all non-relativistic calculations at larger $n_e$, which is understood as arising from the more accurate characterisation of the electron wave function at very small distances in the relativistic calculation (see, e.g., ref.~\cite{Roberts:2015lga}). Comparing the results at small electron number ($n_e<3$), we find that our result and that of QEDark are about a factor of two higher than Pandey et~al. The $n_e=1$ and $n_e=2$ results (equivalent to low electron kinetic energy) are most sensitive to the effective potential used in the calculation. The Pandey et~al calculation only includes the classical potential energy in their calculation of the continuum wave function. In contrast, our result includes the potential energy and an effective exchange potential. Rerunning our calculations with $k_x=0$ in eq.~\eqref{eq:VHX} (so that only the classical potential energy contributes), we find that the scattering rate falls just below the Pandey et~al result for $n_e<3$. This confirms that the $n_e=1$ and $n_e=2$ rates are sensitive to the effective potential, so differences between our result and Pandey et~al should be expected since we use different models of the effective potential.

\bibliography{refs}

\begin{thebibliography}{147}%
\makeatletter
\providecommand \@ifxundefined [1]{%
 \@ifx{#1\undefined}
}%
\providecommand \@ifnum [1]{%
 \ifnum #1\expandafter \@firstoftwo
 \else \expandafter \@secondoftwo
 \fi
}%
\providecommand \@ifx [1]{%
 \ifx #1\expandafter \@firstoftwo
 \else \expandafter \@secondoftwo
 \fi
}%
\providecommand \natexlab [1]{#1}%
\providecommand \enquote  [1]{``#1''}%
\providecommand \bibnamefont  [1]{#1}%
\providecommand \bibfnamefont [1]{#1}%
\providecommand \citenamefont [1]{#1}%
\providecommand \href@noop [0]{\@secondoftwo}%
\providecommand \href [0]{\begingroup \@sanitize@url \@href}%
\providecommand \@href[1]{\@@startlink{#1}\@@href}%
\providecommand \@@href[1]{\endgroup#1\@@endlink}%
\providecommand \@sanitize@url [0]{\catcode `\\12\catcode `\$12\catcode
  `\&12\catcode `\#12\catcode `\^12\catcode `\_12\catcode `\%12\relax}%
\providecommand \@@startlink[1]{}%
\providecommand \@@endlink[0]{}%
\providecommand \url  [0]{\begingroup\@sanitize@url \@url }%
\providecommand \@url [1]{\endgroup\@href {#1}{\urlprefix }}%
\providecommand \urlprefix  [0]{URL }%
\providecommand \Eprint [0]{\href }%
\providecommand \doibase [0]{https://doi.org/}%
\providecommand \selectlanguage [0]{\@gobble}%
\providecommand \bibinfo  [0]{\@secondoftwo}%
\providecommand \bibfield  [0]{\@secondoftwo}%
\providecommand \translation [1]{[#1]}%
\providecommand \BibitemOpen [0]{}%
\providecommand \bibitemStop [0]{}%
\providecommand \bibitemNoStop [0]{.\EOS\space}%
\providecommand \EOS [0]{\spacefactor3000\relax}%
\providecommand \BibitemShut  [1]{\csname bibitem#1\endcsname}%
\let\auto@bib@innerbib\@empty
\bibitem [{APP()}]{APPEC}%
  \BibitemOpen
  \href@noop {} {}\bibinfo {note} {European Astroparticle Physics Strategy
  2017-2026, \url{https://www.appec.org/roadmap}}\BibitemShut {NoStop}%
\bibitem [{\citenamefont {Bertone}\ and\ \citenamefont
  {Hooper}(2018)}]{Bertone:2016nfn}%
  \BibitemOpen
  \bibfield  {author} {\bibinfo {author} {\bibfnamefont {G.}~\bibnamefont
  {Bertone}}\ and\ \bibinfo {author} {\bibfnamefont {D.}~\bibnamefont
  {Hooper}},\ }\href {https://doi.org/10.1103/RevModPhys.90.045002} {\bibfield
  {journal} {\bibinfo  {journal} {Rev. Mod. Phys.}\ }\textbf {\bibinfo {volume}
  {90}},\ \bibinfo {pages} {045002} (\bibinfo {year} {2018})},\ \Eprint
  {https://arxiv.org/abs/1605.04909} {arXiv:1605.04909 [astro-ph.CO]}
  \BibitemShut {NoStop}%
\bibitem [{\citenamefont {Bertone}\ \emph {et~al.}(2005)\citenamefont
  {Bertone}, \citenamefont {Hooper},\ and\ \citenamefont
  {Silk}}]{Bertone:2004pz}%
  \BibitemOpen
  \bibfield  {author} {\bibinfo {author} {\bibfnamefont {G.}~\bibnamefont
  {Bertone}}, \bibinfo {author} {\bibfnamefont {D.}~\bibnamefont {Hooper}},\
  and\ \bibinfo {author} {\bibfnamefont {J.}~\bibnamefont {Silk}},\ }\href
  {https://doi.org/10.1016/j.physrep.2004.08.031} {\bibfield  {journal}
  {\bibinfo  {journal} {Phys. Rept.}\ }\textbf {\bibinfo {volume} {405}},\
  \bibinfo {pages} {279} (\bibinfo {year} {2005})},\ \Eprint
  {https://arxiv.org/abs/hep-ph/0404175} {arXiv:hep-ph/0404175} \BibitemShut
  {NoStop}%
\bibitem [{\citenamefont {Edwards}\ \emph {et~al.}(2018)\citenamefont
  {Edwards}, \citenamefont {Kavanagh},\ and\ \citenamefont
  {Weniger}}]{Edwards:2018lsl}%
  \BibitemOpen
  \bibfield  {author} {\bibinfo {author} {\bibfnamefont {T.~D.~P.}\
  \bibnamefont {Edwards}}, \bibinfo {author} {\bibfnamefont {B.~J.}\
  \bibnamefont {Kavanagh}},\ and\ \bibinfo {author} {\bibfnamefont
  {C.}~\bibnamefont {Weniger}},\ }\href
  {https://doi.org/10.1103/PhysRevLett.121.181101} {\bibfield  {journal}
  {\bibinfo  {journal} {Phys. Rev. Lett.}\ }\textbf {\bibinfo {volume} {121}},\
  \bibinfo {pages} {181101} (\bibinfo {year} {2018})},\ \Eprint
  {https://arxiv.org/abs/1805.04117} {arXiv:1805.04117 [hep-ph]} \BibitemShut
  {NoStop}%
\bibitem [{\citenamefont {Billard}\ \emph {et~al.}(2022)\citenamefont {Billard}
  \emph {et~al.}}]{Billard:2021uyg}%
  \BibitemOpen
  \bibfield  {author} {\bibinfo {author} {\bibfnamefont {J.}~\bibnamefont
  {Billard}} \emph {et~al.},\ }\href {https://doi.org/10.1088/1361-6633/ac5754}
  {\bibfield  {journal} {\bibinfo  {journal} {Rept. Prog. Phys.}\ }\textbf
  {\bibinfo {volume} {85}},\ \bibinfo {pages} {056201} (\bibinfo {year}
  {2022})},\ \Eprint {https://arxiv.org/abs/2104.07634} {arXiv:2104.07634
  [hep-ex]} \BibitemShut {NoStop}%
\bibitem [{\citenamefont {Baudis}(2016)}]{Baudis:2015mpa}%
  \BibitemOpen
  \bibfield  {author} {\bibinfo {author} {\bibfnamefont {L.}~\bibnamefont
  {Baudis}},\ }\href {https://doi.org/10.1002/andp.201500114} {\bibfield
  {journal} {\bibinfo  {journal} {Annalen Phys.}\ }\textbf {\bibinfo {volume}
  {528}},\ \bibinfo {pages} {74} (\bibinfo {year} {2016})},\ \Eprint
  {https://arxiv.org/abs/1509.00869} {arXiv:1509.00869 [astro-ph.CO]}
  \BibitemShut {NoStop}%
\bibitem [{\citenamefont {Marrodan~Undagoitia}\ and\ \citenamefont
  {Rauch}(2016)}]{Undagoitia:2015gya}%
  \BibitemOpen
  \bibfield  {author} {\bibinfo {author} {\bibfnamefont {T.}~\bibnamefont
  {Marrodan~Undagoitia}}\ and\ \bibinfo {author} {\bibfnamefont
  {L.}~\bibnamefont {Rauch}},\ }\href
  {https://doi.org/10.1088/0954-3899/43/1/013001} {\bibfield  {journal}
  {\bibinfo  {journal} {J. Phys. G}\ }\textbf {\bibinfo {volume} {43}},\
  \bibinfo {pages} {013001} (\bibinfo {year} {2016})},\ \Eprint
  {https://arxiv.org/abs/1509.08767} {arXiv:1509.08767 [physics.ins-det]}
  \BibitemShut {NoStop}%
\bibitem [{\citenamefont {Schumann}(2019)}]{Schumann:2019eaa}%
  \BibitemOpen
  \bibfield  {author} {\bibinfo {author} {\bibfnamefont {M.}~\bibnamefont
  {Schumann}},\ }\href {https://doi.org/10.1088/1361-6471/ab2ea5} {\bibfield
  {journal} {\bibinfo  {journal} {J. Phys. G}\ }\textbf {\bibinfo {volume}
  {46}},\ \bibinfo {pages} {103003} (\bibinfo {year} {2019})},\ \Eprint
  {https://arxiv.org/abs/1903.03026} {arXiv:1903.03026 [astro-ph.CO]}
  \BibitemShut {NoStop}%
\bibitem [{\citenamefont {Battaglieri}\ \emph {et~al.}(2017)\citenamefont
  {Battaglieri} \emph {et~al.}}]{Battaglieri:2017aum}%
  \BibitemOpen
  \bibfield  {author} {\bibinfo {author} {\bibfnamefont {M.}~\bibnamefont
  {Battaglieri}} \emph {et~al.},\ }in\ \href@noop {} {\emph {\bibinfo
  {booktitle} {{U.S. Cosmic Visions: New Ideas in Dark Matter}}}}\ (\bibinfo
  {year} {2017})\ \Eprint {https://arxiv.org/abs/1707.04591} {arXiv:1707.04591
  [hep-ph]} \BibitemShut {NoStop}%
\bibitem [{\citenamefont {Arnaud}\ \emph
  {et~al.}(2018{\natexlab{a}})\citenamefont {Arnaud} \emph
  {et~al.}}]{Arnaud:2017usi}%
  \BibitemOpen
  \bibfield  {author} {\bibinfo {author} {\bibfnamefont {Q.}~\bibnamefont
  {Arnaud}} \emph {et~al.} (\bibinfo {collaboration} {EDELWEISS}),\ }\href
  {https://doi.org/10.1103/PhysRevD.97.022003} {\bibfield  {journal} {\bibinfo
  {journal} {Phys. Rev. D}\ }\textbf {\bibinfo {volume} {97}},\ \bibinfo
  {pages} {022003} (\bibinfo {year} {2018}{\natexlab{a}})},\ \Eprint
  {https://arxiv.org/abs/1707.04308} {arXiv:1707.04308 [physics.ins-det]}
  \BibitemShut {NoStop}%
\bibitem [{\citenamefont {Agnese}\ \emph {et~al.}(2019)\citenamefont {Agnese}
  \emph {et~al.}}]{Agnese:2018gze}%
  \BibitemOpen
  \bibfield  {author} {\bibinfo {author} {\bibfnamefont {R.}~\bibnamefont
  {Agnese}} \emph {et~al.} (\bibinfo {collaboration} {SuperCDMS}),\ }\href
  {https://doi.org/10.1103/PhysRevD.99.062001} {\bibfield  {journal} {\bibinfo
  {journal} {Phys. Rev. D}\ }\textbf {\bibinfo {volume} {99}},\ \bibinfo
  {pages} {062001} (\bibinfo {year} {2019})},\ \Eprint
  {https://arxiv.org/abs/1808.09098} {arXiv:1808.09098 [astro-ph.CO]}
  \BibitemShut {NoStop}%
\bibitem [{\citenamefont {Abdelhameed}\ \emph
  {et~al.}(2019{\natexlab{a}})\citenamefont {Abdelhameed} \emph
  {et~al.}}]{Abdelhameed:2019szb}%
  \BibitemOpen
  \bibfield  {author} {\bibinfo {author} {\bibfnamefont {A.~H.}\ \bibnamefont
  {Abdelhameed}} \emph {et~al.} (\bibinfo {collaboration} {CRESST}),\ }\href
  {https://doi.org/10.1140/epjc/s10052-019-7126-4} {\bibfield  {journal}
  {\bibinfo  {journal} {Eur. Phys. J. C}\ }\textbf {\bibinfo {volume} {79}},\
  \bibinfo {pages} {630} (\bibinfo {year} {2019}{\natexlab{a}})},\ \Eprint
  {https://arxiv.org/abs/1902.07587} {arXiv:1902.07587 [astro-ph.IM]}
  \BibitemShut {NoStop}%
\bibitem [{\citenamefont {Abdelhameed}\ \emph
  {et~al.}(2019{\natexlab{b}})\citenamefont {Abdelhameed} \emph
  {et~al.}}]{Abdelhameed:2019hmk}%
  \BibitemOpen
  \bibfield  {author} {\bibinfo {author} {\bibfnamefont {A.~H.}\ \bibnamefont
  {Abdelhameed}} \emph {et~al.} (\bibinfo {collaboration} {CRESST}),\ }\href
  {https://doi.org/10.1103/PhysRevD.100.102002} {\bibfield  {journal} {\bibinfo
   {journal} {Phys. Rev. D}\ }\textbf {\bibinfo {volume} {100}},\ \bibinfo
  {pages} {102002} (\bibinfo {year} {2019}{\natexlab{b}})},\ \Eprint
  {https://arxiv.org/abs/1904.00498} {arXiv:1904.00498 [astro-ph.CO]}
  \BibitemShut {NoStop}%
\bibitem [{\citenamefont {Alkhatib}\ \emph {et~al.}(2021)\citenamefont
  {Alkhatib} \emph {et~al.}}]{Alkhatib:2020slm}%
  \BibitemOpen
  \bibfield  {author} {\bibinfo {author} {\bibfnamefont {I.}~\bibnamefont
  {Alkhatib}} \emph {et~al.} (\bibinfo {collaboration} {SuperCDMS}),\ }\href
  {https://doi.org/10.1103/PhysRevLett.127.061801} {\bibfield  {journal}
  {\bibinfo  {journal} {Phys. Rev. Lett.}\ }\textbf {\bibinfo {volume} {127}},\
  \bibinfo {pages} {061801} (\bibinfo {year} {2021})},\ \Eprint
  {https://arxiv.org/abs/2007.14289} {arXiv:2007.14289 [hep-ex]} \BibitemShut
  {NoStop}%
\bibitem [{\citenamefont {Aguilar-Arevalo}\ \emph {et~al.}(2020)\citenamefont
  {Aguilar-Arevalo} \emph {et~al.}}]{Aguilar-Arevalo:2020oii}%
  \BibitemOpen
  \bibfield  {author} {\bibinfo {author} {\bibfnamefont {A.}~\bibnamefont
  {Aguilar-Arevalo}} \emph {et~al.} (\bibinfo {collaboration} {DAMIC}),\ }\href
  {https://doi.org/10.1103/PhysRevLett.125.241803} {\bibfield  {journal}
  {\bibinfo  {journal} {Phys. Rev. Lett.}\ }\textbf {\bibinfo {volume} {125}},\
  \bibinfo {pages} {241803} (\bibinfo {year} {2020})},\ \Eprint
  {https://arxiv.org/abs/2007.15622} {arXiv:2007.15622 [astro-ph.CO]}
  \BibitemShut {NoStop}%
\bibitem [{\citenamefont {Kouvaris}\ and\ \citenamefont
  {Pradler}(2017)}]{Kouvaris:2016afs}%
  \BibitemOpen
  \bibfield  {author} {\bibinfo {author} {\bibfnamefont {C.}~\bibnamefont
  {Kouvaris}}\ and\ \bibinfo {author} {\bibfnamefont {J.}~\bibnamefont
  {Pradler}},\ }\href {https://doi.org/10.1103/PhysRevLett.118.031803}
  {\bibfield  {journal} {\bibinfo  {journal} {Phys. Rev. Lett.}\ }\textbf
  {\bibinfo {volume} {118}},\ \bibinfo {pages} {031803} (\bibinfo {year}
  {2017})},\ \Eprint {https://arxiv.org/abs/1607.01789} {arXiv:1607.01789
  [hep-ph]} \BibitemShut {NoStop}%
\bibitem [{\citenamefont {McCabe}(2017)}]{McCabe:2017rln}%
  \BibitemOpen
  \bibfield  {author} {\bibinfo {author} {\bibfnamefont {C.}~\bibnamefont
  {McCabe}},\ }\href {https://doi.org/10.1103/PhysRevD.96.043010} {\bibfield
  {journal} {\bibinfo  {journal} {Phys. Rev. D}\ }\textbf {\bibinfo {volume}
  {96}},\ \bibinfo {pages} {043010} (\bibinfo {year} {2017})},\ \Eprint
  {https://arxiv.org/abs/1702.04730} {arXiv:1702.04730 [hep-ph]} \BibitemShut
  {NoStop}%
\bibitem [{\citenamefont {Ibe}\ \emph {et~al.}(2018)\citenamefont {Ibe},
  \citenamefont {Nakano}, \citenamefont {Shoji},\ and\ \citenamefont
  {Suzuki}}]{Ibe:2017yqa}%
  \BibitemOpen
  \bibfield  {author} {\bibinfo {author} {\bibfnamefont {M.}~\bibnamefont
  {Ibe}}, \bibinfo {author} {\bibfnamefont {W.}~\bibnamefont {Nakano}},
  \bibinfo {author} {\bibfnamefont {Y.}~\bibnamefont {Shoji}},\ and\ \bibinfo
  {author} {\bibfnamefont {K.}~\bibnamefont {Suzuki}},\ }\href
  {https://doi.org/10.1007/JHEP03(2018)194} {\bibfield  {journal} {\bibinfo
  {journal} {JHEP}\ }\textbf {\bibinfo {volume} {03}},\ \bibinfo {pages}
  {194}},\ \Eprint {https://arxiv.org/abs/1707.07258} {arXiv:1707.07258
  [hep-ph]} \BibitemShut {NoStop}%
\bibitem [{\citenamefont {Dolan}\ \emph {et~al.}(2018)\citenamefont {Dolan},
  \citenamefont {Kahlhoefer},\ and\ \citenamefont {McCabe}}]{Dolan:2017xbu}%
  \BibitemOpen
  \bibfield  {author} {\bibinfo {author} {\bibfnamefont {M.~J.}\ \bibnamefont
  {Dolan}}, \bibinfo {author} {\bibfnamefont {F.}~\bibnamefont {Kahlhoefer}},\
  and\ \bibinfo {author} {\bibfnamefont {C.}~\bibnamefont {McCabe}},\ }\href
  {https://doi.org/10.1103/PhysRevLett.121.101801} {\bibfield  {journal}
  {\bibinfo  {journal} {Phys. Rev. Lett.}\ }\textbf {\bibinfo {volume} {121}},\
  \bibinfo {pages} {101801} (\bibinfo {year} {2018})},\ \Eprint
  {https://arxiv.org/abs/1711.09906} {arXiv:1711.09906 [hep-ph]} \BibitemShut
  {NoStop}%
\bibitem [{\citenamefont {Akerib}\ \emph {et~al.}(2019)\citenamefont {Akerib}
  \emph {et~al.}}]{Akerib:2018hck}%
  \BibitemOpen
  \bibfield  {author} {\bibinfo {author} {\bibfnamefont {D.~S.}\ \bibnamefont
  {Akerib}} \emph {et~al.} (\bibinfo {collaboration} {LUX}),\ }\href
  {https://doi.org/10.1103/PhysRevLett.122.131301} {\bibfield  {journal}
  {\bibinfo  {journal} {Phys. Rev. Lett.}\ }\textbf {\bibinfo {volume} {122}},\
  \bibinfo {pages} {131301} (\bibinfo {year} {2019})},\ \Eprint
  {https://arxiv.org/abs/1811.11241} {arXiv:1811.11241 [astro-ph.CO]}
  \BibitemShut {NoStop}%
\bibitem [{\citenamefont {Armengaud}\ \emph {et~al.}(2019)\citenamefont
  {Armengaud} \emph {et~al.}}]{Armengaud:2019kfj}%
  \BibitemOpen
  \bibfield  {author} {\bibinfo {author} {\bibfnamefont {E.}~\bibnamefont
  {Armengaud}} \emph {et~al.} (\bibinfo {collaboration} {EDELWEISS}),\ }\href
  {https://doi.org/10.1103/PhysRevD.99.082003} {\bibfield  {journal} {\bibinfo
  {journal} {Phys. Rev. D}\ }\textbf {\bibinfo {volume} {99}},\ \bibinfo
  {pages} {082003} (\bibinfo {year} {2019})},\ \Eprint
  {https://arxiv.org/abs/1901.03588} {arXiv:1901.03588 [astro-ph.GA]}
  \BibitemShut {NoStop}%
\bibitem [{\citenamefont {Bell}\ \emph {et~al.}(2020)\citenamefont {Bell},
  \citenamefont {Dent}, \citenamefont {Newstead}, \citenamefont {Sabharwal},\
  and\ \citenamefont {Weiler}}]{Bell:2019egg}%
  \BibitemOpen
  \bibfield  {author} {\bibinfo {author} {\bibfnamefont {N.~F.}\ \bibnamefont
  {Bell}}, \bibinfo {author} {\bibfnamefont {J.~B.}\ \bibnamefont {Dent}},
  \bibinfo {author} {\bibfnamefont {J.~L.}\ \bibnamefont {Newstead}}, \bibinfo
  {author} {\bibfnamefont {S.}~\bibnamefont {Sabharwal}},\ and\ \bibinfo
  {author} {\bibfnamefont {T.~J.}\ \bibnamefont {Weiler}},\ }\href
  {https://doi.org/10.1103/PhysRevD.101.015012} {\bibfield  {journal} {\bibinfo
   {journal} {Phys. Rev. D}\ }\textbf {\bibinfo {volume} {101}},\ \bibinfo
  {pages} {015012} (\bibinfo {year} {2020})},\ \Eprint
  {https://arxiv.org/abs/1905.00046} {arXiv:1905.00046 [hep-ph]} \BibitemShut
  {NoStop}%
\bibitem [{\citenamefont {Liu}\ \emph {et~al.}(2019)\citenamefont {Liu} \emph
  {et~al.}}]{Liu:2019kzq}%
  \BibitemOpen
  \bibfield  {author} {\bibinfo {author} {\bibfnamefont {Z.~Z.}\ \bibnamefont
  {Liu}} \emph {et~al.} (\bibinfo {collaboration} {CDEX}),\ }\href
  {https://doi.org/10.1103/PhysRevLett.123.161301} {\bibfield  {journal}
  {\bibinfo  {journal} {Phys. Rev. Lett.}\ }\textbf {\bibinfo {volume} {123}},\
  \bibinfo {pages} {161301} (\bibinfo {year} {2019})},\ \Eprint
  {https://arxiv.org/abs/1905.00354} {arXiv:1905.00354 [hep-ex]} \BibitemShut
  {NoStop}%
\bibitem [{\citenamefont {Baxter}\ \emph {et~al.}(2020)\citenamefont {Baxter},
  \citenamefont {Kahn},\ and\ \citenamefont {Krnjaic}}]{Baxter:2019pnz}%
  \BibitemOpen
  \bibfield  {author} {\bibinfo {author} {\bibfnamefont {D.}~\bibnamefont
  {Baxter}}, \bibinfo {author} {\bibfnamefont {Y.}~\bibnamefont {Kahn}},\ and\
  \bibinfo {author} {\bibfnamefont {G.}~\bibnamefont {Krnjaic}},\ }\href
  {https://doi.org/10.1103/PhysRevD.101.076014} {\bibfield  {journal} {\bibinfo
   {journal} {Phys. Rev. D}\ }\textbf {\bibinfo {volume} {101}},\ \bibinfo
  {pages} {076014} (\bibinfo {year} {2020})},\ \Eprint
  {https://arxiv.org/abs/1908.00012} {arXiv:1908.00012 [hep-ph]} \BibitemShut
  {NoStop}%
\bibitem [{\citenamefont {Essig}\ \emph {et~al.}(2020)\citenamefont {Essig},
  \citenamefont {Pradler}, \citenamefont {Sholapurkar},\ and\ \citenamefont
  {Yu}}]{Essig:2019xkx}%
  \BibitemOpen
  \bibfield  {author} {\bibinfo {author} {\bibfnamefont {R.}~\bibnamefont
  {Essig}}, \bibinfo {author} {\bibfnamefont {J.}~\bibnamefont {Pradler}},
  \bibinfo {author} {\bibfnamefont {M.}~\bibnamefont {Sholapurkar}},\ and\
  \bibinfo {author} {\bibfnamefont {T.-T.}\ \bibnamefont {Yu}},\ }\href
  {https://doi.org/10.1103/PhysRevLett.124.021801} {\bibfield  {journal}
  {\bibinfo  {journal} {Phys. Rev. Lett.}\ }\textbf {\bibinfo {volume} {124}},\
  \bibinfo {pages} {021801} (\bibinfo {year} {2020})},\ \Eprint
  {https://arxiv.org/abs/1908.10881} {arXiv:1908.10881 [hep-ph]} \BibitemShut
  {NoStop}%
\bibitem [{\citenamefont {Liang}\ \emph {et~al.}(2020)\citenamefont {Liang},
  \citenamefont {Zhang}, \citenamefont {Zheng},\ and\ \citenamefont
  {Zhang}}]{Liang:2019nnx}%
  \BibitemOpen
  \bibfield  {author} {\bibinfo {author} {\bibfnamefont {Z.-L.}\ \bibnamefont
  {Liang}}, \bibinfo {author} {\bibfnamefont {L.}~\bibnamefont {Zhang}},
  \bibinfo {author} {\bibfnamefont {F.}~\bibnamefont {Zheng}},\ and\ \bibinfo
  {author} {\bibfnamefont {P.}~\bibnamefont {Zhang}},\ }\href
  {https://doi.org/10.1103/PhysRevD.102.043007} {\bibfield  {journal} {\bibinfo
   {journal} {Phys. Rev. D}\ }\textbf {\bibinfo {volume} {102}},\ \bibinfo
  {pages} {043007} (\bibinfo {year} {2020})},\ \Eprint
  {https://arxiv.org/abs/1912.13484} {arXiv:1912.13484 [cond-mat.mes-hall]}
  \BibitemShut {NoStop}%
\bibitem [{\citenamefont {Knapen}\ \emph {et~al.}(2021)\citenamefont {Knapen},
  \citenamefont {Kozaczuk},\ and\ \citenamefont {Lin}}]{Knapen:2020aky}%
  \BibitemOpen
  \bibfield  {author} {\bibinfo {author} {\bibfnamefont {S.}~\bibnamefont
  {Knapen}}, \bibinfo {author} {\bibfnamefont {J.}~\bibnamefont {Kozaczuk}},\
  and\ \bibinfo {author} {\bibfnamefont {T.}~\bibnamefont {Lin}},\ }\href
  {https://doi.org/10.1103/PhysRevLett.127.081805} {\bibfield  {journal}
  {\bibinfo  {journal} {Phys. Rev. Lett.}\ }\textbf {\bibinfo {volume} {127}},\
  \bibinfo {pages} {081805} (\bibinfo {year} {2021})},\ \Eprint
  {https://arxiv.org/abs/2011.09496} {arXiv:2011.09496 [hep-ph]} \BibitemShut
  {NoStop}%
\bibitem [{\citenamefont {Araujo}\ \emph {et~al.}(2022)\citenamefont {Araujo}
  \emph {et~al.}}]{Araujo:2022wjh}%
  \BibitemOpen
  \bibfield  {author} {\bibinfo {author} {\bibfnamefont {H.~M.}\ \bibnamefont
  {Araujo}} \emph {et~al.},\ }\Eprint {https://arxiv.org/abs/2207.08284}
  {arXiv:2207.08284 [hep-ex]}  (\bibinfo {year} {2022})\BibitemShut {NoStop}%
\bibitem [{\citenamefont {Cox}\ \emph {et~al.}(2022)\citenamefont {Cox},
  \citenamefont {Dolan}, \citenamefont {McCabe},\ and\ \citenamefont
  {Quiney}}]{Cox:2022ekg}%
  \BibitemOpen
  \bibfield  {author} {\bibinfo {author} {\bibfnamefont {P.}~\bibnamefont
  {Cox}}, \bibinfo {author} {\bibfnamefont {M.~J.}\ \bibnamefont {Dolan}},
  \bibinfo {author} {\bibfnamefont {C.}~\bibnamefont {McCabe}},\ and\ \bibinfo
  {author} {\bibfnamefont {H.~M.}\ \bibnamefont {Quiney}},\ }\Eprint
  {https://arxiv.org/abs/2208.12222} {arXiv:2208.12222 [hep-ph]}  (\bibinfo
  {year} {2022})\BibitemShut {NoStop}%
\bibitem [{\citenamefont {Kouvaris}(2015)}]{Kouvaris:2015nsa}%
  \BibitemOpen
  \bibfield  {author} {\bibinfo {author} {\bibfnamefont {C.}~\bibnamefont
  {Kouvaris}},\ }\href {https://doi.org/10.1103/PhysRevD.92.075001} {\bibfield
  {journal} {\bibinfo  {journal} {Phys. Rev. D}\ }\textbf {\bibinfo {volume}
  {92}},\ \bibinfo {pages} {075001} (\bibinfo {year} {2015})},\ \Eprint
  {https://arxiv.org/abs/1506.04316} {arXiv:1506.04316 [hep-ph]} \BibitemShut
  {NoStop}%
\bibitem [{\citenamefont {An}\ \emph {et~al.}(2018)\citenamefont {An},
  \citenamefont {Pospelov}, \citenamefont {Pradler},\ and\ \citenamefont
  {Ritz}}]{An:2017ojc}%
  \BibitemOpen
  \bibfield  {author} {\bibinfo {author} {\bibfnamefont {H.}~\bibnamefont
  {An}}, \bibinfo {author} {\bibfnamefont {M.}~\bibnamefont {Pospelov}},
  \bibinfo {author} {\bibfnamefont {J.}~\bibnamefont {Pradler}},\ and\ \bibinfo
  {author} {\bibfnamefont {A.}~\bibnamefont {Ritz}},\ }\href
  {https://doi.org/10.1103/PhysRevLett.120.141801} {\bibfield  {journal}
  {\bibinfo  {journal} {Phys. Rev. Lett.}\ }\textbf {\bibinfo {volume} {120}},\
  \bibinfo {pages} {141801} (\bibinfo {year} {2018})},\ \bibinfo {note}
  {[Erratum: Phys.Rev.Lett. 121, 259903 (2018)]},\ \Eprint
  {https://arxiv.org/abs/1708.03642} {arXiv:1708.03642 [hep-ph]} \BibitemShut
  {NoStop}%
\bibitem [{\citenamefont {Emken}\ \emph {et~al.}(2018)\citenamefont {Emken},
  \citenamefont {Kouvaris},\ and\ \citenamefont {Nielsen}}]{Emken:2017hnp}%
  \BibitemOpen
  \bibfield  {author} {\bibinfo {author} {\bibfnamefont {T.}~\bibnamefont
  {Emken}}, \bibinfo {author} {\bibfnamefont {C.}~\bibnamefont {Kouvaris}},\
  and\ \bibinfo {author} {\bibfnamefont {N.~G.}\ \bibnamefont {Nielsen}},\
  }\href {https://doi.org/10.1103/PhysRevD.97.063007} {\bibfield  {journal}
  {\bibinfo  {journal} {Phys. Rev. D}\ }\textbf {\bibinfo {volume} {97}},\
  \bibinfo {pages} {063007} (\bibinfo {year} {2018})},\ \Eprint
  {https://arxiv.org/abs/1709.06573} {arXiv:1709.06573 [hep-ph]} \BibitemShut
  {NoStop}%
\bibitem [{\citenamefont {Bringmann}\ and\ \citenamefont
  {Pospelov}(2019)}]{Bringmann:2018cvk}%
  \BibitemOpen
  \bibfield  {author} {\bibinfo {author} {\bibfnamefont {T.}~\bibnamefont
  {Bringmann}}\ and\ \bibinfo {author} {\bibfnamefont {M.}~\bibnamefont
  {Pospelov}},\ }\href {https://doi.org/10.1103/PhysRevLett.122.171801}
  {\bibfield  {journal} {\bibinfo  {journal} {Phys. Rev. Lett.}\ }\textbf
  {\bibinfo {volume} {122}},\ \bibinfo {pages} {171801} (\bibinfo {year}
  {2019})},\ \Eprint {https://arxiv.org/abs/1810.10543} {arXiv:1810.10543
  [hep-ph]} \BibitemShut {NoStop}%
\bibitem [{\citenamefont {Ema}\ \emph {et~al.}(2019)\citenamefont {Ema},
  \citenamefont {Sala},\ and\ \citenamefont {Sato}}]{Ema:2018bih}%
  \BibitemOpen
  \bibfield  {author} {\bibinfo {author} {\bibfnamefont {Y.}~\bibnamefont
  {Ema}}, \bibinfo {author} {\bibfnamefont {F.}~\bibnamefont {Sala}},\ and\
  \bibinfo {author} {\bibfnamefont {R.}~\bibnamefont {Sato}},\ }\href
  {https://doi.org/10.1103/PhysRevLett.122.181802} {\bibfield  {journal}
  {\bibinfo  {journal} {Phys. Rev. Lett.}\ }\textbf {\bibinfo {volume} {122}},\
  \bibinfo {pages} {181802} (\bibinfo {year} {2019})},\ \Eprint
  {https://arxiv.org/abs/1811.00520} {arXiv:1811.00520 [hep-ph]} \BibitemShut
  {NoStop}%
\bibitem [{\citenamefont {Alvey}\ \emph {et~al.}(2019)\citenamefont {Alvey},
  \citenamefont {Campos}, \citenamefont {Fairbairn},\ and\ \citenamefont
  {You}}]{Alvey:2019zaa}%
  \BibitemOpen
  \bibfield  {author} {\bibinfo {author} {\bibfnamefont {J.}~\bibnamefont
  {Alvey}}, \bibinfo {author} {\bibfnamefont {M.}~\bibnamefont {Campos}},
  \bibinfo {author} {\bibfnamefont {M.}~\bibnamefont {Fairbairn}},\ and\
  \bibinfo {author} {\bibfnamefont {T.}~\bibnamefont {You}},\ }\href
  {https://doi.org/10.1103/PhysRevLett.123.261802} {\bibfield  {journal}
  {\bibinfo  {journal} {Phys. Rev. Lett.}\ }\textbf {\bibinfo {volume} {123}},\
  \bibinfo {pages} {261802} (\bibinfo {year} {2019})},\ \Eprint
  {https://arxiv.org/abs/1905.05776} {arXiv:1905.05776 [hep-ph]} \BibitemShut
  {NoStop}%
\bibitem [{\citenamefont {Cappiello}\ and\ \citenamefont
  {Beacom}(2019)}]{Cappiello:2019qsw}%
  \BibitemOpen
  \bibfield  {author} {\bibinfo {author} {\bibfnamefont {C.}~\bibnamefont
  {Cappiello}}\ and\ \bibinfo {author} {\bibfnamefont {J.~F.}\ \bibnamefont
  {Beacom}},\ }\href {https://doi.org/10.1103/PhysRevD.100.103011} {\bibfield
  {journal} {\bibinfo  {journal} {Phys. Rev. D}\ }\textbf {\bibinfo {volume}
  {100}},\ \bibinfo {pages} {103011} (\bibinfo {year} {2019})},\ \Eprint
  {https://arxiv.org/abs/1906.11283} {arXiv:1906.11283 [hep-ph]} \BibitemShut
  {NoStop}%
\bibitem [{\citenamefont {Herrera}\ and\ \citenamefont
  {Ibarra}(2021)}]{Herrera:2021puj}%
  \BibitemOpen
  \bibfield  {author} {\bibinfo {author} {\bibfnamefont {G.}~\bibnamefont
  {Herrera}}\ and\ \bibinfo {author} {\bibfnamefont {A.}~\bibnamefont
  {Ibarra}},\ }\href {https://doi.org/10.1016/j.physletb.2021.136551}
  {\bibfield  {journal} {\bibinfo  {journal} {Phys. Lett. B}\ }\textbf
  {\bibinfo {volume} {820}},\ \bibinfo {pages} {136551} (\bibinfo {year}
  {2021})},\ \Eprint {https://arxiv.org/abs/2104.04445} {arXiv:2104.04445
  [hep-ph]} \BibitemShut {NoStop}%
\bibitem [{\citenamefont {Essig}\ \emph
  {et~al.}(2012{\natexlab{a}})\citenamefont {Essig}, \citenamefont {Mardon},\
  and\ \citenamefont {Volansky}}]{Essig:2011nj}%
  \BibitemOpen
  \bibfield  {author} {\bibinfo {author} {\bibfnamefont {R.}~\bibnamefont
  {Essig}}, \bibinfo {author} {\bibfnamefont {J.}~\bibnamefont {Mardon}},\ and\
  \bibinfo {author} {\bibfnamefont {T.}~\bibnamefont {Volansky}},\ }\href
  {https://doi.org/10.1103/PhysRevD.85.076007} {\bibfield  {journal} {\bibinfo
  {journal} {Phys. Rev. D}\ }\textbf {\bibinfo {volume} {85}},\ \bibinfo
  {pages} {076007} (\bibinfo {year} {2012}{\natexlab{a}})},\ \Eprint
  {https://arxiv.org/abs/1108.5383} {arXiv:1108.5383 [hep-ph]} \BibitemShut
  {NoStop}%
\bibitem [{\citenamefont {Graham}\ \emph {et~al.}(2012)\citenamefont {Graham},
  \citenamefont {Kaplan}, \citenamefont {Rajendran},\ and\ \citenamefont
  {Walters}}]{Graham:2012su}%
  \BibitemOpen
  \bibfield  {author} {\bibinfo {author} {\bibfnamefont {P.~W.}\ \bibnamefont
  {Graham}}, \bibinfo {author} {\bibfnamefont {D.~E.}\ \bibnamefont {Kaplan}},
  \bibinfo {author} {\bibfnamefont {S.}~\bibnamefont {Rajendran}},\ and\
  \bibinfo {author} {\bibfnamefont {M.~T.}\ \bibnamefont {Walters}},\ }\href
  {https://doi.org/10.1016/j.dark.2012.09.001} {\bibfield  {journal} {\bibinfo
  {journal} {Phys. Dark Univ.}\ }\textbf {\bibinfo {volume} {1}},\ \bibinfo
  {pages} {32} (\bibinfo {year} {2012})},\ \Eprint
  {https://arxiv.org/abs/1203.2531} {arXiv:1203.2531 [hep-ph]} \BibitemShut
  {NoStop}%
\bibitem [{\citenamefont {Hochberg}\ \emph
  {et~al.}(2016{\natexlab{a}})\citenamefont {Hochberg}, \citenamefont {Zhao},\
  and\ \citenamefont {Zurek}}]{Hochberg:2015pha}%
  \BibitemOpen
  \bibfield  {author} {\bibinfo {author} {\bibfnamefont {Y.}~\bibnamefont
  {Hochberg}}, \bibinfo {author} {\bibfnamefont {Y.}~\bibnamefont {Zhao}},\
  and\ \bibinfo {author} {\bibfnamefont {K.~M.}\ \bibnamefont {Zurek}},\ }\href
  {https://doi.org/10.1103/PhysRevLett.116.011301} {\bibfield  {journal}
  {\bibinfo  {journal} {Phys. Rev. Lett.}\ }\textbf {\bibinfo {volume} {116}},\
  \bibinfo {pages} {011301} (\bibinfo {year} {2016}{\natexlab{a}})},\ \Eprint
  {https://arxiv.org/abs/1504.07237} {arXiv:1504.07237 [hep-ph]} \BibitemShut
  {NoStop}%
\bibitem [{\citenamefont {Essig}\ \emph {et~al.}(2016)\citenamefont {Essig},
  \citenamefont {Fernandez-Serra}, \citenamefont {Mardon}, \citenamefont
  {Soto}, \citenamefont {Volansky},\ and\ \citenamefont {Yu}}]{Essig:2015cda}%
  \BibitemOpen
  \bibfield  {author} {\bibinfo {author} {\bibfnamefont {R.}~\bibnamefont
  {Essig}}, \bibinfo {author} {\bibfnamefont {M.}~\bibnamefont
  {Fernandez-Serra}}, \bibinfo {author} {\bibfnamefont {J.}~\bibnamefont
  {Mardon}}, \bibinfo {author} {\bibfnamefont {A.}~\bibnamefont {Soto}},
  \bibinfo {author} {\bibfnamefont {T.}~\bibnamefont {Volansky}},\ and\
  \bibinfo {author} {\bibfnamefont {T.-T.}\ \bibnamefont {Yu}},\ }\href
  {https://doi.org/10.1007/JHEP05(2016)046} {\bibfield  {journal} {\bibinfo
  {journal} {JHEP}\ }\textbf {\bibinfo {volume} {05}},\ \bibinfo {pages}
  {046}},\ \Eprint {https://arxiv.org/abs/1509.01598} {arXiv:1509.01598
  [hep-ph]} \BibitemShut {NoStop}%
\bibitem [{\citenamefont {Hochberg}\ \emph
  {et~al.}(2016{\natexlab{b}})\citenamefont {Hochberg}, \citenamefont {Pyle},
  \citenamefont {Zhao},\ and\ \citenamefont {Zurek}}]{Hochberg:2015fth}%
  \BibitemOpen
  \bibfield  {author} {\bibinfo {author} {\bibfnamefont {Y.}~\bibnamefont
  {Hochberg}}, \bibinfo {author} {\bibfnamefont {M.}~\bibnamefont {Pyle}},
  \bibinfo {author} {\bibfnamefont {Y.}~\bibnamefont {Zhao}},\ and\ \bibinfo
  {author} {\bibfnamefont {K.~M.}\ \bibnamefont {Zurek}},\ }\href
  {https://doi.org/10.1007/JHEP08(2016)057} {\bibfield  {journal} {\bibinfo
  {journal} {JHEP}\ }\textbf {\bibinfo {volume} {08}},\ \bibinfo {pages}
  {057}},\ \Eprint {https://arxiv.org/abs/1512.04533} {arXiv:1512.04533
  [hep-ph]} \BibitemShut {NoStop}%
\bibitem [{\citenamefont {Derenzo}\ \emph {et~al.}(2017)\citenamefont
  {Derenzo}, \citenamefont {Essig}, \citenamefont {Massari}, \citenamefont
  {Soto},\ and\ \citenamefont {Yu}}]{Derenzo:2016fse}%
  \BibitemOpen
  \bibfield  {author} {\bibinfo {author} {\bibfnamefont {S.}~\bibnamefont
  {Derenzo}}, \bibinfo {author} {\bibfnamefont {R.}~\bibnamefont {Essig}},
  \bibinfo {author} {\bibfnamefont {A.}~\bibnamefont {Massari}}, \bibinfo
  {author} {\bibfnamefont {A.}~\bibnamefont {Soto}},\ and\ \bibinfo {author}
  {\bibfnamefont {T.-T.}\ \bibnamefont {Yu}},\ }\href
  {https://doi.org/10.1103/PhysRevD.96.016026} {\bibfield  {journal} {\bibinfo
  {journal} {Phys. Rev. D}\ }\textbf {\bibinfo {volume} {96}},\ \bibinfo
  {pages} {016026} (\bibinfo {year} {2017})},\ \Eprint
  {https://arxiv.org/abs/1607.01009} {arXiv:1607.01009 [hep-ph]} \BibitemShut
  {NoStop}%
\bibitem [{\citenamefont {Essig}\ \emph
  {et~al.}(2017{\natexlab{a}})\citenamefont {Essig}, \citenamefont {Mardon},
  \citenamefont {Slone},\ and\ \citenamefont {Volansky}}]{Essig:2016crl}%
  \BibitemOpen
  \bibfield  {author} {\bibinfo {author} {\bibfnamefont {R.}~\bibnamefont
  {Essig}}, \bibinfo {author} {\bibfnamefont {J.}~\bibnamefont {Mardon}},
  \bibinfo {author} {\bibfnamefont {O.}~\bibnamefont {Slone}},\ and\ \bibinfo
  {author} {\bibfnamefont {T.}~\bibnamefont {Volansky}},\ }\href
  {https://doi.org/10.1103/PhysRevD.95.056011} {\bibfield  {journal} {\bibinfo
  {journal} {Phys. Rev. D}\ }\textbf {\bibinfo {volume} {95}},\ \bibinfo
  {pages} {056011} (\bibinfo {year} {2017}{\natexlab{a}})},\ \Eprint
  {https://arxiv.org/abs/1608.02940} {arXiv:1608.02940 [hep-ph]} \BibitemShut
  {NoStop}%
\bibitem [{\citenamefont {Cavoto}\ \emph {et~al.}(2018)\citenamefont {Cavoto},
  \citenamefont {Luchetta},\ and\ \citenamefont {Polosa}}]{Cavoto:2017otc}%
  \BibitemOpen
  \bibfield  {author} {\bibinfo {author} {\bibfnamefont {G.}~\bibnamefont
  {Cavoto}}, \bibinfo {author} {\bibfnamefont {F.}~\bibnamefont {Luchetta}},\
  and\ \bibinfo {author} {\bibfnamefont {A.~D.}\ \bibnamefont {Polosa}},\
  }\href {https://doi.org/10.1016/j.physletb.2017.11.064} {\bibfield  {journal}
  {\bibinfo  {journal} {Phys. Lett. B}\ }\textbf {\bibinfo {volume} {776}},\
  \bibinfo {pages} {338} (\bibinfo {year} {2018})},\ \Eprint
  {https://arxiv.org/abs/1706.02487} {arXiv:1706.02487 [hep-ph]} \BibitemShut
  {NoStop}%
\bibitem [{\citenamefont {Hochberg}\ \emph {et~al.}(2018)\citenamefont
  {Hochberg}, \citenamefont {Kahn}, \citenamefont {Lisanti}, \citenamefont
  {Zurek}, \citenamefont {Grushin}, \citenamefont {Ilan}, \citenamefont
  {Griffin}, \citenamefont {Liu}, \citenamefont {Weber},\ and\ \citenamefont
  {Neaton}}]{Hochberg:2017wce}%
  \BibitemOpen
  \bibfield  {author} {\bibinfo {author} {\bibfnamefont {Y.}~\bibnamefont
  {Hochberg}}, \bibinfo {author} {\bibfnamefont {Y.}~\bibnamefont {Kahn}},
  \bibinfo {author} {\bibfnamefont {M.}~\bibnamefont {Lisanti}}, \bibinfo
  {author} {\bibfnamefont {K.~M.}\ \bibnamefont {Zurek}}, \bibinfo {author}
  {\bibfnamefont {A.~G.}\ \bibnamefont {Grushin}}, \bibinfo {author}
  {\bibfnamefont {R.}~\bibnamefont {Ilan}}, \bibinfo {author} {\bibfnamefont
  {S.~M.}\ \bibnamefont {Griffin}}, \bibinfo {author} {\bibfnamefont {Z.-F.}\
  \bibnamefont {Liu}}, \bibinfo {author} {\bibfnamefont {S.~F.}\ \bibnamefont
  {Weber}},\ and\ \bibinfo {author} {\bibfnamefont {J.~B.}\ \bibnamefont
  {Neaton}},\ }\href {https://doi.org/10.1103/PhysRevD.97.015004} {\bibfield
  {journal} {\bibinfo  {journal} {Phys. Rev. D}\ }\textbf {\bibinfo {volume}
  {97}},\ \bibinfo {pages} {015004} (\bibinfo {year} {2018})},\ \Eprint
  {https://arxiv.org/abs/1708.08929} {arXiv:1708.08929 [hep-ph]} \BibitemShut
  {NoStop}%
\bibitem [{\citenamefont {Kurinsky}\ \emph {et~al.}(2019)\citenamefont
  {Kurinsky}, \citenamefont {Yu}, \citenamefont {Hochberg},\ and\ \citenamefont
  {Cabrera}}]{Kurinsky:2019pgb}%
  \BibitemOpen
  \bibfield  {author} {\bibinfo {author} {\bibfnamefont {N.~A.}\ \bibnamefont
  {Kurinsky}}, \bibinfo {author} {\bibfnamefont {T.~C.}\ \bibnamefont {Yu}},
  \bibinfo {author} {\bibfnamefont {Y.}~\bibnamefont {Hochberg}},\ and\
  \bibinfo {author} {\bibfnamefont {B.}~\bibnamefont {Cabrera}},\ }\href
  {https://doi.org/10.1103/PhysRevD.99.123005} {\bibfield  {journal} {\bibinfo
  {journal} {Phys. Rev. D}\ }\textbf {\bibinfo {volume} {99}},\ \bibinfo
  {pages} {123005} (\bibinfo {year} {2019})},\ \Eprint
  {https://arxiv.org/abs/1901.07569} {arXiv:1901.07569 [hep-ex]} \BibitemShut
  {NoStop}%
\bibitem [{\citenamefont {Essig}\ \emph {et~al.}(2019)\citenamefont {Essig},
  \citenamefont {P\'erez-R\'\i{}os}, \citenamefont {Ramani},\ and\
  \citenamefont {Slone}}]{Essig:2019kfe}%
  \BibitemOpen
  \bibfield  {author} {\bibinfo {author} {\bibfnamefont {R.}~\bibnamefont
  {Essig}}, \bibinfo {author} {\bibfnamefont {J.}~\bibnamefont
  {P\'erez-R\'\i{}os}}, \bibinfo {author} {\bibfnamefont {H.}~\bibnamefont
  {Ramani}},\ and\ \bibinfo {author} {\bibfnamefont {O.}~\bibnamefont
  {Slone}},\ }\href {https://doi.org/10.1103/PhysRevResearch.1.033105}
  {\bibfield  {journal} {\bibinfo  {journal} {Phys. Rev. Research.}\ }\textbf
  {\bibinfo {volume} {1}},\ \bibinfo {pages} {033105} (\bibinfo {year}
  {2019})},\ \Eprint {https://arxiv.org/abs/1907.07682} {arXiv:1907.07682
  [hep-ph]} \BibitemShut {NoStop}%
\bibitem [{\citenamefont {Geilhufe}\ \emph {et~al.}(2020)\citenamefont
  {Geilhufe}, \citenamefont {Kahlhoefer},\ and\ \citenamefont
  {Winkler}}]{Geilhufe:2019ndy}%
  \BibitemOpen
  \bibfield  {author} {\bibinfo {author} {\bibfnamefont {R.~M.}\ \bibnamefont
  {Geilhufe}}, \bibinfo {author} {\bibfnamefont {F.}~\bibnamefont
  {Kahlhoefer}},\ and\ \bibinfo {author} {\bibfnamefont {M.~W.}\ \bibnamefont
  {Winkler}},\ }\href {https://doi.org/10.1103/PhysRevD.101.055005} {\bibfield
  {journal} {\bibinfo  {journal} {Phys. Rev. D}\ }\textbf {\bibinfo {volume}
  {101}},\ \bibinfo {pages} {055005} (\bibinfo {year} {2020})},\ \Eprint
  {https://arxiv.org/abs/1910.02091} {arXiv:1910.02091 [hep-ph]} \BibitemShut
  {NoStop}%
\bibitem [{\citenamefont {Blanco}\ \emph {et~al.}(2021)\citenamefont {Blanco},
  \citenamefont {Kahn}, \citenamefont {Lillard},\ and\ \citenamefont
  {McDermott}}]{Blanco:2021hlm}%
  \BibitemOpen
  \bibfield  {author} {\bibinfo {author} {\bibfnamefont {C.}~\bibnamefont
  {Blanco}}, \bibinfo {author} {\bibfnamefont {Y.}~\bibnamefont {Kahn}},
  \bibinfo {author} {\bibfnamefont {B.}~\bibnamefont {Lillard}},\ and\ \bibinfo
  {author} {\bibfnamefont {S.~D.}\ \bibnamefont {McDermott}},\ }\href
  {https://doi.org/10.1103/PhysRevD.104.036011} {\bibfield  {journal} {\bibinfo
   {journal} {Phys. Rev. D}\ }\textbf {\bibinfo {volume} {104}},\ \bibinfo
  {pages} {036011} (\bibinfo {year} {2021})},\ \Eprint
  {https://arxiv.org/abs/2103.08601} {arXiv:2103.08601 [hep-ph]} \BibitemShut
  {NoStop}%
\bibitem [{\citenamefont {Kahn}\ and\ \citenamefont
  {Lin}(2022)}]{Kahn:2021ttr}%
  \BibitemOpen
  \bibfield  {author} {\bibinfo {author} {\bibfnamefont {Y.}~\bibnamefont
  {Kahn}}\ and\ \bibinfo {author} {\bibfnamefont {T.}~\bibnamefont {Lin}},\
  }\href {https://doi.org/10.1088/1361-6633/ac5f63} {\bibfield  {journal}
  {\bibinfo  {journal} {Rept. Prog. Phys.}\ }\textbf {\bibinfo {volume} {85}},\
  \bibinfo {pages} {066901} (\bibinfo {year} {2022})},\ \Eprint
  {https://arxiv.org/abs/2108.03239} {arXiv:2108.03239 [hep-ph]} \BibitemShut
  {NoStop}%
\bibitem [{\citenamefont {Essig}\ \emph
  {et~al.}(2012{\natexlab{b}})\citenamefont {Essig}, \citenamefont
  {Manalaysay}, \citenamefont {Mardon}, \citenamefont {Sorensen},\ and\
  \citenamefont {Volansky}}]{Essig:2012yx}%
  \BibitemOpen
  \bibfield  {author} {\bibinfo {author} {\bibfnamefont {R.}~\bibnamefont
  {Essig}}, \bibinfo {author} {\bibfnamefont {A.}~\bibnamefont {Manalaysay}},
  \bibinfo {author} {\bibfnamefont {J.}~\bibnamefont {Mardon}}, \bibinfo
  {author} {\bibfnamefont {P.}~\bibnamefont {Sorensen}},\ and\ \bibinfo
  {author} {\bibfnamefont {T.}~\bibnamefont {Volansky}},\ }\href
  {https://doi.org/10.1103/PhysRevLett.109.021301} {\bibfield  {journal}
  {\bibinfo  {journal} {Phys. Rev. Lett.}\ }\textbf {\bibinfo {volume} {109}},\
  \bibinfo {pages} {021301} (\bibinfo {year} {2012}{\natexlab{b}})},\ \Eprint
  {https://arxiv.org/abs/1206.2644} {arXiv:1206.2644 [astro-ph.CO]}
  \BibitemShut {NoStop}%
\bibitem [{\citenamefont {Essig}\ \emph
  {et~al.}(2017{\natexlab{b}})\citenamefont {Essig}, \citenamefont {Volansky},\
  and\ \citenamefont {Yu}}]{Essig:2017kqs}%
  \BibitemOpen
  \bibfield  {author} {\bibinfo {author} {\bibfnamefont {R.}~\bibnamefont
  {Essig}}, \bibinfo {author} {\bibfnamefont {T.}~\bibnamefont {Volansky}},\
  and\ \bibinfo {author} {\bibfnamefont {T.-T.}\ \bibnamefont {Yu}},\ }\href
  {https://doi.org/10.1103/PhysRevD.96.043017} {\bibfield  {journal} {\bibinfo
  {journal} {Phys. Rev. D}\ }\textbf {\bibinfo {volume} {96}},\ \bibinfo
  {pages} {043017} (\bibinfo {year} {2017}{\natexlab{b}})},\ \Eprint
  {https://arxiv.org/abs/1703.00910} {arXiv:1703.00910 [hep-ph]} \BibitemShut
  {NoStop}%
\bibitem [{\citenamefont {Agnes}\ \emph {et~al.}(2018)\citenamefont {Agnes}
  \emph {et~al.}}]{Agnes:2018oej}%
  \BibitemOpen
  \bibfield  {author} {\bibinfo {author} {\bibfnamefont {P.}~\bibnamefont
  {Agnes}} \emph {et~al.} (\bibinfo {collaboration} {DarkSide}),\ }\href
  {https://doi.org/10.1103/PhysRevLett.121.111303} {\bibfield  {journal}
  {\bibinfo  {journal} {Phys. Rev. Lett.}\ }\textbf {\bibinfo {volume} {121}},\
  \bibinfo {pages} {111303} (\bibinfo {year} {2018})},\ \Eprint
  {https://arxiv.org/abs/1802.06998} {arXiv:1802.06998 [astro-ph.CO]}
  \BibitemShut {NoStop}%
\bibitem [{\citenamefont {Agnese}\ \emph {et~al.}(2018)\citenamefont {Agnese}
  \emph {et~al.}}]{Agnese:2018col}%
  \BibitemOpen
  \bibfield  {author} {\bibinfo {author} {\bibfnamefont {R.}~\bibnamefont
  {Agnese}} \emph {et~al.} (\bibinfo {collaboration} {SuperCDMS}),\ }\href
  {https://doi.org/10.1103/PhysRevLett.121.051301} {\bibfield  {journal}
  {\bibinfo  {journal} {Phys. Rev. Lett.}\ }\textbf {\bibinfo {volume} {121}},\
  \bibinfo {pages} {051301} (\bibinfo {year} {2018})},\ \bibinfo {note}
  {[Erratum: Phys.Rev.Lett. 122, 069901 (2019)]},\ \Eprint
  {https://arxiv.org/abs/1804.10697} {arXiv:1804.10697 [hep-ex]} \BibitemShut
  {NoStop}%
\bibitem [{\citenamefont {Aprile}\ \emph {et~al.}(2019)\citenamefont {Aprile}
  \emph {et~al.}}]{Aprile:2019xxb}%
  \BibitemOpen
  \bibfield  {author} {\bibinfo {author} {\bibfnamefont {E.}~\bibnamefont
  {Aprile}} \emph {et~al.} (\bibinfo {collaboration} {XENON}),\ }\href
  {https://doi.org/10.1103/PhysRevLett.123.251801} {\bibfield  {journal}
  {\bibinfo  {journal} {Phys. Rev. Lett.}\ }\textbf {\bibinfo {volume} {123}},\
  \bibinfo {pages} {251801} (\bibinfo {year} {2019})},\ \Eprint
  {https://arxiv.org/abs/1907.11485} {arXiv:1907.11485 [hep-ex]} \BibitemShut
  {NoStop}%
\bibitem [{\citenamefont {Aguilar-Arevalo}\ \emph {et~al.}(2019)\citenamefont
  {Aguilar-Arevalo} \emph {et~al.}}]{Aguilar-Arevalo:2019wdi}%
  \BibitemOpen
  \bibfield  {author} {\bibinfo {author} {\bibfnamefont {A.}~\bibnamefont
  {Aguilar-Arevalo}} \emph {et~al.} (\bibinfo {collaboration} {DAMIC}),\ }\href
  {https://doi.org/10.1103/PhysRevLett.123.181802} {\bibfield  {journal}
  {\bibinfo  {journal} {Phys. Rev. Lett.}\ }\textbf {\bibinfo {volume} {123}},\
  \bibinfo {pages} {181802} (\bibinfo {year} {2019})},\ \Eprint
  {https://arxiv.org/abs/1907.12628} {arXiv:1907.12628 [astro-ph.CO]}
  \BibitemShut {NoStop}%
\bibitem [{\citenamefont {Blanco}\ \emph {et~al.}(2020)\citenamefont {Blanco},
  \citenamefont {Collar}, \citenamefont {Kahn},\ and\ \citenamefont
  {Lillard}}]{Blanco:2019lrf}%
  \BibitemOpen
  \bibfield  {author} {\bibinfo {author} {\bibfnamefont {C.}~\bibnamefont
  {Blanco}}, \bibinfo {author} {\bibfnamefont {J.~I.}\ \bibnamefont {Collar}},
  \bibinfo {author} {\bibfnamefont {Y.}~\bibnamefont {Kahn}},\ and\ \bibinfo
  {author} {\bibfnamefont {B.}~\bibnamefont {Lillard}},\ }\href
  {https://doi.org/10.1103/PhysRevD.101.056001} {\bibfield  {journal} {\bibinfo
   {journal} {Phys. Rev. D}\ }\textbf {\bibinfo {volume} {101}},\ \bibinfo
  {pages} {056001} (\bibinfo {year} {2020})},\ \Eprint
  {https://arxiv.org/abs/1912.02822} {arXiv:1912.02822 [hep-ph]} \BibitemShut
  {NoStop}%
\bibitem [{\citenamefont {Arnaud}\ \emph {et~al.}(2020)\citenamefont {Arnaud}
  \emph {et~al.}}]{Arnaud:2020svb}%
  \BibitemOpen
  \bibfield  {author} {\bibinfo {author} {\bibfnamefont {Q.}~\bibnamefont
  {Arnaud}} \emph {et~al.} (\bibinfo {collaboration} {EDELWEISS}),\ }\href
  {https://doi.org/10.1103/PhysRevLett.125.141301} {\bibfield  {journal}
  {\bibinfo  {journal} {Phys. Rev. Lett.}\ }\textbf {\bibinfo {volume} {125}},\
  \bibinfo {pages} {141301} (\bibinfo {year} {2020})},\ \Eprint
  {https://arxiv.org/abs/2003.01046} {arXiv:2003.01046 [astro-ph.GA]}
  \BibitemShut {NoStop}%
\bibitem [{\citenamefont {Barak}\ \emph {et~al.}(2020)\citenamefont {Barak}
  \emph {et~al.}}]{Barak:2020fql}%
  \BibitemOpen
  \bibfield  {author} {\bibinfo {author} {\bibfnamefont {L.}~\bibnamefont
  {Barak}} \emph {et~al.} (\bibinfo {collaboration} {SENSEI}),\ }\href
  {https://doi.org/10.1103/PhysRevLett.125.171802} {\bibfield  {journal}
  {\bibinfo  {journal} {Phys. Rev. Lett.}\ }\textbf {\bibinfo {volume} {125}},\
  \bibinfo {pages} {171802} (\bibinfo {year} {2020})},\ \Eprint
  {https://arxiv.org/abs/2004.11378} {arXiv:2004.11378 [astro-ph.CO]}
  \BibitemShut {NoStop}%
\bibitem [{\citenamefont {Giomataris}\ and\ \citenamefont
  {Vergados}(2004)}]{Giomataris:2003bp}%
  \BibitemOpen
  \bibfield  {author} {\bibinfo {author} {\bibfnamefont {Y.}~\bibnamefont
  {Giomataris}}\ and\ \bibinfo {author} {\bibfnamefont {J.~D.}\ \bibnamefont
  {Vergados}},\ }\href {https://doi.org/10.1016/j.nima.2004.04.223} {\bibfield
  {journal} {\bibinfo  {journal} {Nucl. Instrum. Meth. A}\ }\textbf {\bibinfo
  {volume} {530}},\ \bibinfo {pages} {330} (\bibinfo {year} {2004})},\ \Eprint
  {https://arxiv.org/abs/hep-ex/0303045} {arXiv:hep-ex/0303045} \BibitemShut
  {NoStop}%
\bibitem [{\citenamefont {Giomataris}\ \emph {et~al.}(2008)\citenamefont
  {Giomataris} \emph {et~al.}}]{Giomataris_2008}%
  \BibitemOpen
  \bibfield  {author} {\bibinfo {author} {\bibfnamefont {I.}~\bibnamefont
  {Giomataris}} \emph {et~al.},\ }\href
  {https://doi.org/10.1088/1748-0221/3/09/p09007} {\bibfield  {journal}
  {\bibinfo  {journal} {{Journal of Instrumentation}}\ }\textbf {\bibinfo
  {volume} {3}},\ \bibinfo {pages} {P09007} (\bibinfo {year}
  {2008})}\BibitemShut {NoStop}%
\bibitem [{\citenamefont {Arnaud}\ \emph
  {et~al.}(2018{\natexlab{b}})\citenamefont {Arnaud} \emph
  {et~al.}}]{Arnaud:2017bjh}%
  \BibitemOpen
  \bibfield  {author} {\bibinfo {author} {\bibfnamefont {Q.}~\bibnamefont
  {Arnaud}} \emph {et~al.} (\bibinfo {collaboration} {NEWS-G}),\ }\href
  {https://doi.org/10.1016/j.astropartphys.2017.10.009} {\bibfield  {journal}
  {\bibinfo  {journal} {Astropart. Phys.}\ }\textbf {\bibinfo {volume} {97}},\
  \bibinfo {pages} {54} (\bibinfo {year} {2018}{\natexlab{b}})},\ \Eprint
  {https://arxiv.org/abs/1706.04934} {arXiv:1706.04934 [astro-ph.IM]}
  \BibitemShut {NoStop}%
\bibitem [{\citenamefont {Balogh}\ \emph {et~al.}(2021)\citenamefont {Balogh}
  \emph {et~al.}}]{Balogh:2020nmo}%
  \BibitemOpen
  \bibfield  {author} {\bibinfo {author} {\bibfnamefont {L.}~\bibnamefont
  {Balogh}} \emph {et~al.} (\bibinfo {collaboration} {NEWS-G}),\ }\href
  {https://doi.org/10.1016/j.nima.2020.164844} {\bibfield  {journal} {\bibinfo
  {journal} {Nucl. Instrum. Meth. A}\ }\textbf {\bibinfo {volume} {988}},\
  \bibinfo {pages} {164844} (\bibinfo {year} {2021})},\ \Eprint
  {https://arxiv.org/abs/2008.03153} {arXiv:2008.03153 [physics.ins-det]}
  \BibitemShut {NoStop}%
\bibitem [{\citenamefont {Balogh}\ \emph {et~al.}(2022)\citenamefont {Balogh}
  \emph {et~al.}}]{NEWS-G:2022kon}%
  \BibitemOpen
  \bibfield  {author} {\bibinfo {author} {\bibfnamefont {L.}~\bibnamefont
  {Balogh}} \emph {et~al.} (\bibinfo {collaboration} {NEWS-G}),\ }\Eprint
  {https://arxiv.org/abs/2205.15433} {arXiv:2205.15433 [physics.ins-det]}
  (\bibinfo {year} {2022})\BibitemShut {NoStop}%
\bibitem [{\citenamefont {Balogh}\ \emph {et~al.}(2023)\citenamefont {Balogh}
  \emph {et~al.}}]{Balogh:2023hba}%
  \BibitemOpen
  \bibfield  {author} {\bibinfo {author} {\bibfnamefont {L.}~\bibnamefont
  {Balogh}} \emph {et~al.},\ }\Eprint {https://arxiv.org/abs/2301.05183}
  {arXiv:2301.05183 [hep-ex]}  (\bibinfo {year} {2023})\BibitemShut {NoStop}%
\bibitem [{\citenamefont {Giganon}\ \emph {et~al.}(2017)\citenamefont {Giganon}
  \emph {et~al.}}]{Giganon:2017isb}%
  \BibitemOpen
  \bibfield  {author} {\bibinfo {author} {\bibfnamefont {A.}~\bibnamefont
  {Giganon}} \emph {et~al.},\ }\href
  {https://doi.org/10.1088/1748-0221/12/12/P12031} {\bibfield  {journal}
  {\bibinfo  {journal} {JINST}\ }\textbf {\bibinfo {volume} {12}}\bibfield
  {number} {\bibinfo  {number} { (12)},\ \bibinfo {pages} {P12031}},\ }\Eprint
  {https://arxiv.org/abs/1707.09254} {arXiv:1707.09254 [physics.ins-det]}
  \BibitemShut {NoStop}%
\bibitem [{\citenamefont {Giomataris}\ \emph {et~al.}(2020)\citenamefont
  {Giomataris} \emph {et~al.}}]{Giomataris:2020rna}%
  \BibitemOpen
  \bibfield  {author} {\bibinfo {author} {\bibfnamefont {I.}~\bibnamefont
  {Giomataris}} \emph {et~al.},\ }\href
  {https://doi.org/10.1088/1748-0221/15/11/P11023} {\bibfield  {journal}
  {\bibinfo  {journal} {JINST}\ }\textbf {\bibinfo {volume} {15}}\bibfield
  {number} {\bibinfo  {number} { (11)},\ \bibinfo {pages} {11}},\ }\Eprint
  {https://arxiv.org/abs/2003.01068} {arXiv:2003.01068 [physics.ins-det]}
  \BibitemShut {NoStop}%
\bibitem [{\citenamefont {Katsioulas}\ \emph
  {et~al.}(2022{\natexlab{a}})\citenamefont {Katsioulas}, \citenamefont
  {Knights}, \citenamefont {Manthos}, \citenamefont {Matthews}, \citenamefont
  {Neep}, \citenamefont {Nikolopoulos},\ and\ \citenamefont
  {Ward}}]{Katsioulas:2022cqe}%
  \BibitemOpen
  \bibfield  {author} {\bibinfo {author} {\bibfnamefont {I.}~\bibnamefont
  {Katsioulas}}, \bibinfo {author} {\bibfnamefont {P.}~\bibnamefont {Knights}},
  \bibinfo {author} {\bibfnamefont {I.}~\bibnamefont {Manthos}}, \bibinfo
  {author} {\bibfnamefont {J.}~\bibnamefont {Matthews}}, \bibinfo {author}
  {\bibfnamefont {T.}~\bibnamefont {Neep}}, \bibinfo {author} {\bibfnamefont
  {K.}~\bibnamefont {Nikolopoulos}},\ and\ \bibinfo {author} {\bibfnamefont
  {R.}~\bibnamefont {Ward}},\ }\href
  {https://doi.org/10.1088/1748-0221/17/08/C08025} {\bibfield  {journal}
  {\bibinfo  {journal} {JINST}\ }\textbf {\bibinfo {volume} {17}}\bibinfo
  {number} { (08)},\ \bibinfo {pages} {C08025}}\BibitemShut {NoStop}%
\bibitem [{\citenamefont {Bougamont}\ \emph {et~al.}(2012)\citenamefont
  {Bougamont} \emph {et~al.}}]{Bougamont:2010mj}%
  \BibitemOpen
\bibfield  {number} {  }\bibfield  {author} {\bibinfo {author} {\bibfnamefont
  {E.}~\bibnamefont {Bougamont}} \emph {et~al.},\ }\href
  {https://doi.org/10.4236/jmp.2012.31008} {\bibfield  {journal} {\bibinfo
  {journal} {JMP}\ }\textbf {\bibinfo {volume} {3}},\ \bibinfo {pages} {57}
  (\bibinfo {year} {2012})},\ \Eprint {https://arxiv.org/abs/1010.4132}
  {arXiv:1010.4132 [physics.ins-det]} \BibitemShut {NoStop}%
\bibitem [{\citenamefont {Arnaud}\ \emph {et~al.}(2019)\citenamefont {Arnaud}
  \emph {et~al.}}]{Arnaud:2019nyp}%
  \BibitemOpen
  \bibfield  {author} {\bibinfo {author} {\bibfnamefont {Q.}~\bibnamefont
  {Arnaud}} \emph {et~al.} (\bibinfo {collaboration} {NEWS-G}),\ }\href
  {https://doi.org/10.1103/PhysRevD.99.102003} {\bibfield  {journal} {\bibinfo
  {journal} {Phys. Rev. D}\ }\textbf {\bibinfo {volume} {99}},\ \bibinfo
  {pages} {102003} (\bibinfo {year} {2019})},\ \Eprint
  {https://arxiv.org/abs/1902.08960} {arXiv:1902.08960 [physics.ins-det]}
  \BibitemShut {NoStop}%
\bibitem [{\citenamefont {Vergados}\ and\ \citenamefont
  {Giomataris}(2007)}]{Vergados:2005ny}%
  \BibitemOpen
  \bibfield  {author} {\bibinfo {author} {\bibfnamefont {J.~D.}\ \bibnamefont
  {Vergados}}\ and\ \bibinfo {author} {\bibfnamefont {Y.}~\bibnamefont
  {Giomataris}},\ }\href {https://doi.org/10.1134/S1063778807010164} {\bibfield
   {journal} {\bibinfo  {journal} {Phys. Atom. Nucl.}\ }\textbf {\bibinfo
  {volume} {70}},\ \bibinfo {pages} {140} (\bibinfo {year} {2007})},\ \Eprint
  {https://arxiv.org/abs/astro-ph/0511470} {arXiv:astro-ph/0511470}
  \BibitemShut {NoStop}%
\bibitem [{\citenamefont {Meregaglia}\ \emph {et~al.}(2018)\citenamefont
  {Meregaglia} \emph {et~al.}}]{Meregaglia:2017nhx}%
  \BibitemOpen
  \bibfield  {author} {\bibinfo {author} {\bibfnamefont {A.}~\bibnamefont
  {Meregaglia}} \emph {et~al.},\ }\href
  {https://doi.org/10.1088/1748-0221/13/01/P01009} {\bibfield  {journal}
  {\bibinfo  {journal} {JINST}\ }\textbf {\bibinfo {volume} {13}}\bibfield
  {number} {\bibinfo  {number} { (01)},\ \bibinfo {pages} {P01009}},\ }\Eprint
  {https://arxiv.org/abs/1710.04536} {arXiv:1710.04536 [physics.ins-det]}
  \BibitemShut {NoStop}%
\bibitem [{\citenamefont {Knoll}(2000)}]{Knollbook1981}%
  \BibitemOpen
  \bibfield  {author} {\bibinfo {author} {\bibfnamefont {G.~F.}\ \bibnamefont
  {Knoll}},\ }\href@noop {} {\emph {\bibinfo {title} {Radiation detection and
  measurement, Third edition}}}\ (\bibinfo  {publisher} {John Wiley \& Sons,
  Inc.},\ \bibinfo {year} {2000})\BibitemShut {NoStop}%
\bibitem [{\citenamefont {Roberts}\ \emph
  {et~al.}(2016{\natexlab{a}})\citenamefont {Roberts}, \citenamefont {Dzuba},
  \citenamefont {Flambaum}, \citenamefont {Pospelov},\ and\ \citenamefont
  {Stadnik}}]{Roberts:2016xfw}%
  \BibitemOpen
  \bibfield  {author} {\bibinfo {author} {\bibfnamefont {B.}~\bibnamefont
  {Roberts}}, \bibinfo {author} {\bibfnamefont {V.}~\bibnamefont {Dzuba}},
  \bibinfo {author} {\bibfnamefont {V.}~\bibnamefont {Flambaum}}, \bibinfo
  {author} {\bibfnamefont {M.}~\bibnamefont {Pospelov}},\ and\ \bibinfo
  {author} {\bibfnamefont {Y.}~\bibnamefont {Stadnik}},\ }\href
  {https://doi.org/10.1103/PhysRevD.93.115037} {\bibfield  {journal} {\bibinfo
  {journal} {Phys. Rev. D}\ }\textbf {\bibinfo {volume} {93}},\ \bibinfo
  {pages} {115037} (\bibinfo {year} {2016}{\natexlab{a}})},\ \Eprint
  {https://arxiv.org/abs/1604.04559} {arXiv:1604.04559 [hep-ph]} \BibitemShut
  {NoStop}%
\bibitem [{\citenamefont {Pandey}\ \emph {et~al.}(2020)\citenamefont {Pandey}
  \emph {et~al.}}]{Pandey:2018esq}%
  \BibitemOpen
  \bibfield  {author} {\bibinfo {author} {\bibfnamefont {M.~K.}\ \bibnamefont
  {Pandey}} \emph {et~al.},\ }\href
  {https://doi.org/10.1103/PhysRevD.102.123025} {\bibfield  {journal} {\bibinfo
   {journal} {Phys. Rev. D}\ }\textbf {\bibinfo {volume} {102}},\ \bibinfo
  {pages} {123025} (\bibinfo {year} {2020})},\ \Eprint
  {https://arxiv.org/abs/1812.11759} {arXiv:1812.11759 [hep-ph]} \BibitemShut
  {NoStop}%
\bibitem [{Git()}]{Git_results}%
  \BibitemOpen
  \href@noop {} {\bibinfo {title} {Atomic and molecular orbitals and form
  factors on {GitHub}}},\ \bibinfo {howpublished}
  {\url{https://github.com/mccabech/AtMolDM}}\BibitemShut {NoStop}%
\bibitem [{\citenamefont {Reid}\ and\ \citenamefont
  {Brunthaler}(2004)}]{Reid:2004rd}%
  \BibitemOpen
  \bibfield  {author} {\bibinfo {author} {\bibfnamefont {M.~J.}\ \bibnamefont
  {Reid}}\ and\ \bibinfo {author} {\bibfnamefont {A.}~\bibnamefont
  {Brunthaler}},\ }\href {https://doi.org/10.1086/424960} {\bibfield  {journal}
  {\bibinfo  {journal} {Astrophys. J.}\ }\textbf {\bibinfo {volume} {616}},\
  \bibinfo {pages} {872} (\bibinfo {year} {2004})},\ \Eprint
  {https://arxiv.org/abs/astro-ph/0408107} {arXiv:astro-ph/0408107}
  \BibitemShut {NoStop}%
\bibitem [{\citenamefont {{Gravity Collaboration}}\ \emph
  {et~al.}(2019)\citenamefont {{Gravity Collaboration}}, \citenamefont
  {{Abuter}} \emph {et~al.}}]{Gravity2019}%
  \BibitemOpen
  \bibfield  {author} {\bibinfo {author} {\bibnamefont {{Gravity
  Collaboration}}}, \bibinfo {author} {\bibfnamefont {R.}~\bibnamefont
  {{Abuter}}}, \emph {et~al.},\ }\href
  {https://doi.org/10.1051/0004-6361/201935656} {\bibfield  {journal} {\bibinfo
   {journal} {Astronomy \& Astrophysics}\ }\textbf {\bibinfo {volume} {625}},\
  \bibinfo {eid} {L10} (\bibinfo {year} {2019})},\ \Eprint
  {https://arxiv.org/abs/1904.05721} {arXiv:1904.05721 [astro-ph.GA]}
  \BibitemShut {NoStop}%
\bibitem [{\citenamefont {{Abuter}}\ \emph {et~al.}(2021)\citenamefont
  {{Abuter}} \emph {et~al.}}]{Gravity2021}%
  \BibitemOpen
  \bibfield  {author} {\bibinfo {author} {\bibfnamefont {R.}~\bibnamefont
  {{Abuter}}} \emph {et~al.} (\bibinfo {collaboration} {Gravity
  Collaboration}),\ }\href {https://doi.org/10.1051/0004-6361/202040208}
  {\bibfield  {journal} {\bibinfo  {journal} {Astronomy \& Astrophysics}\
  }\textbf {\bibinfo {volume} {647}},\ \bibinfo {eid} {A59} (\bibinfo {year}
  {2021})},\ \Eprint {https://arxiv.org/abs/2101.12098} {arXiv:2101.12098
  [astro-ph.GA]} \BibitemShut {NoStop}%
\bibitem [{\citenamefont {Baxter}\ \emph {et~al.}(2021)\citenamefont {Baxter}
  \emph {et~al.}}]{Baxter:2021pqo}%
  \BibitemOpen
  \bibfield  {author} {\bibinfo {author} {\bibfnamefont {D.}~\bibnamefont
  {Baxter}} \emph {et~al.},\ }\href
  {https://doi.org/10.1140/epjc/s10052-021-09655-y} {\bibfield  {journal}
  {\bibinfo  {journal} {Eur. Phys. J. C}\ }\textbf {\bibinfo {volume} {81}},\
  \bibinfo {pages} {907} (\bibinfo {year} {2021})},\ \Eprint
  {https://arxiv.org/abs/2105.00599} {arXiv:2105.00599 [hep-ex]} \BibitemShut
  {NoStop}%
\bibitem [{\citenamefont {{Deason}}\ \emph {et~al.}(2019)\citenamefont
  {{Deason}} \emph {et~al.}}]{Deason2019}%
  \BibitemOpen
  \bibfield  {author} {\bibinfo {author} {\bibfnamefont {A.~J.}\ \bibnamefont
  {{Deason}}} \emph {et~al.},\ }\href {https://doi.org/10.1093/mnras/stz623}
  {\bibfield  {journal} {\bibinfo  {journal} {MNRAS}\ }\textbf {\bibinfo
  {volume} {485}},\ \bibinfo {pages} {3514} (\bibinfo {year} {2019})},\ \Eprint
  {https://arxiv.org/abs/1901.02016} {arXiv:1901.02016 [astro-ph.GA]}
  \BibitemShut {NoStop}%
\bibitem [{\citenamefont {Schoenrich}\ \emph {et~al.}(2010)\citenamefont
  {Schoenrich}, \citenamefont {Binney},\ and\ \citenamefont
  {Dehnen}}]{Schoenrich:2009bx}%
  \BibitemOpen
  \bibfield  {author} {\bibinfo {author} {\bibfnamefont {R.}~\bibnamefont
  {Schoenrich}}, \bibinfo {author} {\bibfnamefont {J.}~\bibnamefont {Binney}},\
  and\ \bibinfo {author} {\bibfnamefont {W.}~\bibnamefont {Dehnen}},\ }\href
  {https://doi.org/10.1111/j.1365-2966.2010.16253.x} {\bibfield  {journal}
  {\bibinfo  {journal} {Mon. Not. Roy. Astron. Soc.}\ }\textbf {\bibinfo
  {volume} {403}},\ \bibinfo {pages} {1829} (\bibinfo {year} {2010})},\ \Eprint
  {https://arxiv.org/abs/0912.3693} {arXiv:0912.3693 [astro-ph.GA]}
  \BibitemShut {NoStop}%
\bibitem [{\citenamefont {McCabe}(2014)}]{McCabe:2013kea}%
  \BibitemOpen
  \bibfield  {author} {\bibinfo {author} {\bibfnamefont {C.}~\bibnamefont
  {McCabe}},\ }\href {https://doi.org/10.1088/1475-7516/2014/02/027} {\bibfield
   {journal} {\bibinfo  {journal} {JCAP}\ }\textbf {\bibinfo {volume} {02}},\
  \bibinfo {pages} {027}},\ \Eprint {https://arxiv.org/abs/1312.1355}
  {arXiv:1312.1355 [astro-ph.CO]} \BibitemShut {NoStop}%
\bibitem [{\citenamefont {O'Hare}\ \emph {et~al.}(2018)\citenamefont {O'Hare},
  \citenamefont {McCabe}, \citenamefont {Evans}, \citenamefont {Myeong},\ and\
  \citenamefont {Belokurov}}]{OHare:2018trr}%
  \BibitemOpen
  \bibfield  {author} {\bibinfo {author} {\bibfnamefont {C.~A.}\ \bibnamefont
  {O'Hare}}, \bibinfo {author} {\bibfnamefont {C.}~\bibnamefont {McCabe}},
  \bibinfo {author} {\bibfnamefont {N.~W.}\ \bibnamefont {Evans}}, \bibinfo
  {author} {\bibfnamefont {G.}~\bibnamefont {Myeong}},\ and\ \bibinfo {author}
  {\bibfnamefont {V.}~\bibnamefont {Belokurov}},\ }\href
  {https://doi.org/10.1103/PhysRevD.98.103006} {\bibfield  {journal} {\bibinfo
  {journal} {Phys. Rev. D}\ }\textbf {\bibinfo {volume} {98}},\ \bibinfo
  {pages} {103006} (\bibinfo {year} {2018})},\ \Eprint
  {https://arxiv.org/abs/1807.09004} {arXiv:1807.09004 [astro-ph.CO]}
  \BibitemShut {NoStop}%
\bibitem [{\citenamefont {Evans}\ \emph {et~al.}(2019)\citenamefont {Evans},
  \citenamefont {O'Hare},\ and\ \citenamefont {McCabe}}]{Evans:2018bqy}%
  \BibitemOpen
  \bibfield  {author} {\bibinfo {author} {\bibfnamefont {N.~W.}\ \bibnamefont
  {Evans}}, \bibinfo {author} {\bibfnamefont {C.~A.}\ \bibnamefont {O'Hare}},\
  and\ \bibinfo {author} {\bibfnamefont {C.}~\bibnamefont {McCabe}},\ }\href
  {https://doi.org/10.1103/PhysRevD.99.023012} {\bibfield  {journal} {\bibinfo
  {journal} {Phys. Rev. D}\ }\textbf {\bibinfo {volume} {99}},\ \bibinfo
  {pages} {023012} (\bibinfo {year} {2019})},\ \Eprint
  {https://arxiv.org/abs/1810.11468} {arXiv:1810.11468 [astro-ph.GA]}
  \BibitemShut {NoStop}%
\bibitem [{\citenamefont {O'Hare}\ \emph {et~al.}(2020)\citenamefont {O'Hare},
  \citenamefont {Evans}, \citenamefont {McCabe}, \citenamefont {Myeong},\ and\
  \citenamefont {Belokurov}}]{OHare:2019qxc}%
  \BibitemOpen
  \bibfield  {author} {\bibinfo {author} {\bibfnamefont {C.~A.}\ \bibnamefont
  {O'Hare}}, \bibinfo {author} {\bibfnamefont {N.~W.}\ \bibnamefont {Evans}},
  \bibinfo {author} {\bibfnamefont {C.}~\bibnamefont {McCabe}}, \bibinfo
  {author} {\bibfnamefont {G.}~\bibnamefont {Myeong}},\ and\ \bibinfo {author}
  {\bibfnamefont {V.}~\bibnamefont {Belokurov}},\ }\href
  {https://doi.org/10.1103/PhysRevD.101.023006} {\bibfield  {journal} {\bibinfo
   {journal} {Phys. Rev. D}\ }\textbf {\bibinfo {volume} {101}},\ \bibinfo
  {pages} {023006} (\bibinfo {year} {2020})},\ \Eprint
  {https://arxiv.org/abs/1909.04684} {arXiv:1909.04684 [astro-ph.GA]}
  \BibitemShut {NoStop}%
\bibitem [{\citenamefont {Buch}\ \emph {et~al.}(2020)\citenamefont {Buch},
  \citenamefont {Buen-Abad}, \citenamefont {Fan},\ and\ \citenamefont
  {Leung}}]{Buch:2020xyt}%
  \BibitemOpen
  \bibfield  {author} {\bibinfo {author} {\bibfnamefont {J.}~\bibnamefont
  {Buch}}, \bibinfo {author} {\bibfnamefont {M.~A.}\ \bibnamefont {Buen-Abad}},
  \bibinfo {author} {\bibfnamefont {J.}~\bibnamefont {Fan}},\ and\ \bibinfo
  {author} {\bibfnamefont {J.~S.~C.}\ \bibnamefont {Leung}},\ }\href
  {https://doi.org/10.1103/PhysRevD.102.083010} {\bibfield  {journal} {\bibinfo
   {journal} {Phys. Rev. D}\ }\textbf {\bibinfo {volume} {102}},\ \bibinfo
  {pages} {083010} (\bibinfo {year} {2020})},\ \Eprint
  {https://arxiv.org/abs/2007.13750} {arXiv:2007.13750 [hep-ph]} \BibitemShut
  {NoStop}%
\bibitem [{\citenamefont {Radick}\ \emph {et~al.}(2021)\citenamefont {Radick},
  \citenamefont {Taki},\ and\ \citenamefont {Yu}}]{Radick:2020qip}%
  \BibitemOpen
  \bibfield  {author} {\bibinfo {author} {\bibfnamefont {A.}~\bibnamefont
  {Radick}}, \bibinfo {author} {\bibfnamefont {A.-M.}\ \bibnamefont {Taki}},\
  and\ \bibinfo {author} {\bibfnamefont {T.-T.}\ \bibnamefont {Yu}},\ }\href
  {https://doi.org/10.1088/1475-7516/2021/02/004} {\bibfield  {journal}
  {\bibinfo  {journal} {JCAP}\ }\textbf {\bibinfo {volume} {02}},\ \bibinfo
  {pages} {004}},\ \Eprint {https://arxiv.org/abs/2011.02493} {arXiv:2011.02493
  [hep-ph]} \BibitemShut {NoStop}%
\bibitem [{\citenamefont {Sun}\ \emph {et~al.}(2017)\citenamefont {Sun} \emph
  {et~al.}}]{PYSCF2017}%
  \BibitemOpen
  \bibfield  {author} {\bibinfo {author} {\bibfnamefont {Q.}~\bibnamefont
  {Sun}} \emph {et~al.},\ }\href {https://doi.org/10.1002/wcms.1340} {\bibfield
   {journal} {\bibinfo  {journal} {WIREs Comput Mol Sci}\ }\textbf {\bibinfo
  {volume} {8}},\ \bibinfo {pages} {e1340} (\bibinfo {year}
  {2017})}\BibitemShut {NoStop}%
\bibitem [{\citenamefont {Sun}\ \emph {et~al.}(2020)\citenamefont {Sun} \emph
  {et~al.}}]{PYSCF2020}%
  \BibitemOpen
  \bibfield  {author} {\bibinfo {author} {\bibfnamefont {Q.}~\bibnamefont
  {Sun}} \emph {et~al.},\ }\href {https://doi.org/10.1063/5.0006074} {\bibfield
   {journal} {\bibinfo  {journal} {The Journal of Chemical Physics}\ }\textbf
  {\bibinfo {volume} {153}},\ \bibinfo {pages} {024109} (\bibinfo {year}
  {2020})}\BibitemShut {NoStop}%
\bibitem [{\citenamefont {Roberts}\ \emph
  {et~al.}(2016{\natexlab{b}})\citenamefont {Roberts}, \citenamefont
  {Flambaum},\ and\ \citenamefont {Gribakin}}]{Roberts:2015lga}%
  \BibitemOpen
  \bibfield  {author} {\bibinfo {author} {\bibfnamefont {B.}~\bibnamefont
  {Roberts}}, \bibinfo {author} {\bibfnamefont {V.}~\bibnamefont {Flambaum}},\
  and\ \bibinfo {author} {\bibfnamefont {G.}~\bibnamefont {Gribakin}},\ }\href
  {https://doi.org/10.1103/PhysRevLett.116.023201} {\bibfield  {journal}
  {\bibinfo  {journal} {Phys. Rev. Lett.}\ }\textbf {\bibinfo {volume} {116}},\
  \bibinfo {pages} {023201} (\bibinfo {year} {2016}{\natexlab{b}})},\ \Eprint
  {https://arxiv.org/abs/1509.09044} {arXiv:1509.09044 [physics.atom-ph]}
  \BibitemShut {NoStop}%
\bibitem [{\citenamefont {Roberts}\ and\ \citenamefont
  {Flambaum}(2019)}]{Roberts:2019chv}%
  \BibitemOpen
  \bibfield  {author} {\bibinfo {author} {\bibfnamefont {B.}~\bibnamefont
  {Roberts}}\ and\ \bibinfo {author} {\bibfnamefont {V.}~\bibnamefont
  {Flambaum}},\ }\href {https://doi.org/10.1103/PhysRevD.100.063017} {\bibfield
   {journal} {\bibinfo  {journal} {Phys. Rev. D}\ }\textbf {\bibinfo {volume}
  {100}},\ \bibinfo {pages} {063017} (\bibinfo {year} {2019})},\ \Eprint
  {https://arxiv.org/abs/1904.07127} {arXiv:1904.07127 [hep-ph]} \BibitemShut
  {NoStop}%
\bibitem [{\citenamefont {Kennedy}\ and\ \citenamefont
  {Manson}(1972)}]{Kennedy1972}%
  \BibitemOpen
  \bibfield  {author} {\bibinfo {author} {\bibfnamefont {D.~J.}\ \bibnamefont
  {Kennedy}}\ and\ \bibinfo {author} {\bibfnamefont {S.~T.}\ \bibnamefont
  {Manson}},\ }\href {https://doi.org/10.1103/PhysRevA.5.227} {\bibfield
  {journal} {\bibinfo  {journal} {Phys. Rev. A}\ }\textbf {\bibinfo {volume}
  {5}},\ \bibinfo {pages} {227} (\bibinfo {year} {1972})}\BibitemShut {NoStop}%
\bibitem [{\citenamefont {Woon}\ and\ \citenamefont
  {Dunning}(1994)}]{aug-cc-pv5Z-Woon1994}%
  \BibitemOpen
  \bibfield  {author} {\bibinfo {author} {\bibfnamefont {D.~E.}\ \bibnamefont
  {Woon}}\ and\ \bibinfo {author} {\bibfnamefont {T.~H.}\ \bibnamefont
  {Dunning}},\ }\href {https://doi.org/10.1063/1.466439} {\bibfield  {journal}
  {\bibinfo  {journal} {The Journal of Chemical Physics}\ }\textbf {\bibinfo
  {volume} {100}},\ \bibinfo {pages} {2975} (\bibinfo {year}
  {1994})}\BibitemShut {NoStop}%
\bibitem [{\citenamefont
  {Dunning}(1989)}]{aug-cc-pv5Z-Ne-doi:10.1063/1.456153}%
  \BibitemOpen
  \bibfield  {author} {\bibinfo {author} {\bibfnamefont {T.~H.}\ \bibnamefont
  {Dunning}},\ }\href {https://doi.org/10.1063/1.456153} {\bibfield  {journal}
  {\bibinfo  {journal} {The Journal of Chemical Physics}\ }\textbf {\bibinfo
  {volume} {90}},\ \bibinfo {pages} {1007} (\bibinfo {year}
  {1989})}\BibitemShut {NoStop}%
\bibitem [{\citenamefont {Ceolin}\ \emph {et~al.}(2013)\citenamefont {Ceolin},
  \citenamefont {de~Berr{\^e}do},\ and\ \citenamefont
  {Jorge}}]{Xe_qzp_Ceolin2013}%
  \BibitemOpen
  \bibfield  {author} {\bibinfo {author} {\bibfnamefont {G.~A.}\ \bibnamefont
  {Ceolin}}, \bibinfo {author} {\bibfnamefont {R.~C.}\ \bibnamefont
  {de~Berr{\^e}do}},\ and\ \bibinfo {author} {\bibfnamefont {F.~E.}\
  \bibnamefont {Jorge}},\ }\href {https://doi.org/10.1007/s00214-013-1339-7}
  {\bibfield  {journal} {\bibinfo  {journal} {Theoretical Chemistry Accounts}\
  }\textbf {\bibinfo {volume} {132}},\ \bibinfo {pages} {1339} (\bibinfo {year}
  {2013})}\BibitemShut {NoStop}%
\bibitem [{\citenamefont {Ditchfield}\ \emph {et~al.}(1971)\citenamefont
  {Ditchfield}, \citenamefont {Hehre},\ and\ \citenamefont
  {Pople}}]{ditchfield1971a}%
  \BibitemOpen
  \bibfield  {author} {\bibinfo {author} {\bibfnamefont {R.}~\bibnamefont
  {Ditchfield}}, \bibinfo {author} {\bibfnamefont {W.~J.}\ \bibnamefont
  {Hehre}},\ and\ \bibinfo {author} {\bibfnamefont {J.~A.}\ \bibnamefont
  {Pople}},\ }\href {https://doi.org/10.1063/1.1674902} {\bibfield  {journal}
  {\bibinfo  {journal} {J. Chem. Phys.}\ }\textbf {\bibinfo {volume} {54}},\
  \bibinfo {pages} {724} (\bibinfo {year} {1971})}\BibitemShut {NoStop}%
\bibitem [{\citenamefont {Hehre}\ \emph {et~al.}(1972)\citenamefont {Hehre},
  \citenamefont {Ditchfield},\ and\ \citenamefont {Pople}}]{hehre1972a}%
  \BibitemOpen
  \bibfield  {author} {\bibinfo {author} {\bibfnamefont {W.~J.}\ \bibnamefont
  {Hehre}}, \bibinfo {author} {\bibfnamefont {R.}~\bibnamefont {Ditchfield}},\
  and\ \bibinfo {author} {\bibfnamefont {J.~A.}\ \bibnamefont {Pople}},\ }\href
  {https://doi.org/10.1063/1.1677527} {\bibfield  {journal} {\bibinfo
  {journal} {J. Chem. Phys.}\ }\textbf {\bibinfo {volume} {56}},\ \bibinfo
  {pages} {2257} (\bibinfo {year} {1972})}\BibitemShut {NoStop}%
\bibitem [{\citenamefont {Hariharan}\ and\ \citenamefont
  {Pople}(1973)}]{hariharan1973a}%
  \BibitemOpen
  \bibfield  {author} {\bibinfo {author} {\bibfnamefont {P.~C.}\ \bibnamefont
  {Hariharan}}\ and\ \bibinfo {author} {\bibfnamefont {J.~A.}\ \bibnamefont
  {Pople}},\ }\href {https://doi.org/10.1007/bf00533485} {\bibfield  {journal}
  {\bibinfo  {journal} {Theor. Chim. Acta}\ }\textbf {\bibinfo {volume} {28}},\
  \bibinfo {pages} {213} (\bibinfo {year} {1973})}\BibitemShut {NoStop}%
\bibitem [{\citenamefont {Feller}(1996)}]{feller1996a}%
  \BibitemOpen
  \bibfield  {author} {\bibinfo {author} {\bibfnamefont {D.}~\bibnamefont
  {Feller}},\ }\href
  {https://doi.org/10.1002/(SICI)1096-987X(199610)17:13<1571::AID-JCC9>3.0.CO;2-P}
  {\bibfield  {journal} {\bibinfo  {journal} {J. Comput. Chem.}\ }\textbf
  {\bibinfo {volume} {17}},\ \bibinfo {pages} {1571} (\bibinfo {year}
  {1996})}\BibitemShut {NoStop}%
\bibitem [{\citenamefont {Schuchardt}\ \emph {et~al.}(2007)\citenamefont
  {Schuchardt}, \citenamefont {Didier}, \citenamefont {Elsethagen},
  \citenamefont {Sun}, \citenamefont {Gurumoorthi}, \citenamefont {Chase},
  \citenamefont {Li},\ and\ \citenamefont {Windus}}]{schuchardt2007a}%
  \BibitemOpen
  \bibfield  {author} {\bibinfo {author} {\bibfnamefont {K.~L.}\ \bibnamefont
  {Schuchardt}}, \bibinfo {author} {\bibfnamefont {B.~T.}\ \bibnamefont
  {Didier}}, \bibinfo {author} {\bibfnamefont {T.}~\bibnamefont {Elsethagen}},
  \bibinfo {author} {\bibfnamefont {L.}~\bibnamefont {Sun}}, \bibinfo {author}
  {\bibfnamefont {V.}~\bibnamefont {Gurumoorthi}}, \bibinfo {author}
  {\bibfnamefont {J.}~\bibnamefont {Chase}}, \bibinfo {author} {\bibfnamefont
  {J.}~\bibnamefont {Li}},\ and\ \bibinfo {author} {\bibfnamefont {T.~L.}\
  \bibnamefont {Windus}},\ }\href {https://doi.org/10.1021/ci600510j}
  {\bibfield  {journal} {\bibinfo  {journal} {J. Chem. Inf. Model.}\ }\textbf
  {\bibinfo {volume} {47}},\ \bibinfo {pages} {1045} (\bibinfo {year}
  {2007})}\BibitemShut {NoStop}%
\bibitem [{\citenamefont {Pritchard}\ \emph {et~al.}(2019)\citenamefont
  {Pritchard}, \citenamefont {Altarawy}, \citenamefont {Didier}, \citenamefont
  {Gibsom},\ and\ \citenamefont {Windus}}]{pritchard2019a}%
  \BibitemOpen
  \bibfield  {author} {\bibinfo {author} {\bibfnamefont {B.~P.}\ \bibnamefont
  {Pritchard}}, \bibinfo {author} {\bibfnamefont {D.}~\bibnamefont {Altarawy}},
  \bibinfo {author} {\bibfnamefont {B.}~\bibnamefont {Didier}}, \bibinfo
  {author} {\bibfnamefont {T.~D.}\ \bibnamefont {Gibsom}},\ and\ \bibinfo
  {author} {\bibfnamefont {T.~L.}\ \bibnamefont {Windus}},\ }\href
  {https://doi.org/10.1021/acs.jcim.9b00725} {\bibfield  {journal} {\bibinfo
  {journal} {J. Chem. Inf. Model.}\ }\textbf {\bibinfo {volume} {59}},\
  \bibinfo {pages} {4814} (\bibinfo {year} {2019})}\BibitemShut {NoStop}%
\bibitem [{\citenamefont {{NIST Computational Chemistry Comparison and
  Benchmark Database}}( III)}]{NISTCCCBDB}%
  \BibitemOpen
  \bibfield  {author} {\bibinfo {author} {\bibnamefont {{NIST Computational
  Chemistry Comparison and Benchmark Database}}},\ }\bibfield  {journal}
  {\bibinfo  {journal} {NIST Standard Reference Database Number 101}\ }\textbf
  {\bibinfo {volume} {http://cccbdb.nist.gov/}},\ \href
  {https://doi.org/10.18434/T47C7Z} {10.18434/T47C7Z} (\bibinfo {year} {Release
  21, August 2020, Editor: Russell D. Johnson III})\BibitemShut {NoStop}%
\bibitem [{\citenamefont {Thomas}(1970)}]{Darrah1970}%
  \BibitemOpen
  \bibfield  {author} {\bibinfo {author} {\bibfnamefont {T.~D.}\ \bibnamefont
  {Thomas}},\ }\href {https://doi.org/10.1063/1.1673140} {\bibfield  {journal}
  {\bibinfo  {journal} {The Journal of Chemical Physics}\ }\textbf {\bibinfo
  {volume} {52}},\ \bibinfo {pages} {1373} (\bibinfo {year}
  {1970})}\BibitemShut {NoStop}%
\bibitem [{\citenamefont {Bieri}\ and\ \citenamefont
  {Asbrink}(1980)}]{BIERI1980}%
  \BibitemOpen
  \bibfield  {author} {\bibinfo {author} {\bibfnamefont {G.}~\bibnamefont
  {Bieri}}\ and\ \bibinfo {author} {\bibfnamefont {L.}~\bibnamefont
  {Asbrink}},\ }\href
  {https://doi.org/https://doi.org/10.1016/0368-2048(80)85013-4} {\bibfield
  {journal} {\bibinfo  {journal} {Journal of Electron Spectroscopy and Related
  Phenomena}\ }\textbf {\bibinfo {volume} {20}},\ \bibinfo {pages} {149}
  (\bibinfo {year} {1980})}\BibitemShut {NoStop}%
\bibitem [{\citenamefont {Deng}\ \emph {et~al.}(2001)\citenamefont {Deng} \emph
  {et~al.}}]{Deng2001}%
  \BibitemOpen
  \bibfield  {author} {\bibinfo {author} {\bibfnamefont {J.~K.}\ \bibnamefont
  {Deng}} \emph {et~al.},\ }\href {https://doi.org/10.1063/1.1321313}
  {\bibfield  {journal} {\bibinfo  {journal} {The Journal of Chemical Physics}\
  }\textbf {\bibinfo {volume} {114}},\ \bibinfo {pages} {882} (\bibinfo {year}
  {2001})}\BibitemShut {NoStop}%
\bibitem [{\citenamefont {Williams}(2009)}]{lbl_xray_booklet}%
  \BibitemOpen
  \bibfield  {author} {\bibinfo {author} {\bibfnamefont {G.}~\bibnamefont
  {Williams}},\ }\href@noop {} {\emph {\bibinfo {title} {X-Ray Data Booklet,
  Sec.(1.1)}}}\ (\bibinfo  {publisher} {Lawrence Berkeley National
  Laboratory},\ \bibinfo {address} {Lawrence Berkeley National Laboratory,
  Berkeley CA 94720, U.S.A.},\ \bibinfo {year} {2009})\BibitemShut {NoStop}%
\bibitem [{\citenamefont {Bunge}\ \emph {et~al.}(1993)\citenamefont {Bunge},
  \citenamefont {Barrientos},\ and\ \citenamefont {Bunge}}]{BUNGE1993113}%
  \BibitemOpen
  \bibfield  {author} {\bibinfo {author} {\bibfnamefont {C.}~\bibnamefont
  {Bunge}}, \bibinfo {author} {\bibfnamefont {J.}~\bibnamefont {Barrientos}},\
  and\ \bibinfo {author} {\bibfnamefont {A.}~\bibnamefont {Bunge}},\ }\href
  {https://doi.org/https://doi.org/10.1006/adnd.1993.1003} {\bibfield
  {journal} {\bibinfo  {journal} {Atomic Data and Nuclear Data Tables}\
  }\textbf {\bibinfo {volume} {53}},\ \bibinfo {pages} {113 } (\bibinfo {year}
  {1993})}\BibitemShut {NoStop}%
\bibitem [{\citenamefont {Fischer}(2000)}]{ASTP_FISCHER2000635}%
  \BibitemOpen
  \bibfield  {author} {\bibinfo {author} {\bibfnamefont {C.~F.}\ \bibnamefont
  {Fischer}},\ }\href
  {https://doi.org/https://doi.org/10.1016/S0010-4655(00)00009-6} {\bibfield
  {journal} {\bibinfo  {journal} {Computer Physics Communications}\ }\textbf
  {\bibinfo {volume} {128}},\ \bibinfo {pages} {635 } (\bibinfo {year}
  {2000})}\BibitemShut {NoStop}%
\bibitem [{\citenamefont {Brundle}\ \emph {et~al.}(1970)\citenamefont
  {Brundle}, \citenamefont {Robin},\ and\ \citenamefont {Basch}}]{Brundle1970}%
  \BibitemOpen
  \bibfield  {author} {\bibinfo {author} {\bibfnamefont {C.~R.}\ \bibnamefont
  {Brundle}}, \bibinfo {author} {\bibfnamefont {M.~B.}\ \bibnamefont {Robin}},\
  and\ \bibinfo {author} {\bibfnamefont {H.}~\bibnamefont {Basch}},\ }\href
  {https://doi.org/10.1063/1.1674313} {\bibfield  {journal} {\bibinfo
  {journal} {The Journal of Chemical Physics}\ }\textbf {\bibinfo {volume}
  {53}},\ \bibinfo {pages} {2196} (\bibinfo {year} {1970})}\BibitemShut
  {NoStop}%
\bibitem [{\citenamefont {Hamel}\ \emph {et~al.}(2002)\citenamefont {Hamel},
  \citenamefont {Duffy}, \citenamefont {Casida},\ and\ \citenamefont
  {Salahub}}]{HAMEL2002}%
  \BibitemOpen
  \bibfield  {author} {\bibinfo {author} {\bibfnamefont {S.}~\bibnamefont
  {Hamel}}, \bibinfo {author} {\bibfnamefont {P.}~\bibnamefont {Duffy}},
  \bibinfo {author} {\bibfnamefont {M.~E.}\ \bibnamefont {Casida}},\ and\
  \bibinfo {author} {\bibfnamefont {D.~R.}\ \bibnamefont {Salahub}},\ }\href
  {https://doi.org/https://doi.org/10.1016/S0368-2048(02)00032-4} {\bibfield
  {journal} {\bibinfo  {journal} {Journal of Electron Spectroscopy and Related
  Phenomena}\ }\textbf {\bibinfo {volume} {123}},\ \bibinfo {pages} {345}
  (\bibinfo {year} {2002})}\BibitemShut {NoStop}%
\bibitem [{\citenamefont {Grant}(2007)}]{GrantBook}%
  \BibitemOpen
  \bibinfo {editor} {\bibfnamefont {I.~P.}\ \bibnamefont {Grant}},\ ed.,\ \href
  {https://doi.org/10.1007/978-0-387-35069-1} {\emph {\bibinfo {title}
  {Relativistic Quantum Theory of Atoms and Molecules}}}\ (\bibinfo
  {publisher} {Springer New York},\ \bibinfo {year} {2007})\BibitemShut
  {NoStop}%
\bibitem [{\citenamefont {Jonsson}\ \emph {et~al.}(2007)\citenamefont
  {Jonsson}, \citenamefont {He}, \citenamefont {Froese~Fischer},\ and\
  \citenamefont {Grant}}]{Jonsson:2007}%
  \BibitemOpen
  \bibfield  {author} {\bibinfo {author} {\bibfnamefont {P.}~\bibnamefont
  {Jonsson}}, \bibinfo {author} {\bibfnamefont {X.}~\bibnamefont {He}},
  \bibinfo {author} {\bibfnamefont {C.}~\bibnamefont {Froese~Fischer}},\ and\
  \bibinfo {author} {\bibfnamefont {I.~P.}\ \bibnamefont {Grant}},\ }\href
  {https://doi.org/https://doi.org/10.1016/j.cpc.2007.06.002} {\bibfield
  {journal} {\bibinfo  {journal} {Comput. Phys. Commun.}\ }\textbf {\bibinfo
  {volume} {177}},\ \bibinfo {pages} {597} (\bibinfo {year}
  {2007})}\BibitemShut {NoStop}%
\bibitem [{\citenamefont {Froese~Fischer}\ \emph {et~al.}(2019)\citenamefont
  {Froese~Fischer}, \citenamefont {Gaigalas}, \citenamefont {Jönsson},\ and\
  \citenamefont {Bieroń}}]{Fischer:2019}%
  \BibitemOpen
  \bibfield  {author} {\bibinfo {author} {\bibfnamefont {C.}~\bibnamefont
  {Froese~Fischer}}, \bibinfo {author} {\bibfnamefont {G.}~\bibnamefont
  {Gaigalas}}, \bibinfo {author} {\bibfnamefont {P.}~\bibnamefont {Jönsson}},\
  and\ \bibinfo {author} {\bibfnamefont {J.}~\bibnamefont {Bieroń}},\ }\href
  {https://doi.org/https://doi.org/10.1016/j.cpc.2018.10.032} {\bibfield
  {journal} {\bibinfo  {journal} {Comput. Phys. Commun.}\ }\textbf {\bibinfo
  {volume} {237}},\ \bibinfo {pages} {184} (\bibinfo {year}
  {2019})}\BibitemShut {NoStop}%
\bibitem [{\citenamefont {Quiney}\ \emph {et~al.}(1998)\citenamefont {Quiney},
  \citenamefont {Skaane},\ and\ \citenamefont {Grant}}]{Quiney:1998}%
  \BibitemOpen
  \bibfield  {author} {\bibinfo {author} {\bibfnamefont {H.~M.}\ \bibnamefont
  {Quiney}}, \bibinfo {author} {\bibfnamefont {H.}~\bibnamefont {Skaane}},\
  and\ \bibinfo {author} {\bibfnamefont {I.~P.}\ \bibnamefont {Grant}},\ }\href
  {https://doi.org/https://doi.org/10.1016/S0065-3276(08)60405-0} {\bibfield
  {journal} {\bibinfo  {journal} {Adv. Quantum Chem.}\ }\textbf {\bibinfo
  {volume} {32}},\ \bibinfo {pages} {1} (\bibinfo {year} {1998})}\BibitemShut
  {NoStop}%
\bibitem [{\citenamefont {{Wolfram Research{,} Inc.}}(2021)}]{Mathematica}%
  \BibitemOpen
  \bibfield  {author} {\bibinfo {author} {\bibnamefont {{Wolfram Research{,}
  Inc.}}},\ }\href {https://www.wolfram.com/mathematica} {\bibinfo {title}
  {Mathematica, {V}ersion 12.3.1}} (\bibinfo {year} {2021})\BibitemShut
  {NoStop}%
\bibitem [{\citenamefont {Bates}\ and\ \citenamefont
  {Seaton}(1949)}]{Bates1949}%
  \BibitemOpen
  \bibfield  {author} {\bibinfo {author} {\bibfnamefont {D.~R.}\ \bibnamefont
  {Bates}}\ and\ \bibinfo {author} {\bibfnamefont {M.~J.}\ \bibnamefont
  {Seaton}},\ }\href {https://doi.org/10.1093/mnras/109.6.698} {\bibfield
  {journal} {\bibinfo  {journal} {Monthly Notices of the Royal Astronomical
  Society}\ }\textbf {\bibinfo {volume} {109}},\ \bibinfo {pages} {698}
  (\bibinfo {year} {1949})}\BibitemShut {NoStop}%
\bibitem [{\citenamefont {Amusia}\ and\ \citenamefont
  {Chernysheva}(1997)}]{Amusiabook1997}%
  \BibitemOpen
  \bibfield  {author} {\bibinfo {author} {\bibfnamefont {M.~Y.}\ \bibnamefont
  {Amusia}}\ and\ \bibinfo {author} {\bibfnamefont {L.~V.}\ \bibnamefont
  {Chernysheva}},\ }\href {https://doi.org/10.1201/9781003040859} {\emph
  {\bibinfo {title} {Computation of Atomic Processes}}}\ (\bibinfo  {publisher}
  {IoP Publishing Ltd},\ \bibinfo {year} {1997})\BibitemShut {NoStop}%
\bibitem [{\citenamefont {Cowan}(1967)}]{Cowan1967}%
  \BibitemOpen
  \bibfield  {author} {\bibinfo {author} {\bibfnamefont {R.~D.}\ \bibnamefont
  {Cowan}},\ }\href {https://doi.org/10.1103/PhysRev.163.54} {\bibfield
  {journal} {\bibinfo  {journal} {Phys. Rev.}\ }\textbf {\bibinfo {volume}
  {163}},\ \bibinfo {pages} {54} (\bibinfo {year} {1967})}\BibitemShut
  {NoStop}%
\bibitem [{\citenamefont {Cowan}(1981)}]{Cowanbook1981}%
  \BibitemOpen
  \bibfield  {author} {\bibinfo {author} {\bibfnamefont {R.~D.}\ \bibnamefont
  {Cowan}},\ }\href@noop {} {\emph {\bibinfo {title} {The theory of atomic
  structure and spectra}}}\ (\bibinfo  {publisher} {University of California
  Press},\ \bibinfo {year} {1981})\BibitemShut {NoStop}%
\bibitem [{\citenamefont {Condon}\ and\ \citenamefont
  {Odabasi}(1980)}]{Condonbook1980}%
  \BibitemOpen
  \bibfield  {author} {\bibinfo {author} {\bibfnamefont {E.}~\bibnamefont
  {Condon}}\ and\ \bibinfo {author} {\bibfnamefont {H.}~\bibnamefont
  {Odabasi}},\ }\href@noop {} {\emph {\bibinfo {title} {Atomic Structure}}}\
  (\bibinfo  {publisher} {Cambridge University Press},\ \bibinfo {year}
  {1980})\BibitemShut {NoStop}%
\bibitem [{\citenamefont {Catena}\ \emph {et~al.}(2020)\citenamefont {Catena},
  \citenamefont {Emken}, \citenamefont {Spaldin},\ and\ \citenamefont
  {Tarantino}}]{Catena:2019gfa}%
  \BibitemOpen
  \bibfield  {author} {\bibinfo {author} {\bibfnamefont {R.}~\bibnamefont
  {Catena}}, \bibinfo {author} {\bibfnamefont {T.}~\bibnamefont {Emken}},
  \bibinfo {author} {\bibfnamefont {N.~A.}\ \bibnamefont {Spaldin}},\ and\
  \bibinfo {author} {\bibfnamefont {W.}~\bibnamefont {Tarantino}},\ }\href
  {https://doi.org/10.1103/PhysRevResearch.2.033195} {\bibfield  {journal}
  {\bibinfo  {journal} {Phys. Rev. Res.}\ }\textbf {\bibinfo {volume} {2}},\
  \bibinfo {pages} {033195} (\bibinfo {year} {2020})},\ \Eprint
  {https://arxiv.org/abs/1912.08204} {arXiv:1912.08204 [hep-ph]} \BibitemShut
  {NoStop}%
\bibitem [{\citenamefont {Jain}(1986)}]{Jain1986}%
  \BibitemOpen
  \bibfield  {author} {\bibinfo {author} {\bibfnamefont {A.}~\bibnamefont
  {Jain}},\ }\href {https://doi.org/10.1103/PhysRevA.34.3707} {\bibfield
  {journal} {\bibinfo  {journal} {Phys. Rev. A}\ }\textbf {\bibinfo {volume}
  {34}},\ \bibinfo {pages} {3707} (\bibinfo {year} {1986})}\BibitemShut
  {NoStop}%
\bibitem [{\citenamefont {Baluja}\ \emph {et~al.}(1992)\citenamefont {Baluja},
  \citenamefont {Jain}, \citenamefont {Martino},\ and\ \citenamefont
  {Gianturco}}]{Baluja_1992}%
  \BibitemOpen
  \bibfield  {author} {\bibinfo {author} {\bibfnamefont {K.~L.}\ \bibnamefont
  {Baluja}}, \bibinfo {author} {\bibfnamefont {A.}~\bibnamefont {Jain}},
  \bibinfo {author} {\bibfnamefont {V.~D.}\ \bibnamefont {Martino}},\ and\
  \bibinfo {author} {\bibfnamefont {F.~A.}\ \bibnamefont {Gianturco}},\ }\href
  {https://doi.org/10.1209/0295-5075/17/2/010} {\bibfield  {journal} {\bibinfo
  {journal} {Europhysics Letters ({EPL})}\ }\textbf {\bibinfo {volume} {17}},\
  \bibinfo {pages} {139} (\bibinfo {year} {1992})}\BibitemShut {NoStop}%
\bibitem [{\citenamefont {Das}\ \emph {et~al.}(2014)\citenamefont {Das},
  \citenamefont {Stauffer},\ and\ \citenamefont {Srivastava}}]{Das2014}%
  \BibitemOpen
  \bibfield  {author} {\bibinfo {author} {\bibfnamefont {T.}~\bibnamefont
  {Das}}, \bibinfo {author} {\bibfnamefont {A.~D.}\ \bibnamefont {Stauffer}},\
  and\ \bibinfo {author} {\bibfnamefont {R.}~\bibnamefont {Srivastava}},\
  }\href {https://doi.org/10.1140/epjd/e2014-40713-7} {\bibfield  {journal}
  {\bibinfo  {journal} {The European Physical Journal D}\ }\textbf {\bibinfo
  {volume} {68}},\ \bibinfo {pages} {102} (\bibinfo {year} {2014})}\BibitemShut
  {NoStop}%
\bibitem [{\citenamefont {Er-Jun}\ \emph {et~al.}(2007)\citenamefont {Er-Jun},
  \citenamefont {Yu-Gang}, \citenamefont {Xiang-Zhou}, \citenamefont {De-Qing},
  \citenamefont {Wen-Qing},\ and\ \citenamefont {Wen-Dong}}]{Er_Jun_2007}%
  \BibitemOpen
  \bibfield  {author} {\bibinfo {author} {\bibfnamefont {M.}~\bibnamefont
  {Er-Jun}}, \bibinfo {author} {\bibfnamefont {M.}~\bibnamefont {Yu-Gang}},
  \bibinfo {author} {\bibfnamefont {C.}~\bibnamefont {Xiang-Zhou}}, \bibinfo
  {author} {\bibfnamefont {F.}~\bibnamefont {De-Qing}}, \bibinfo {author}
  {\bibfnamefont {S.}~\bibnamefont {Wen-Qing}},\ and\ \bibinfo {author}
  {\bibfnamefont {T.}~\bibnamefont {Wen-Dong}},\ }\href
  {https://doi.org/10.1088/1009-1963/16/11/033} {\bibfield  {journal} {\bibinfo
   {journal} {Chinese Physics}\ }\textbf {\bibinfo {volume} {16}},\ \bibinfo
  {pages} {3339} (\bibinfo {year} {2007})}\BibitemShut {NoStop}%
\bibitem [{\citenamefont {Mahato}\ \emph {et~al.}(2019)\citenamefont {Mahato},
  \citenamefont {Sharma}, \citenamefont {Stauffer},\ and\ \citenamefont
  {Srivastava}}]{Mahato2019}%
  \BibitemOpen
  \bibfield  {author} {\bibinfo {author} {\bibfnamefont {D.}~\bibnamefont
  {Mahato}}, \bibinfo {author} {\bibfnamefont {L.}~\bibnamefont {Sharma}},
  \bibinfo {author} {\bibfnamefont {A.~D.}\ \bibnamefont {Stauffer}},\ and\
  \bibinfo {author} {\bibfnamefont {R.}~\bibnamefont {Srivastava}},\ }\href
  {https://doi.org/10.1140/epjd/e2019-100060-5} {\bibfield  {journal} {\bibinfo
   {journal} {The European Physical Journal D}\ }\textbf {\bibinfo {volume}
  {73}},\ \bibinfo {pages} {189} (\bibinfo {year} {2019})}\BibitemShut
  {NoStop}%
\bibitem [{\citenamefont {Galassi}\ \emph {et~al.}(2021)\citenamefont {Galassi}
  \emph {et~al.}}]{GSL}%
  \BibitemOpen
  \bibfield  {author} {\bibinfo {author} {\bibfnamefont {M.}~\bibnamefont
  {Galassi}} \emph {et~al.},\ }\href {https://doi.org/10.5281/zenodo.1146014}
  {\bibinfo {title} {{GNU Scientific Library Reference Manual (3rd Ed.)}}}
  (\bibinfo {year} {2021})\BibitemShut {NoStop}%
\bibitem [{\citenamefont {Johnson}(2020)}]{Cubature}%
  \BibitemOpen
  \bibfield  {author} {\bibinfo {author} {\bibfnamefont {S.}~\bibnamefont
  {Johnson}},\ }\href {https://github.com/stevengj/cubature} {\bibinfo {title}
  {{Cubature v1.0.4}}} (\bibinfo {year} {2020})\BibitemShut {NoStop}%
\bibitem [{\citenamefont {Tange}(2022)}]{tange_2022_6891516}%
  \BibitemOpen
  \bibfield  {author} {\bibinfo {author} {\bibfnamefont {O.}~\bibnamefont
  {Tange}},\ }\href {https://doi.org/10.5281/zenodo.6891516} {\bibinfo {title}
  {{GNU Parallel 20220722 (`Roe vs Wade')}}} (\bibinfo {year}
  {2022})\BibitemShut {NoStop}%
\bibitem [{\citenamefont {Kudryavtsev}\ \emph {et~al.}(2003)\citenamefont
  {Kudryavtsev}, \citenamefont {Spooner},\ and\ \citenamefont
  {McMillan}}]{Kudryavtsev:2003aua}%
  \BibitemOpen
  \bibfield  {author} {\bibinfo {author} {\bibfnamefont {V.~A.}\ \bibnamefont
  {Kudryavtsev}}, \bibinfo {author} {\bibfnamefont {N.~J.~C.}\ \bibnamefont
  {Spooner}},\ and\ \bibinfo {author} {\bibfnamefont {J.~E.}\ \bibnamefont
  {McMillan}},\ }\href {https://doi.org/10.1016/S0168-9002(03)00983-5}
  {\bibfield  {journal} {\bibinfo  {journal} {Nucl. Instrum. Meth. A}\ }\textbf
  {\bibinfo {volume} {505}},\ \bibinfo {pages} {688} (\bibinfo {year}
  {2003})},\ \Eprint {https://arxiv.org/abs/hep-ex/0303007}
  {arXiv:hep-ex/0303007} \BibitemShut {NoStop}%
\bibitem [{\citenamefont {Smith}\ \emph {et~al.}(2005)\citenamefont {Smith},
  \citenamefont {Snowden-Ifft}, \citenamefont {Smith}, \citenamefont
  {Luscher},\ and\ \citenamefont {Lewin}}]{Smith:2005se}%
  \BibitemOpen
  \bibfield  {author} {\bibinfo {author} {\bibfnamefont {P.~F.}\ \bibnamefont
  {Smith}}, \bibinfo {author} {\bibfnamefont {D.}~\bibnamefont {Snowden-Ifft}},
  \bibinfo {author} {\bibfnamefont {N.~J.~T.}\ \bibnamefont {Smith}}, \bibinfo
  {author} {\bibfnamefont {R.}~\bibnamefont {Luscher}},\ and\ \bibinfo {author}
  {\bibfnamefont {J.~D.}\ \bibnamefont {Lewin}},\ }\href
  {https://doi.org/10.1016/j.astropartphys.2004.09.005} {\bibfield  {journal}
  {\bibinfo  {journal} {Astropart. Phys.}\ }\textbf {\bibinfo {volume} {22}},\
  \bibinfo {pages} {409} (\bibinfo {year} {2005})}\BibitemShut {NoStop}%
\bibitem [{\citenamefont {Malczewski}\ \emph {et~al.}(2013)\citenamefont
  {Malczewski}, \citenamefont {Kisiel},\ and\ \citenamefont
  {Dorda}}]{Malczewski:2013lqy}%
  \BibitemOpen
  \bibfield  {author} {\bibinfo {author} {\bibfnamefont {D.}~\bibnamefont
  {Malczewski}}, \bibinfo {author} {\bibfnamefont {J.}~\bibnamefont {Kisiel}},\
  and\ \bibinfo {author} {\bibfnamefont {J.}~\bibnamefont {Dorda}},\ }\href
  {https://doi.org/10.1007/s10967-013-2540-9} {\bibfield  {journal} {\bibinfo
  {journal} {J. Radioanal. Nucl. Chem.}\ }\textbf {\bibinfo {volume} {298}},\
  \bibinfo {pages} {1483} (\bibinfo {year} {2013})}\BibitemShut {NoStop}%
\bibitem [{\citenamefont {Brossard}(2020)}]{brossard:tel-02923528}%
  \BibitemOpen
  \bibfield  {author} {\bibinfo {author} {\bibfnamefont {A.}~\bibnamefont
  {Brossard}},\ }\emph {\bibinfo {title} {{Optimization of spherical
  proportional counter backgrounds and response for low mass dark matter
  search}}},\ \href {https://tel.archives-ouvertes.fr/tel-02923528} {\bibinfo
  {type} {Theses}},\ \bibinfo  {school} {{Universit{\'e} Paris-Saclay; Queen's
  University at Kingston}} (\bibinfo {year} {2020})\BibitemShut {NoStop}%
\bibitem [{\citenamefont {Agostinelli}\ \emph {et~al.}(2003)\citenamefont
  {Agostinelli} \emph {et~al.}}]{GEANT4}%
  \BibitemOpen
  \bibfield  {author} {\bibinfo {author} {\bibfnamefont {S.}~\bibnamefont
  {Agostinelli}} \emph {et~al.} (\bibinfo {collaboration} {GEANT4}),\ }\href
  {https://doi.org/10.1016/S0168-9002(03)01368-8} {\bibfield  {journal}
  {\bibinfo  {journal} {Nucl. Instrum. Meth. A}\ }\textbf {\bibinfo {volume}
  {506}},\ \bibinfo {pages} {250} (\bibinfo {year} {2003})}\BibitemShut
  {NoStop}%
\bibitem [{\citenamefont {Amare}\ \emph {et~al.}(2018)\citenamefont {Amare}
  \emph {et~al.}}]{Amare:2017roa}%
  \BibitemOpen
  \bibfield  {author} {\bibinfo {author} {\bibfnamefont {J.}~\bibnamefont
  {Amare}} \emph {et~al.},\ }\href
  {https://doi.org/10.1016/j.astropartphys.2017.11.004} {\bibfield  {journal}
  {\bibinfo  {journal} {Astropart. Phys.}\ }\textbf {\bibinfo {volume} {97}},\
  \bibinfo {pages} {96} (\bibinfo {year} {2018})},\ \Eprint
  {https://arxiv.org/abs/1706.05818} {arXiv:1706.05818 [physics.ins-det]}
  \BibitemShut {NoStop}%
\bibitem [{\citenamefont {Adari}\ \emph {et~al.}(2022)\citenamefont {Adari}
  \emph {et~al.}}]{Fuss:2022fxe}%
  \BibitemOpen
  \bibfield  {author} {\bibinfo {author} {\bibfnamefont {P.}~\bibnamefont
  {Adari}} \emph {et~al.},\ }\href
  {https://doi.org/10.21468/SciPostPhysProc.9.001} {\bibfield  {journal}
  {\bibinfo  {journal} {SciPost Phys. Proc.}\ }\textbf {\bibinfo {volume}
  {9}},\ \bibinfo {pages} {001} (\bibinfo {year} {2022})},\ \Eprint
  {https://arxiv.org/abs/2202.05097} {arXiv:2202.05097 [astro-ph.IM]}
  \BibitemShut {NoStop}%
\bibitem [{\citenamefont {Combecher}(1980)}]{Combecher1980}%
  \BibitemOpen
  \bibfield  {author} {\bibinfo {author} {\bibfnamefont {D.}~\bibnamefont
  {Combecher}},\ }\href {https://doi.org/10.2307/3575293} {\bibfield  {journal}
  {\bibinfo  {journal} {Radiation Research}\ }\textbf {\bibinfo {volume}
  {84}},\ \bibinfo {pages} {189} (\bibinfo {year} {1980})}\BibitemShut
  {NoStop}%
\bibitem [{\citenamefont {Parks}\ \emph {et~al.}(1972)\citenamefont {Parks},
  \citenamefont {Hurst}, \citenamefont {Stewart},\ and\ \citenamefont
  {Weidner}}]{Parks1972}%
  \BibitemOpen
  \bibfield  {author} {\bibinfo {author} {\bibfnamefont {J.~E.}\ \bibnamefont
  {Parks}}, \bibinfo {author} {\bibfnamefont {G.~S.}\ \bibnamefont {Hurst}},
  \bibinfo {author} {\bibfnamefont {T.~E.}\ \bibnamefont {Stewart}},\ and\
  \bibinfo {author} {\bibfnamefont {H.~L.}\ \bibnamefont {Weidner}},\ }\href
  {https://doi.org/10.1063/1.1678247} {\bibfield  {journal} {\bibinfo
  {journal} {The Journal of Chemical Physics}\ }\textbf {\bibinfo {volume}
  {57}},\ \bibinfo {pages} {5467} (\bibinfo {year} {1972})}\BibitemShut
  {NoStop}%
\bibitem [{\citenamefont {Dalgarno}\ \emph {et~al.}(1999)\citenamefont
  {Dalgarno}, \citenamefont {Yan},\ and\ \citenamefont {Liu}}]{Dalgarno1999}%
  \BibitemOpen
  \bibfield  {author} {\bibinfo {author} {\bibfnamefont {A.}~\bibnamefont
  {Dalgarno}}, \bibinfo {author} {\bibfnamefont {M.}~\bibnamefont {Yan}},\ and\
  \bibinfo {author} {\bibfnamefont {W.}~\bibnamefont {Liu}},\ }\href
  {https://doi.org/10.1086/313267} {\bibfield  {journal} {\bibinfo  {journal}
  {The Astrophysical Journal Supplement Series}\ }\textbf {\bibinfo {volume}
  {125}},\ \bibinfo {pages} {237} (\bibinfo {year} {1999})}\BibitemShut
  {NoStop}%
\bibitem [{\citenamefont {Katsioulas}\ \emph
  {et~al.}(2022{\natexlab{b}})\citenamefont {Katsioulas}, \citenamefont
  {Knights},\ and\ \citenamefont {Nikolopoulos}}]{Katsioulas:2021sgl}%
  \BibitemOpen
  \bibfield  {author} {\bibinfo {author} {\bibfnamefont {I.}~\bibnamefont
  {Katsioulas}}, \bibinfo {author} {\bibfnamefont {P.}~\bibnamefont
  {Knights}},\ and\ \bibinfo {author} {\bibfnamefont {K.}~\bibnamefont
  {Nikolopoulos}},\ }\href
  {https://doi.org/10.1016/j.astropartphys.2022.102707} {\bibfield  {journal}
  {\bibinfo  {journal} {Astropart. Phys.}\ }\textbf {\bibinfo {volume} {141}},\
  \bibinfo {pages} {102707} (\bibinfo {year} {2022}{\natexlab{b}})},\ \Eprint
  {https://arxiv.org/abs/2105.01414} {arXiv:2105.01414 [hep-ex]} \BibitemShut
  {NoStop}%
\bibitem [{\citenamefont {Vinagre}\ and\ \citenamefont
  {Conde}(2000)}]{Vinagre2000}%
  \BibitemOpen
  \bibfield  {author} {\bibinfo {author} {\bibfnamefont {F.~L.~R.}\
  \bibnamefont {Vinagre}}\ and\ \bibinfo {author} {\bibfnamefont {C.~A.~N.}\
  \bibnamefont {Conde}},\ }\href {https://doi.org/10.1063/1.1290710} {\bibfield
   {journal} {\bibinfo  {journal} {Journal of Applied Physics}\ }\textbf
  {\bibinfo {volume} {88}},\ \bibinfo {pages} {5426} (\bibinfo {year}
  {2000})}\BibitemShut {NoStop}%
\bibitem [{\citenamefont {Cowan}\ \emph {et~al.}(2011)\citenamefont {Cowan},
  \citenamefont {Cranmer}, \citenamefont {Gross},\ and\ \citenamefont
  {Vitells}}]{Cowan:2010js}%
  \BibitemOpen
  \bibfield  {author} {\bibinfo {author} {\bibfnamefont {G.}~\bibnamefont
  {Cowan}}, \bibinfo {author} {\bibfnamefont {K.}~\bibnamefont {Cranmer}},
  \bibinfo {author} {\bibfnamefont {E.}~\bibnamefont {Gross}},\ and\ \bibinfo
  {author} {\bibfnamefont {O.}~\bibnamefont {Vitells}},\ }\href
  {https://doi.org/10.1140/epjc/s10052-011-1554-0} {\bibfield  {journal}
  {\bibinfo  {journal} {Eur. Phys. J. C}\ }\textbf {\bibinfo {volume} {71}},\
  \bibinfo {pages} {1554} (\bibinfo {year} {2011})},\ \bibinfo {note}
  {[Erratum: Eur.Phys.J.C 73, 2501 (2013)]},\ \Eprint
  {https://arxiv.org/abs/1007.1727} {arXiv:1007.1727 [physics.data-an]}
  \BibitemShut {NoStop}%
\bibitem [{\citenamefont {Cheng}\ \emph {et~al.}(2021)\citenamefont {Cheng}
  \emph {et~al.}}]{PandaX-II:2021nsg}%
  \BibitemOpen
  \bibfield  {author} {\bibinfo {author} {\bibfnamefont {C.}~\bibnamefont
  {Cheng}} \emph {et~al.} (\bibinfo {collaboration} {PandaX-II}),\ }\href
  {https://doi.org/10.1103/PhysRevLett.126.211803} {\bibfield  {journal}
  {\bibinfo  {journal} {Phys. Rev. Lett.}\ }\textbf {\bibinfo {volume} {126}},\
  \bibinfo {pages} {211803} (\bibinfo {year} {2021})},\ \Eprint
  {https://arxiv.org/abs/2101.07479} {arXiv:2101.07479 [hep-ex]} \BibitemShut
  {NoStop}%
\bibitem [{\citenamefont {Abramowitz}\ and\ \citenamefont
  {Stegun}(1972)}]{Abramowitzbook}%
  \BibitemOpen
  \bibfield  {author} {\bibinfo {author} {\bibfnamefont {M.}~\bibnamefont
  {Abramowitz}}\ and\ \bibinfo {author} {\bibfnamefont {I.~A.}\ \bibnamefont
  {Stegun}},\ }\href@noop {} {\emph {\bibinfo {title} {Handbook of Mathematical
  Functions with Formulas, Graphs, and Mathematical Tables}}}\ (\bibinfo
  {publisher} {United States Department of Commerce, National Bureau of
  Standards (NBS)},\ \bibinfo {year} {1972})\BibitemShut {NoStop}%
\bibitem [{QED()}]{QEDark_Git}%
  \BibitemOpen
  \href@noop {} {\bibinfo {title} {{QEDark} package}},\ \bibinfo {howpublished}
  {\url{https://github.com/tientienyu/QEdark}}\BibitemShut {NoStop}%
\end{thebibliography}%
\bibliographystyle{apsrev4-2}

\end{document}